\documentclass[60pt]{article}

\usepackage{amsmath,amssymb,amsthm,amscd,mathtools, graphicx, mathtext,color,wrapfig, verbatim, slashed, float}
\usepackage{tikz-cd}
\usepackage{bbm}

\usepackage{mathrsfs}
\usepackage{pifont}
\usepackage{graphicx}
\usepackage[font={small,sl}]{caption}
\usepackage[all]{xy}
\usepackage{tikz}
\usetikzlibrary{matrix,arrows}

\usepackage[numbers,sort&compress]{natbib}
\usepackage[enableskew]{youngtab}
\usepackage{ytableau}
\usepackage{thmtools}

\newcommand{\svdots}{%
  \vbox{
    \scriptsize \baselineskip 2pt \lineskiplimit 0pt
    \hbox {.}\hbox {.}\hbox {.}\kern-0.75pt
  }%
}

\makeatletter

\@addtoreset{equation}{section}
\makeatother

\newcommand{\beq}{\begin{equation}}
\newcommand{\eeq}{\end{equation}}

\newcommand{\Vertex}{\mathsf{Vertex}}

\newcommand{\nc}{\newcommand}
\nc{\mc}{\mathcal}
\DeclareMathAlphabet{\mathpzc}{OT1}{pzc}{m}{it}

\newcommand{\scF}{\ensuremath{\mathcal{F}}}

\newcommand{\scQ}{\ensuremath{\mathcal{Q}}}

\newcommand{\scX}{\ensuremath{\mathcal{X}}}

\newcommand{\fg}{\mathfrak{g}}

\newcommand{\fq}{\mathfrak{q}}

\newcommand{\fC}{\mathfrak{C}}

\newcommand{\Lfgh}{\widehat{^L\mathfrak{g}}}
\newtheorem{theorem}{Theorem}
\newtheorem{conjecture}[theorem]{Conjecture}
\newenvironment{theorem*}
 {\expandafter\def\expandafter\thetheorem\expandafter{\thetheorem*}\theorem}
 {\endtheorem}

\begin{document}
\baselineskip=28pt  % a la harvmac
\baselineskip 0.7cm
\setcounter{tocdepth}{2}

\begin{titlepage}

%% Set the number of the title with 0

% change the footnote symbol
\renewcommand{\thefootnote}{\fnsymbol{footnote}}

\vskip 1.0cm

\begin{center}
{\LARGE \bf
Knot Categorification from
\vskip 0.5 cm
Mirror Symmetry
\vskip 0.5cm}
{\it \large
%of WRT Knot Invariants\\ 
Part I:  Coherent Sheaves
\vskip 0.5cm }
{\large
Mina Aganagic}

\medskip

\vskip 0.5cm

{\it
Center for Theoretical Physics, University of California, Berkeley\\
Department of Mathematics, University of California, Berkeley\\
%$^3$Department of Mathematics, Columbia University
}

\end{center}

\vskip 0.5cm

%-----------------------------------------
\centerline{{\bf Abstract}}
\medskip
We derive two geometric approaches to categorification of quantum invariants of links associated to an arbitrary compact simple Lie group $^L{G}$.  In part I, we describe the first approach, based on an equivariant derived category of coherent sheaves on ${\cal X}$, the moduli space of singular $G$-monopoles, where $G$ is related to $^LG$ by Langlands duality. In part II, we describe the second approach, based on the derived category of a Fukaya-Seidel category of a Calabi-Yau $Y$ with potential $W$. The two approaches are related by a version of mirror symmetry, which plays a crucial role in the story.  In part III, we explain the string theory origin of these results, and the relation to an approach due to Witten.

\noindent\end{titlepage}
\setcounter{page}{1} % don't number title page

\setcounter{section}{0}
%%%%%%%%%\\
\tableofcontents 
\newpage

\section{Introduction}

The problem of categorifying quantum knot invariants associated to a Lie algebra $^L\fg$ has been around since
Khovanov's pioneering work \cite{Kh} on categorification of the Jones polynomial. The problem is to find a unified approach to categorification of the corresponding $U_{\fq}(^L\fg)$  quantum group knot invariants. Ideally, such approach will have connection to physics, or at least, to geometry.\footnote{It is an open question to understand the relation to the algebraic approach of \cite{webster}.}

This is the first of the sequence of three papers in which we put forward two such approaches, and explain how they emerge from string theory. Both approaches are based in geometry, not algebra. Perhaps the most important aspect of the two approaches is the fact that, unlike in typical approaches to categorification where one comes up with a category and then works to prove that its decategorification leads to invariants one aimed to categorify, in both of our approaches, the second step is manifest.  

In this paper, we will discuss the first of these two approaches. This approach is based on derived categories of coherent sheaves of a variety
${\cal X}$ which may be described as the moduli space of monopoles, as the intersection of slices in affine Grassmannian, or as the Coulomb branch of a certain 3d ${\cal N}=4$ quiver gauge theory. The second paper in the series \cite{A2} deals with the second approach, related to the first two dimensional equivariant mirror symmetry, appropriately defined. Mirror symmetry, and techniques developed to study it play a key role in both of our approaches. The first approach shares basic flavors of earlier works of Kamnitzer and Cautis \cite{CK1, CK2}, for knots colored by the defining representation of $^L\fg = \mathfrak{sl}_n$. The second approach is a cousin of the approach by Seidel and Smith \cite{SS}, for $^L\fg = \mathfrak{sl}_2$, who pioneered such geometric approaches, but produced a theory not sufficiently rich.
The third paper in the series \cite{A3} explains the string theory origin our two approaches, and relation to another approach to the same problem, due to Witten \cite{WF, WK, WJ, GW}.  
All three approaches have the same string theory origin, so what emerges is a unified framework for knot categorification.

\subsection{The first approach}

Link invariants based on the $U_{\fq}(^L\fg)$ quantum group are matrix elements of monodromy matrices of a Knizhnik-Zamolodchikov (KZ) equation, whose solutions are conformal blocks of the affine Lie algebra $\Lfgh_{\kappa}$.
 %The monodromy matrices are given in terms of $U_{\fq}(^L\fg)$ quantum group $R$-matrices, thanks so works of \cite{}.
The starting point for us is a geometric interpretation of the KZ equation and it solutions. In most of this paper, $^L\fg$ is simply laced as the general case requires an extra step \cite{A2, A3}.

\subsubsection{}

In the first approach, the Knizhnik-Zamolodchikov equation is realized geometrically as the quantum differential equation of ${\cal X}$, a result proven recently in \cite{Danilenko}. 
It follows that solutions of the KZ equation, the conformal blocks of ${\Lfgh}_{\kappa}$, arise from geometry as Givental's $J$-functions of ${\cal X}$ \cite{G}, also known as the cohomological vertex functions, in terminology of \cite{OK, OICM}.
Vertex functions are defined in terms of equivariant counts of holomorphic maps from the domain curve ${\rm D}$, which is best thought of as infinitely long cigar, to ${\cal X}$, with suitable boundary conditions at infinity and insertions of point observables at the tip. The equivariant action scales the symplectic form of ${\cal X}$ by a parameter related to $\kappa$. 
\subsubsection{}
 
The vertex function has a physical interpretation as the partition function of the supersymmetric sigma model with target ${\cal X}$. In the interior of ${\rm D}$ one has an A-type twist, but at the boundary at infinity, one imposes a B-type boundary condition. The choice of the boundary condition is a B-brane on ${\cal X}$, or more precisely, of an object 
$$
{\cal F} \in D^bCoh_{\rm T}({\cal X}), 
$$
of the derived category of ${\rm T}$-equivariant coherent sheaves on ${\cal X}$. While the vertex function originates from the A-model on ${\cal X}$, it computes a (generalized) central charge of the B-brane ${\cal F}$ defining the boundary condition on $\partial {\rm D}$. 

\subsubsection{}

From the sigma model interpretation of conformal blocks it follows that the matrix elements of the $U_{\fq}(^L\fg)$ braiding matrix
are amplitudes of the topological B-model on an annulus, with target ${\cal X}$ and a pair of B-type branes  ${\cal G}$, and ${\mathscr B} {\cal F}$ on the two boundaries. The branes ${\cal F}$ and ${\cal G}$ determine the ``in"
 and ``out" conformal blocks as their vertex functions, and ${\mathscr B}$ is the derived equivalence functor corresponding to the braid. 
By cutting open the annulus into a strip, it also follows that the graded Hom 
\beq\label{bas}Hom^{*, *}({\cal G}, {\mathscr B} {\cal F}),
\eeq
computed in ${\mathscr D}_{\cal X}=D^bCoh_{\rm T}({\cal X})$, categorifies the $U_{\fq}(^L\fg)$ braiding matrix element. The bigrading on the Hom's comes from the usual homological grading of the derived category, and from the equivariant grading.  
The precise statement is theorem \ref{t:three}, which 
may be viewed as a
corollary of a theorem of \cite{BO} identifying the derived auto-equivalence functor ${\mathscr B}$ of ${\mathscr D}_{\cal X}$ with the
categorical
lift of the monodromy of the quantum differential equation of ${\cal X}$.
The $U_{\fq}(^L\fg)$ braiding matrix element is the Euler characteristic $\chi({\cal G},  {\mathscr B} {\cal F})$ of \eqref{bas}, which depends only on the K-theory classes of the branes ${\cal G}$ and ${\mathscr B} {\cal F}$.

\subsubsection{} 
In principle, there are many different braid group actions on ${\mathscr D}_{\cal X}$. The supersymmetric sigma model origin of the construction fixes the which derived equivalence 
functor ${\mathscr B}$ we get.

%as a functor that categorifies the action of monodromy of the quantum differential equation on ${K_{{\rm T}}({\cal X})}$.

We conjecture that the functor ${\mathscr B}$ comes from variation of stability conditions along the path $B$, with central charge function ${\cal Z}^0: K({\cal X}) \rightarrow {\mathbb C}$ which is a close, but simpler, cousin, of the vertex function of ${\cal X}$. The stability condition that comes from ${\cal Z}^0$  is known as the Pi-stability \cite{Pi1, Pi2, Douglas}. In general, understanding variations of Pi stability is difficult, since the central charge function has very complicated dependence on Kahler moduli. In our case, this simplifies dramatically since ${\cal X}$ is holomorphic symplectic. 

The fact that central charge has conformal field theory origin will help us construct, fairly explicitly, the derived equivalences ${\mathscr B}$ which come from braiding a pair of vertex operators. A famous result in conformal field theory is that fusion diagonalizes braiding. The action of ${\mathscr B}$ on the derived category ${\mathscr D}_{\cal X}$ cannot be diagonalized. Instead, it leads to a filtration of ${\mathscr D}_{\cal X}$ on which ${\mathscr B}$ acts by a perverse equivalence of Chuang and Rouquier \cite{CR}, and which we spell out explicitly. 

\subsubsection{}
By representing a link $K$ as a braid whose  top and bottom are closed off by a sequence of cups and caps, its quantum invariant becomes a very special element of the corresponding $U_{\fq}(^L\fg)$ braiding matrix. The conformal block that picks out the matrix element describes pairs of vertex operators, colored by complex conjugate representations, which fuse together to copies of identity. It is a specific eigenvector of braiding matrices that exchange the paired endpoints. This special conformal block originates from a very special brane ${\cal U}\in {\mathscr D}_{\cal X}$, which we identify. The brane ${\cal U}$ is an eigen-sheaf of braiding functors and the structure sheaf of a vanishing cycle in ${\cal X}$.

In this way, to every link $K$, we associate homology groups $H^{*,*}({\cal U}, {\mathscr B} {\cal U})$ which categorify the corresponding $U_{\fq}(^L\fg)$ invariant, by theorem \ref{t:three}. For example, ``mirror symmetry" relating $U_{\fq}(^L\fg)$ invariants of a link and its mirror image, follows from a basic property of ${\mathscr D}_{\cal X}$, which is Serre duality. The homology groups $H^{*,*}({\cal U}, {\mathscr B} {\cal U})$ are themselves invariants of link isotopy. This is theorem \ref{t:four} whose proof relies on the special nature of derived equivalences in our case, namely, that they are perverse equivalences and come from variation of stability condition on ${\mathscr D}_{\cal X}$, with respect to ${\cal Z}^0$.  
%\subsubsection{}

One limitation of our approach is that we restrict to knots colored by minuscule representations of $^L\fg$. Otherwise, ${\cal X}$ develops singularities due to monopole bubbling. 
%Our framework contains hints of what to do in the general case.
  
\subsection{Organization}

In section \ref{s-one}, we review the relation of conformal blocks to knot invariants.  In section \ref{s-two}, we describe the geometric realization of conformal blocks, based on vertex functions of the equivariant A-model on ${\cal X}$. In section \ref{s-s}, we describe how the physical interpretation of the vertex function, coming from the supersymmetric sigma model on ${\cal X}$ leads to categorification of $U_{\fq}(^L\fg)$ braid invariants, based on ${\mathscr D}_{\cal X}$, the category of B-branes on ${\cal X}$. In section \ref{s-three}, we use the sigma model origin of conformal blocks to work out the action of braiding on ${\mathscr D}_{\cal X}$. In section \ref{s-four} we apply this to get a construction of homological knot invariants. In the appendix, we review the relation of affine Grassmannians to monopole moduli spaces, and collect some other results about the slices in the affine Grassmannians needed in the text. 

%A disclaimer: While we have tried to write a paper whose ideas can be followed by both mathematicians and physicists, this is a physics paper. Theorems whose proofs are rigorous only by physics standards are marked with $^*$. 

 \subsection{Acknowledgments}
This project grew out of previous joint work with Andrei Okounkov.  Andrei played a crucial role by, among other things, explaining to me his work with Bezrukavnikov. Without discussions with him, this paper would not have been the same. I am also grateful to Vivek Shende for many discussions and help with the second paper in the series, and for collaboration on a joint project with Michael McBreen; to Mohammed Abouzaid, who explained the pitfalls of previous symplectic geometry based approaches to the problem; to Ivan Danilenko for making available an advanced copy of his thesis; to Dimitrii Galakhov for collaboration in early stages of \cite{A2}; to Roman Bezrukavnikov, Catharina Stroppel, Ben Webster and Edward Witten fo helpful discussions. My research is supported, in part, by the NSF foundation grant PHY1820912, by the Simons Investigator Award, and by the Berkeley Center for Theoretical Physics.

\newpage
 
\section{Conformal Blocks and Knot Invariants}\label{s-one}

Chern-Simons theory with gauge group based on Lie algebra $^L\fg$, at level $k$, is the physical framework which leads to quantum link invariants. This was discovered in a paper by Witten \cite{Jones} which also explains that, underlying Chern-Simons theory is a two dimensional conformal field theory with affine $\Lfgh_k$-algebra symmetry.
The quantum link invariants one gets in this way turn out to be matrix elements of $R$-matrices of the
$U_{\fq}(^L\!\fg)$ quantum group \cite{RT}. The relation between conformal field theory and quantum link invariants serves as the starting point for our paper. In this section we will review the relevant aspects of it. 

\subsection{${\Lfgh}_{k}$ conformal blocks and the KZ equation}
Let ${\cal A}$ be a Riemann surface, which we take to be 
\beq\label{RS}
{\cal A}\cong {\mathbb C}^{\times} \cong \textup{infinite cylinder},
\eeq
with coordinate $y$ and punctures at $y=0$, and $y=\infty$. Pick a collection of $n$ generic points on ${\cal A}$ with coordinates $y=a_1, \ldots , a_n$. To a point $y= a_i$ assign a finite dimensional representation $V_{i}$ of $^L\fg$, of highest weight $\mu_i$.

\subsubsection{}
\begin{figure}[!hbtp]
  \centering
   \includegraphics[scale=0.37]{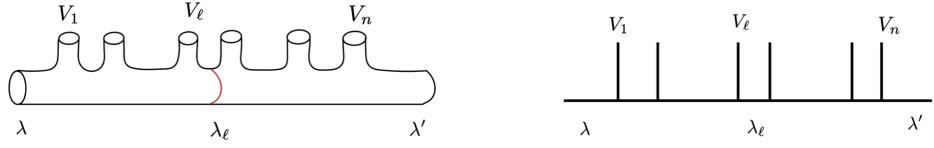}
 \caption{Two representations of a punctured Riemann surface ${\cal A}$. The one on the right spells out the sewing prescriptions, by summing over states in intermediate Verma module representations.}
  \label{f_blocks}
\end{figure}

Conformal blocks of
${\Lfgh}_{k}$ on ${\cal A}$ are chiral correlation functions \cite{KZ}%
\beq\label{electric}
{\cal V}(a_1, \ldots, a_{\ell},\ldots, a_n) = \langle \lambda| \;{ \Phi}_{V_1} (a_{1})\cdots  \Phi_{V_\ell}(a_{\ell}) \cdots { \Phi}_{V_n} (a_{n})    \;|\lambda'\rangle,
\eeq
of vertex operators $\Phi_{V_\ell}(a_{\ell})$, each labeled by the point where it is inserted, and the corresponding representation.
 The states $ | \lambda\rangle$ and $|\lambda' \rangle$ are the highest weight vectors of Verma module representations of $^L\fg$ associated to punctures at $y=0$ and $\infty$. Chiral vertex operators act as intertwiners % \beq\label{chiral}
between pairs of intermediate Verma module representations, see figure \ref{f_blocks}.

Conformal blocks in \eqref{electric} 
%where the level $k$ is related to $\hbar = \exp(2\pi i/\kappa).$
%The state $|\lambda\rangle
%$ is the highest weight vector in a level $k$ Verma module. 
%I%ts weight $\lambda \in {^L\fh}^*$ is an element of the dual of the
%Cartan subalgebra for $^L{\fg}$. 
 take values
% \beq\label{values}
%{\cal V}(a_1,  \ldots ,a_n)  \in (V_{1}\otimes \ldots \otimes {V_{n}})_{\nu},
%\eeq
 in the
subspace of representation 
\beq\label{representation}
V = \bigotimes_{i=1}^n V_i,
\eeq
of a fixed weight $\nu$, satisfying
\beq\label{weightsub}
\textup{weight    } \nu =\lambda-\lambda'.
\eeq
One should view $\nu$ and $\lambda$ as fixed, and then \eqref{weightsub} determines the $\lambda'$, the highest weight of the Verma module at $y=\infty$. For a thorough review see \cite{MS, RCFT, EFK}.

\subsubsection{}\label{ss:KZ}
A good way to think about the conformal blocks in \eqref{electric} is as solutions to a differential equation,
discovered by Knizhnik and Zamolodchikov \cite{KZ}:
\beq\label{KZ}
\kappa\, a_{\ell}{\partial \over \partial a_{\ell}} \,{\cal V} = 
\sum_{j \neq \ell} {r_{\ell i}}(a_\ell/a_j)\, {\cal V}.
\eeq
%`
Above, $\kappa = k+h^{\vee}$, where $h^{\vee}$ is the dual Coxeter number of $^L{\fg}$.
The classical $R$-matrices $r_{ij}$ are
given by
 $$
r_{ij}(a_i/a_j) = \frac{r_{ij} a_i+r_{ji} a_j}{a_i-a_j},
$$
where $r_{ij}$ denotes the action of
\beq\label{smallr}
r= {1\over 2} \sum_a {^L\!\,h}_a\otimes {^L\!\,h}_a + \sum_{\alpha>0}
{^L\!\,e}_{\alpha}\otimes {^L\!\,e}_{-\alpha},
\eeq
in the standard Lie theory notation, on the $i$-th and $j$-th factor in $V$.  The summation in \eqref{KZ} is over all punctures on ${\cal A}$, including those at $y=0$ and $\infty$. 
The matrix $r_{ij}(a_i/a_j)$ is the trigonometric $R$-matrix, as opposed to the rational one, that would have corresponded to ${\cal A} = {\mathbb C}$. The right hand side of \eqref{KZ} acts irreducibly for each fixed weight subspace, since it commutes with gauge transformations that act diagonally on the punctures.

Fix 
a specific ordering ${\vec \mu} = (\mu_5, \mu_2, \mu_7, \ldots)$ of vertex operators  in 
\eqref{electric} or, equivalently, a region of the form 
\beq\label{ordering}
|a_5| < |a_2| <|a_7| <  \dots.
\eeq
Conformal blocks, obtained by sewing chiral vertex operators in the order specified by ${\vec \mu}$ are solutions to the KZ equation analytic in the chamber \eqref{ordering}. 
The KZ equation and its monodromies make sense for any $k \in {\mathbb C}$, so give analytic continuation of Chern-Simons link invariants from integer to complex $k$.

\subsection{$U_{\fq}(^L\fg)$  quantum braid and link invariants}\label{s:knoti}

We get a braid $B$ with $n$ strands colored by representations $V_i$ by varying the positions of vertex operators $a_i = a_i(s)$ as a function of "time" $s \in [0,1]$.  This leads to a monodromy problem, which is to analytically continue the solutions to the KZ equation along the path corresponding to $B$. (We will be referring to parallel transport along any path, not necessarily a closed path, as monodromy.)

%Often it is convenient to arrange the path $B$ to be a closed loop in the configuration space of $n$ points on ${\cal A}$, colored by representations $V_i$.
%
\begin{figure}[!hbtp]
  \centering
   \includegraphics[scale=0.21]{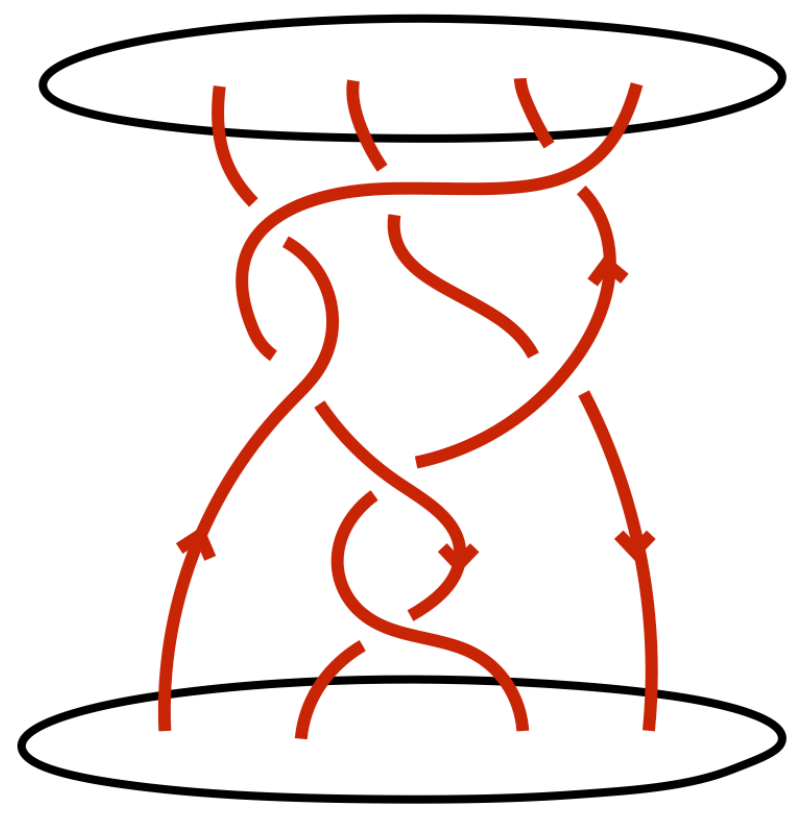}
 \caption{The braid in ${\cal A} \times [0,1]$ obtained by varying the positions of
vertex operators on ${\cal A}.$  Reversing the orientation of a strand acts by replacing the $^L\fg$ representation coloring it by its conjugate.}
  \label{f_braid}
\end{figure}
Monodromy along a path depends only on its homotopy type so the corresponding monodromy matrix ${\mathfrak B}$ is an invariant of the braid $B$ under smooth isotopy, which depends only on $k$, $^L{\fg}$ and the chosen representations. 

\subsubsection{
}

Monodromy problem of the KZ equation was solved in the works of  Tsuchyia and Kanie \cite{TK}, Kohno \cite{Koh}, Drinfeld \cite{Drinfeld} and Kazhdan and Luzsztig \cite{KL}. They showed that monodromy matrices of the $\Lfgh_k$ KZ equation are computed in terms of $R$-matrices of the $U_{\fq}(^L{\fg})$ quantum group. The level $k$ of the affine Lie algebra and the parameter $\fq$ of the quantum group are related as
\beq\label{qk}
\fq= e^{2\pi i\over k+h^{\vee}}.
\eeq
Monodromy of the KZ equation along the braid $B$ determines a $U_{\fq}(^L{\fg})$ matrix ${\mathfrak B}$, 
as a product over the $R$-matrices associated with individual crossings, since braid composition maps to matrix multiplication.

Via the action of monodromies, the space of conformal blocks becomes a module for $U_{\fq}(^L{\fg})$ quantum group. For generic $\kappa \in {\mathbb C}$, the dimension of the $U_{\fq}(^L{\fg})$ representation is the same as the dimension of  $^L\fg$ representation the conformal blocks take values in. In particular, the monodromy acts irreducibly only in the subspace of fixed weight $\nu.$

\subsubsection{}
Any link $K$ can be obtained as a closure of a braid $B$ with $2m$ strands by a collection of $m$ cups and $m$ caps,
at the top and at the bottom. 

\begin{figure}[!hbtp]
  \centering{
   \includegraphics[scale=0.37
  ]{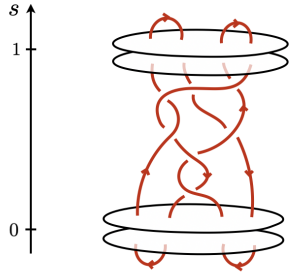}}
 \caption{Figure eight knot as a closure of a braid. }
  \label{f_knot}
\end{figure}  
The quantum invariant of the link $K$ described in this way is the matrix element 
\beq\label{KI}
{\cal J}_{K}({\fq}) = ({\mathfrak U}_1, {\mathfrak B} \,{\mathfrak U}_0), 
\eeq
of the $U_{\fg}(^L\fg)$ braiding matrix ${\mathfrak B}$ between very special conformal blocks ${\mathfrak U}_0$ and ${\mathfrak U}_1$. These conformal blocks  describe pairs of vertex operators, colored by complex conjugate representations, which fuse together and dissapear. %

Conformal blocks where operators $\Phi_{V_i}(a_i)$ and $\Phi_{V_j}(a_j)$ fuse to $\Phi_{V_k}(a_j)$ are
eigenvectors of braiding 
$$
{\mathfrak B}_{ij} :\Phi_{V_i}(a_i) \otimes \Phi_{V_j}(a_j) \rightarrow \Phi_{V_j}(a_j) \otimes \Phi_{V_i}(a_i).
$$
The possible $\Phi_{V_k}(a_j)$ are labeled by distinct representations occurring in tensor product
$$
{V_i} \otimes {V_j}  = \oplus_k  {V_k},
$$
(we suppress the multiplicities, since they are not relevant for us).
 A cap colored by representation $V_i$ corresponds to a conformal block where we bring $\Phi_{V_i}(a_{i})$ and $\Phi_{V_i^\star}(a_{j})$ together and fuse them the identity operator, corresponding to the trivial representation. 

\begin{figure}[!hbtp]
\centering
 \includegraphics[scale=0.13]{inout.pdf}
\caption{Conformal block ${\mathfrak U}$.}
 \label{f_inout}
\end{figure}
The matrix element in \eqref{KI} is per construction an invariant of the braid. The fact that it is also an invariant of the link $K$, independent of the presentation we chose, is a consequence of relations between fusion and braiding in conformal field theory. We will return to this in sections \ref{s-three} and \ref{s-four}.

%
%It is not difficult to see that we can take such closures to be plat closures, corresponds to pairing of consecutive strands, and moreover, we can take the braid $B$ to be a closed loop in configuration space of $2m$ colored points on ${\cal A}$, as in figure \ref{f_knot}.

\subsubsection{}
The choice of the Riemann surface ${\cal A}$ which is an infinite cylinder may seem odd, since to get quantum invariants of links in ${\mathbb R}^3$, working on ${\cal A}$ which is a complex plane would have been simpler. 

For homological invariants of links, taking ${\cal A}$ to be the infinite cylinder turns out to be the natural starting point. This is where the connection to geometry of ${\cal X}$, to mirror symmetry and to string theory most naturally takes place. For this reason, even if we are after invariants of links in ${\mathbb R}^3$, we will want to think of ${\cal A}$ as an infinite cylinder until the very end. The fact that our Riemann surface is a cylinder and not a plane does mean that, for free, we also get homological invariants of links in ${\cal A} \times {\mathbb R} = {\mathbb R}^2\times S^1$.

\subsubsection{}\label{ss-fram}
Quantum knot invariants are invariants of framed knots \cite{Jones}. Framing is a choice of a normal vector field to the knot, which one can picture as a thickening of the knot to a ribbon. All our invariants depend on such a choice.  A change of framing that adds a twist to a strand colored by representation $V_i$ multiplies the invariant ${\cal J}_{K}({\fq })$ by 
\beq\label{framing} e^{\pm 2\pi i h_i} = {\fq}^{\pm {1\over 2} \langle \mu_i, \mu_i +2 ^L\rho\rangle }
\eeq
where $h_i$ is the conformal dimension of the chiral operator $\Phi_{V_i}$, and $^L\rho$ is the Weyl vector of $^L\fg$. The sign depends on the direction of the twist.
The choice of framing implicitly made in writing the monodromy matrices of the KZ equation corresponds to the "vertical" framing \cite{Jones}, where the framing vector field associated to each strand is lifted out of the plane of the paper.  

\subsubsection{}\label{s-group}
So far only the Lie algebra $^L\fg$ of the Chern-Simons gauge group $^LG$ entered. The choice of a group $^LG$ restricts the representations $V_i$ that can appear. 
The highest weight vectors of representations of $^LG$ are elements of the character lattice of $^LG$, which is a sub-lattice of the weight lattice of its Lie algebra $^L{\fg}$. If $^LG$ is simply connected, its character lattice coincides the weight lattice. At the other extreme, if $^LG$ is of adjoint type, its character lattice is the root lattice of $^L{\fg}$.

From now on, we will assume that $^LG$ is simply connected so that its character lattice is as large as possible, and any dominant weight $\mu_i$ of $^L\fg$ can in principle appear as the highest weight of an $^LG$ representation $V_i$. We can get all other types from this by restriction.

 \section{A Geometric Origin of Conformal Blocks}\label{s-two}
 
In this section, we will show that conformal blocks have a geometric interpretation. %We will get all the structure needed for categorification. 
This is a direct consequence of the geometric realization of the Knizhnik-Zamolodchikov equation itself, which was very recently discovered in \cite{Danilenko}. The Knizhnik-Zamolodchikov equation, as written in \eqref{KZ}, is the quantum differential equation of a certain very special manifold ${\cal X}$.  We will learn more from this than just a reinterpretation of conformal blocks in terms of geometry, since the physics origin of this, as we will explain in the next section, is a two-dimensional supersymmetric sigma model with target ${\cal X}$. 
 
The quantum differential equation of a Kahler manifold ${\cal X}$
\beq\label{qdif}
 a_i { \partial \over \partial a_i} {\cal V} -  C_i\star {\cal V} =0
\eeq
is an equation for flat sections of a connection on a vector bundle over the complexified Kahler moduli space of ${\cal X}$, with fibers $H^*({\cal X})$ (for review see \cite{mirrorbook}). It is defined in terms of quantum multiplication by divisors $C_i \in H^2({\cal X})$. The quantum product on $H^*({\cal X})$,
originates from the 
topological A-model of ${\cal X}$, 
\beq\label{qmulti}
\langle \alpha \star \beta, \gamma\rangle = \sum_{d\geq 0, d\in H_2({\cal X})} (\alpha, \beta, \gamma)_d \; a^d,
\eeq
as its three-point function on the sphere with observables corresponding to classes $\alpha, \beta, \gamma \in H^*({\cal X})$ inserted. The first, $d=0$ term is given by the classical cup product on ${\cal X}$, the higher order terms come from instantons. 
The quantum differential equation says that inserting a divisor class corresponds to
differentiating with respect to Kahler moduli, $a_i { \partial \over \partial a_i} a^d = (C_i, d) a^d$. The 
flatness of the connection follows from WWDVV equations \cite{wt, WDVV, Dubrovin}.  
 
Just as Knizhnik-Zamolodchikov equation is central to many questions in representation theory, the quantum differential equation is central to many questions in algebraic geometry, and in mirror symmetry. For us the two equations coincide, so we end up with a deep new connection between geometry and representation theory.

\subsection{The geometry}\label{ss:coulomb}

Let $^L{\fg}$ be a simply laced Lie algebra, so it is of ADE type, and the same as its Langlands dual
$$
^L{\fg} = {\fg}.
$$
We will return to the general case in \cite{A2, A3}. Recall we are taking $^LG$ to be simply connected, and choose $V_i$ to be minuscule representations of $^L{\fg}$. A minuscule representation is also a fundamental representation. The $^L \fg$ representation in which the conformal blocks take value is
\beq\label{rmu}
{V}=\bigotimes_{ i=1}^n \, {V}_i= \bigotimes_{a=1}^{\rm rk} \,V_a^{\otimes
  m_a} \,,
\eeq
where the integers $m_a$ count the number of times the $a$-th fundamental representation of $^L\fg$ appears, and ${\rm rk}={{\rm rk}^L{\fg}}$. The highest weight of the representation $V$ is 
\beq\label{highest w}
  \textup{highest weight         } \mu = \sum_{i=1}^n \mu_i=
\sum_{a=1}^{\rm rk} m_a\, ^Lw_a,
\eeq
where $^Lw_a$ is the highest weight of $a$'th fundamental representation $V_a$.  The weight $\nu$ of the subspace picked out by \eqref{weightsub} can be written as the highest weight $\mu$ minus the weight of the lowering operators
\beq \label{weight}
  \nu  = \sum_{a=1}^{\rm rk} m_a\, ^Lw_a - \sum_{a=1}^{\rm rk} d_a \,^Le_a\,, 
\quad d_a \ge 0 \,, 
\eeq
where $^Le_a$ are the simple positive roots  of 
$^L{\fg}$. 
\subsubsection{}
The manifold ${\cal X}$ we will need is Kahler and holomorphic-symplectic -- it admits a nowhere vanishing holomorphic two-form $\omega^{2,0}$.
It has several descriptions:
%$$
\begin{itemize}
\item[--]  ${\cal X}$ is the moduli space of $G$-monopoles on 
\beq\label{R3}
{\mathbb R}^3 = {\mathbb R} \times {\mathbb C},
\eeq 
with specified Dirac monopole singularities. $G$ is the Lie group related to $^LG$, the Chern-Simons gauge group, by Langlands duality. 
The charge of a singular monopole is an element of $Hom(U(1), T_G)$, where $T_G$ is the Cartan torus of $G$. This is the co-character lattice of $G$. By Langlands duality, the co-character lattice of $G$ is the same as the character lattice of $^LG$. For $^LG$ simply connected, $G$ is of adjoint type and its co-character and co-weight lattices coincide.
Corresponding to every vertex operator $\Phi_{V_i}(a_i)$ in \eqref{electric} is a singular monopole on ${\mathbb R}^3$, of charge $\mu_i$ which is the highest weight of the representation $V_i$, at $\log |y| =\log |a_i|$ on ${\mathbb R}$, and at the origin of ${\mathbb C}$. 
The charges of smooth monopoles
%with gauge group at infinity broken to the maximal torus,
 are classified by
$
\pi_2({G}/{T_G}) = {\pi_1(T_G)}.
$
This coincides with ${\rm Hom}(T_G, U(1))$, the dual of the weight lattice of $G$. The dual is the co-root lattice of ${\fg}$, which is also the root lattice of $^L{\fg}$. The total charge of smooth monopoles is $\mu - \nu = \sum_{a=1}^{\rm rk} d_a \,^Le_a$. 
 The complex dimension of the monopole moduli space is
\beq\label{dimX}{\rm dim}_{\mathbb C}{\cal X}= 2 \langle \rho, \mu-\nu\rangle = 2\sum_{a=1}^{\rm rk} d_a,
\eeq
where $\rho$ is the Weyl vector of $\fg$.
%Since ${\cal X}$ is holomorphic-symplectic, this number must always be even. The fact that it is follows from the fact that the weight lattice, to which the Weyl vector $\rho$ of $\fg$ belongs to, and the co-root lattice of $\fg$, to which the difference $\mu-\nu$ of co-weights belongs to, are dual to each other.\footnote{
%
The (relative) positions of singular monopoles on ${\mathbb R}$ are Kahler moduli of ${\cal X}$, and they are kept fixed.
The vector ${\vec \mu} =(\mu_5, \mu_2, \mu_7, \ldots)$, encoding the singular monopole charges in the order in which they appear on ${\mathbb R}$, labels a chamber in the Kahler moduli of ${\cal X}$. The complex structure moduli of ${\cal X}$ are the relative positions of singular monopoles on ${\mathbb C}$. 
Since all the singular monopoles are at the origin of ${\mathbb C}$, all periods of the holomorphic symplectic form vanish, and the complex structure moduli of ${\cal X}$ are frozen.

From now on, we will use Langlands correspondence to map a co-weight of $\fg$ to a weight of $^L\fg$, etc, and work solely with quantities pertaining to $^L\fg$, with the corresponding inner product $\langle, \rangle$ normalized so the length squared of long roots is $2$. % This way, the right hand side of \eqref{dimX} becomes a product of a co-weight and a root of $^L\fg$. 
While for the most part of this paper $^L{\fg}$ and ${\fg}$ are simply laced and equal, keeping track of the distinction is useful with the non-simply laced cases treated in \cite{A2, A3} in mind.

\item[--]

${\cal X}$ is a resolution of the intersection of slices 
\beq\label{ourX}
{\cal X} = {{\rm Gr}^{\vec \mu}}_{\nu}
\eeq
in the affine Grassmannian 
$${\rm Gr}_G=G((z))/G[[z]]$$ of ${G}$. 
The affine Grassmannian parameterizes the space of possible Hecke modifications of a trivial $G$ bundle on ${\mathbb C}$. The orbit ${\rm Gr}^{\mu_i} = G[[z]]z^{-\mu_i}$ in the affine Grassmannian corresponds to those bundles that can be obtained from a trivial one by a Hecke modification of type $\mu_i$, where $\mu_i$ is a co-character of $G$. As before, $G$ is of adjoint type, so $\mu_i$ is also a co-weight of $\fg$.
The description of moduli space of monopoles on ${\mathbb R}^3 = {\mathbb R} \times {\mathbb C}$ in terms of the affine Grassmannian comes from the fact \cite{KW} that monopole equations describe a family of holomorphic $G$-bundles on ${\mathbb C}$ parameterized by ${\mathbb R}$, where the holomorphic type of the bundle jumps by a Hecke modification of type $\mu_i$, at the location of the corresponding singular monopole, and is constant otherwise. 
The loop variable $z$ of the affine Grassmannian is the coordinate on ${\mathbb C}$. The moduli space of a sequence  ${\vec \mu}$ of such Hecke modifications is parameterized by points of the convolution Grassmannian 
$$
{\rm Gr\,}^{\vec\mu} = {\rm Gr\,}^{\mu_1}\, {\tilde \times}\,{\rm Gr\,}^{\mu_2} {\tilde \times} \ldots {\tilde \times}{\rm Gr\,}^{\mu_n},
$$
defined as (see e.g. \cite{Danilenko})
\beq\label{geomG}{\rm Gr\,}^{\vec\mu}  = \{(L_1, \ldots , L_n)  \in {\rm Gr}^n\,| \;L_0 \stackrel{{ \mu}_1}{ \longrightarrow} {L}_1\stackrel{{ \mu}_2}{ \longrightarrow}  \ldots  \stackrel{{ \mu}_n}{ \longrightarrow} L_n\}.
\eeq
The notation $L_{i-1}\stackrel{{ \mu}_{i}}{ \longrightarrow} L_{i}$ means that  $(L_{i-1}, L_{i})\in  {\rm Gr}\times  {\rm Gr}$ are in the same orbit of diagonal $G((z))$ action as $(z^0, z^{-\mu_{i}})$. If $\mu_i$ is a minuscule weight of $^L\fg$, ${\rm Gr}^{\mu_i}$ is compact and smooth
\beq\label{basic}
{\rm Gr}^{\mu_i} = G/P_i,
\eeq
where $P_i$ is a maximal parabolic subgroup of $G$ corresponding to $\mu_i$.  This is the subgroup containing all negative roots of $\fg$, except for $-{e}_i$, where $e_i$ is the simple positive root dual to $\mu_i$. The dimension of ${\rm Gr}^{\mu_i}$ is  $ 2\langle\rho, \mu_i\rangle$. Since we are taking all $\mu_i$ to be minuscule weights of $^L{\fg}$, the convolution Grassmannian ${\rm Gr\,}^{\vec\mu}$ is compact, smooth and of dimension $ 2\langle\rho, \mu\rangle = \sum_i 2\langle\rho, \mu_i\rangle$.

Bringing all the singular monopoles together, so that $\log |a_i|=\log |a_j|$ for all $i$ and $j$, the convolution Grassmannian ${\rm Gr\,}^{\vec\mu} $ becomes 
\beq\label{projmu}
m_{\vec \mu}: {\rm Gr\,}^{\vec\mu} \rightarrow {\rm Gr\,}^{\mu^\times} = \cup_{\nu\leq \mu} {\rm Gr}^{\nu}.
\eeq
The map $m_{\vec \mu}$ takes $m_{\vec \mu}:(L_1, \ldots , L_n)\rightarrow L_n$, and forgets the resolution. 
${\rm Gr\,}^{\mu^\times}$ is still compact, of the same dimension as $ {\rm Gr\,}^{\vec\mu}$, but it is not smooth for $n\neq 1$.  Instead, it is a union of ${\rm Gr}^{\mu}$, which is smooth but not compact, and lower dimensional orbits. The notation $\nu\leq\mu$ means that $\nu$ is a dominant weight and that $\mu-\nu$ is a sum of positive roots of $^L\fg$.  The lower dimensional orbits describe monopole bubbling phenomena \cite{KW}. The orbit ${\rm Gr}^{\nu} \subset  {\rm Gr\,}^{\mu^\times} $ is a result of monopole bubbling where smooth monopoles, of total charge of $\mu-\nu$, bubble off the charge $\mu$ singular monopole and disappear, leaving behind a monopole of charge $\nu$.

From the space of Hecke modifications we can obtain the moduli space of charge $\nu$ monopoles by considering the transversal slice to ${\rm Gr}^{\nu}$ orbit inside ${\rm Gr}^{\,\mu^{\times}}$, 
\beq\label{slice}
  {\rm Gr}^{\,\mu^{\times}}_{\;\;\;\nu}= {\rm Gr}^{\,\mu^{\times}} \cap {\rm Gr}_{\,\nu}.
\eeq
%The slice is transversal in the \section{•}
%ense that $ {\rm Gr}_{\,\nu} = G_1[z^{-1}] z^{-\nu}$, intersects ${\rm Gr}^{\nu} = G[[z]]z^{-\nu}$ at a single point ($G_1[z^{-1}]$ denotes gauge %transformations that approach identity at $z\rightarrow \infty$). 
Being a transversal slice to an orbit where charge $\mu-\nu$ smooth monopoles bubble off and disappear, corresponds to configurations with total monopole charge $\nu$, which consist of charge $\mu$ singular monopoles, in presence of charge $\nu-\mu$ smooth monopoles. Thus, $  {\rm Gr}^{\,\mu^{\times}}_{\;\;\;\nu}$ is the moduli space of monopoles, at a point in Kahler moduli where all singular monopoles are coincident.
To get ${\cal X}$, we reverse the projection in \eqref{projmu},
$${\cal X}= {{\rm Gr}^{\vec \mu}}_{\nu}:= m_{\vec \mu}^{-1}(   {\rm Gr}^{\,\mu^{\times}}_{\;\;\;\nu}).$$ If all $\mu_i$ are minuscule weights of $^L{\fg}$, ${\cal X}$ is smooth.
Some additional details and summary of conventions are in the appendix.

 \item[--] ${\cal X}$ is the Coulomb branch of a 3d quiver gauge theory with ${\cal N}=4$ supersymmetry, with
$$
\textup{quiver ${\scQ}$} = 
\textup{Dynkin diagram of ${\fg}$} \,.
$$
Here ${\fg}$ is the Lie algebra Langlands dual to $^L{\fg}$. The 3d theory has gauge group and flavor symmetry group
\beq\label{gaugeC}
G_{\scQ} =\prod_{a} U(V_a), \qquad G_{W} = \prod_{a} U(W_a) \,. 
\eeq
The
dimensions of the vector spaces $V_a$ and $W_a$ are 
\beq\label{gaugeCb}
\dim V_a = d_a\,, \quad \dim W_a = m_a.
\eeq
They are determined by the weight space data in \eqref{highest w}, \eqref{weight}. A choice of generic real masses for fundamental hypermultiplets breaks the flavor symmetry $G_W$ to its Cartan. At the same time, we will set to zero the complex mass parameters, so that a $U(1)_V$ subgroup of $SU(2)_V$ vector R-symmetry remains unbroken.  The real masses act as Kahler moduli of ${\cal X}$. Since ${V}_i$ are minuscule, ${\cal X}$ is smooth and has a symmetry corresponding to scaling the holomorphic symplectic form.
% (More precisely, we need to take the limit of the Coulomb branch that involves simultaneously taking $R$ to zero, while taking the gauge coupling of the theory to infinity \cite{Aganagic:2001uw}. This affects the asymptotic form of the metric on ${\cal X}$, but it does not affect its complex structure. )

\end{itemize} 
 
The identification between the Coulomb branch of the 3d quiver gauge theory and the intersection of slices in affine Grassmannian is due to \cite{KW, Nak13a, Nak13b, GC}. To be precise, unless the weights $\mu$ and $\nu$ satisfy the dominance condition $0\leq \nu \leq \mu$, the construction of the Coulomb branch ${\cal X}$ starts from a modification of ${{\rm Gr}^{{\vec \mu}}}_{\nu}$ constructed in \cite{Nak13c}. Knot theory applications require $\nu \geq 0$, so this is not relevant for us. 

\subsubsection{}
Since ${\cal X}$ is Kahler and holomorphic-symplectic, it has hyper-Kahler structure. 
Because all the periods of its holomorphic symplectic form vanish, there is a unique complex structure in which ${\cal X}$ is Kahler, the one we are working in. In this complex structure, ${\cal X}$ has a symmetry that acts by $z\rightarrow {\fq} z$, where $z$ is the coordinate on ${\mathbb C}$ in ${\mathbb R}^3$. The same ${\mathbb C}^{\times}_{\fq}$ symmetry acts on the affine Grassmannian by loop rotations, and scales the holomorphic symplectic form by ${\omega}^{2,0} \rightarrow {\fq}\,{\omega}^{2,0}$. A smooth holomorphic symplectic manifold with these properties is called an equivariant symplectic resolution.  

Our ${\cal X}$ has a larger torus ${\rm T}$ of symmetries, 
\beq\label{torus}
{\rm T}= \Lambda \times {\mathbb C}_{\fq}^{\times},
\eeq
which includes the action of ${\Lambda} = ({\mathbb C}^{\times})^{{\rm rk}\fg}$ on ${\cal X}$ which preserves the holomorphic symplectic form. The action of $\Lambda$ on ${\cal X}$ is induced from the action of the maximal torus of $G$ on the affine Grassmannian. Viewing ${\cal X}$ as the Coulomb branch of the quiver gauge theory, the equivariant parameters of the $\Lambda$-action on ${\cal X}$ are the real FI parameters of the 3d gauge theory.

\subsection{KZ equation as the quantum differential equation}\label{ss:GW}

Since ${\cal X}$ is holomorphic symplectic, quantum multiplication in \eqref{qmulti} is non-trivial (i.e. distinct from classical) only if one works equivariantly with respect to a torus that scales the holomorphic symplectic form. This also reduces the amount of supersymmetry of the sigma model to ${\cal X}$, which features in the next section, by half.

As before, we take all $V_i$'s to be minuscule representations of $^{L} \fg$, so that for generic positions $y_i$ of singular monopoles, ${\cal X}$ is smooth. The following theorem is proven in \cite{Danilenko}.
\begin{theorem}(Danilenko)\label{t:one}
The quantum differential equation of ${\cal X}$, working equivariantly with respect to the ${\rm T}$-action,
 is the Knizhnik-Zamolodchikov equation.
\end{theorem}
All the data on which the KZ equation and the conformal blocks depend on are realized geometrically:
\begin{itemize}
\item[--]
The complexified Kahler moduli of ${\cal X}$ are the positions of vertex operators. If $J\in H^{1,1}({\cal X})$ is the Kahler form on ${\cal X}$, and $B$ is the B-field, there are curve classes $C_{ij}\in H_2({\cal X})$ so that 
\beq\label{km}
\int_{C_{ij}} (J+i B) = \log a_i/a_j.
\eeq
The real Kahler moduli $\int_{C_{ij}} J = \log|a_i/a_j|$ are the positions of vertex operators along the axis of the cylinder ${\cal A}$.  The choice of ordering of vertex operators on ${\cal A}$, specified by the vector ${\vec \mu}$ is the choice of a Kahler cone ${\fC}_{\vec \mu}$ of ${\cal X}$. 
The fact that the $B$ field is periodic is the reason why we needed to take the Riemann surface ${\cal A}$ to be a cylinder, and not a plane.
\item[--]
The weight $\lambda_0=1/\kappa$ of the symplectic form under the ${\mathbb C}_{\fq}^{\times}$ action is related to ${\fq}$ in \eqref{qk} by
%\beq\label{qkb}
$\fq = e^{2\pi i \lambda_0}.$
Turning off the action that scales the holomorphic symplectic form, by setting $\lambda_0$ to zero, means setting $\kappa$ to $\infty$. This is the classical limit of the affine Lie algebra, see \cite{MS}.
\item[--]
The equivariant variables associated to the $\Lambda$-torus action on ${\cal X}$, determine $\lambda$, the highest weight of the Verma module at $0\in {\cal A}$.
%\eeq
\end{itemize}

Just as the Knizhnik-Zamolodchikov equation has a geometric interpretation in terms of ${\cal X}$, so do its solutions.

\subsection{Conformal blocks as vertex functions}

The fact that Knizhnik-Zamolodchikov equation is the quantum differential equation of ${\cal X}$ implies conformal blocks that solve it are the cohomological vertex functions of ${\cal X}$, known also as the $J$-functions. These are some of the most basic objects of Gromov-Witten theory on ${\cal X}$; their pivotal role in the theory was explained in the works of Givental \cite{G, GT}. 

Vertex functions count holomorphic maps from $D$, which is a punctured complex plane, to ${\cal X}$
\beq\label{CxtoX}
{\rm D} = {\mathbb C}^{\times}\; \dasharrow \; {\cal X},
\eeq 
working equivariantly with respect to the torus \eqref{torus}
acting on ${\cal X}$, and ${\mathbb C}^{\times}$ acting on ${\rm D}$. 
A vertex function depends on the choice of insertion at the origin of $D$, and also at infinity.

The choice of insertion at the origin of $D$ is an A-model observable, an equivariant cohomology class 
 $$
 \alpha \in H_{\rm T}^*({\cal X}).
 $$
This makes vertex functions vector valued. The fact that they live in the same vector space as the conformal blocks in \eqref{electric}
is a consequence of geometric Satake correspondence of \cite{Lu,Ginz,Mi} which is an isomorphism of $ H_{\rm T}^*({\cal X})$
and the weight $\nu$ subspace of the representation $V$ in \eqref{representation},
\beq\label{GS}
 H_{\rm T}^*({\cal X})\cong (V_1\otimes \ldots \otimes V_n)_{\nu}.
\eeq
The data at infinity is an insertion of an equivariant K-theory class instead
$$
[{\cal F}] \in K_{\rm T}({\cal X}),
$$ 
as in the works of \cite{Hosono,iritani1, iritani2, KKP}. 
This comes about naturally, as we will explain below, since one should view ${\rm D}$ as an infinite cigar, and the infinity of ${\rm D}$ as a circle
$$
S^1 = \partial {\rm D}.
$$
We will call 
\beq\label{vertexF}
{\cal V}_{\alpha}[{\cal F}] = \Vertex_{\alpha}[{\cal F}],
\eeq
a ``vector" vertex function. The ``vector" refers to the dependence on the cohomology class $\alpha \in  {\rm H}_{\rm T}^*({\cal X})$, inserted at $0\in {\rm D}$.

\subsubsection{}

The vector vertex function has a  simpler cousin, the ``scalar" vertex function 
\beq\label{svf}
{\cal Z}[{\cal F}] = S\Vertex[{\cal F}],
\eeq
where we close off the puncture at the origin of ${\rm D}$ and study maps 
\beq\label{CtoX}
{\rm D} = {\mathbb C}\;\; \dasharrow \;\; {\cal X}.
\eeq 
Since the identity insertion is an element of $H_{\rm T}^*({\cal X})$, there exists a covector ${\rm c}$ such that,
\beq\label{VS}
  c^{\alpha}\,{\cal V}_{\alpha}[{\cal F}] = {\cal Z}[{\cal F}].
\eeq
The scalar and vector vertex functions have the same dependence on the K-theory class $[{\cal F}]$ inserted at infinity, since in relating the two,
the infinity is not touched.  We will sometimes refer to ${\cal Z}[{\cal F}]$ as the solution to the ``scalar Knizhnik-Zamolodchikov equation", or as a ``scalar conformal block". 
\begin{comment}While there is a deeper geometric as well as representation theoretic meaning to this, as explained in \cite{AFO}, for this paper it suffices it to mean the following elementary fact. To the Knizhnik-Zamolodchikov equation, which is a linear differential equation acting on a vector space whose  dimension is that of weight $\nu$ subspace of representation $V,$ there corresponds an equivalent nonlinear differential equation of that order, which we call the scalar Knizhnik-Zamolodchikov equation. Its solutions are the scalar conformal blocks, computed by ${\cal Z}[{\cal F}]$. 
\end{comment}

\subsection{Quantum invariants from geometry}\label{s_BC}

Since positions of vertex operators on ${\cal A}$ are complexified Kahler moduli of ${\cal X}$, a braid $B\in {\cal A} \times[0,1]$ describes a path in the complexified Kahler moduli from ${\cal X} = {\cal X}_{{\vec \mu}_0}$ to ${\cal X}' = {\cal X}_{{\vec \mu}_1} $.  It follows that the monodromy of the KZ equation, acting on the space of conformal blocks, has a geometric interpretation as the monodromy of the quantum differential equation. 

An elementary, but important fact is that monodromy of a differential equation acts on the fundamental solution from the right, if we take the equation itself to act from the left. This means that the action of monodromy on vertex functions $\cal{V}[{\cal F}]$ by,
$${\cal V}[ {\cal F}] \rightarrow 
{\cal V}[{\cal F}'] = {\mathfrak B} {\cal V}[  {\cal F}],$$
comes from the action on K-theory classes $[{\cal F}]\in {K}_{{\rm T}}({\cal X})$ associated with the $S^1$ boundary at infinity of ${\rm D}$.
Monodromy of the quantum differential equation therefore gives a map 
\beq\label{BK}
{\mathfrak B} : K_{\rm T}({\cal X}) \rightarrow K_{\rm T}({\cal X}'),
\eeq
that depends only on the homotopy type of the braid $B$, and takes 
\beq\label{KB}
 [{\cal F}] \in K_{{\rm T}}({\cal X}) \;  \rightarrow \;  [{\cal F}'] = {\mathfrak B}[ {\cal F}] \in K_{{\rm T}}({\cal X}').
\eeq
% 
%It acts on the vector vertex function by
%\beq\label{VB}
%{\cal V}[ {\cal F}] \rightarrow 
%{\cal V}'[{\cal F}'] = {\mathfrak B} {\cal V}[  {\cal F}].
%\eeq
Another consequence of the fact that monodromy acts on the K-theory class at infinity is that it acts in the same way on the vector and scalar vertex functions, so analytic continuation of the scalar vertex function along the path $B$ in Kahler moduli is given by the same monodromy matrix ${\mathfrak B}$ from \eqref{BK}. 

\section{Conformal blocks and derived categories}\label{s-s}

This section describes the physical interpretation of vertex functions,  in terms of a supersymmetric sigma model to ${\cal X}$, and its implications. Most of what we will say in this section will be very general, applicable to any ${\cal X}$ which is an equivariant symplectic resolution \cite{OK}.
% - a holomorphic symplectic manifold which is not singular and which admits an ${\rm T}$-action that scales its holomorphic symplectic form. 
In the next section, using the very special nature of our ${\cal X}$, which starts with fact that its quantum differential equation is the KZ equation, we will be able to make much of this general story very concrete.

As we will recall below,  the vertex function ${\cal V}[{\cal F}]$ is the partition function of the supersymmetric sigma model on $D$ viewed as an infinitely long cigar with an $S^1$ boundary at infinity. In the interior of the cigar, supersymmetry is preserved using an A-type twist. At the origin, one inserts A-model observables which correspond to classes in $H^*_{\rm T}({\cal X})$. The boundary condition, imposed at the $S^1$ at infinity, corresponds to a B-type brane on ${\cal X}$.

Boundary conditions form a category, and the category of B-type branes on ${\cal X}$, working ${\rm T}$-equivariantly is
\beq\label{DT}
{\mathscr D}_{{\cal X}} = D^bCoh_{\rm T}({\cal X}),
\eeq	
the derived category of ${\rm T}$-equivariant coherent sheaves. 
${\mathscr D}_{{\cal X}}$ is a subcategory of the derived category of coherent sheaves on ${\cal X}$, whose objects are complexes of coherent sheaves on ${\cal X}$ which are ${\rm T}$-equivariant.  For example, structure sheaves of ${\rm T}$-fixed points are objects of ${\mathscr D}_{\cal X}$, but structure sheaves of generic points are not. The morphisms of ${\mathscr D}_{\cal X}$ are a {\it refinement} of morphisms of  $D^bCoh({\cal X})$, where we keep track of their equivariant grades.

In physics terms (see \cite{Hori} for an excellent review aimed at physicists), working equivariantly with respect to the ${\rm T}$-action means weak gauging of the ${\rm T}$-symmetry of ${\cal X}$, and of its branes.  Since the ${\rm T}$-action is only weakly gauged, the morphisms of ${\mathscr D}_{\cal X}$ are a refinement of morphisms of the full category.

If $n \in {\mathbb Z}$ is the homological degree, and ${\vec k} \in {\mathbb Z}^{\rm rk T}$ the homological degree, morphisms in degrees $[n]\{{\vec k}\}$ from branes ${\cal F}$ to ${\cal G}$ are elements of the homology group ${Hom}_{{\mathscr D}_{\cal X}}({\cal F}, {\cal G}[n]\{{\vec k}\}) $.

Picking as a boundary condition an equivariant B-type brane
\beq\label{bBbrane}
\scF \in  {\mathscr D}_{\cal X},
\eeq
we get ${\cal V}[{\cal F}]$ as the partition function of the theory on ${\rm D}$. 

\begin{figure}[!hbtp]
  \centering
   \includegraphics[scale=0.25]{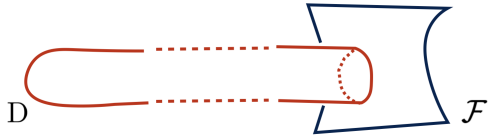}
 \caption{Vertex function ${\cal V}[{\cal F}]$ is the partition function of a supersymmetric sigma model on ${\rm D}$ with brane ${\cal F}$ as the boundary condition at infinity.}
  \label{f_boundary}
\end{figure}
\noindent{}The vertex function ${\cal V}[{\cal F}]$ depends on the choice of the brane $
\scF$ 
only through its K-theory class, $[{\cal F}] \in K_{\rm T}({\cal X})$,  however, the physical sigma model needs an actual object ${\cal F}  \in  {\mathscr D}_{\cal X}$ to serve as the boundary condition on $S^1 =\partial D$. This fact plays a key role in everything that follows.

 \subsection{Vertex function and central charge}\label{cc}

The vector vertex function ${\cal V}[{\cal F}]$, which computes the conformal blocks, is a generalization of the physical ``central charge" ${\cal Z}^0[{\cal F}]$ of the brane ${\cal F} \in {\mathscr D}_{\cal X}$, 
$$
{\cal Z}^0[{\cal F}]: {\rm K}({\cal X}) \rightarrow {\mathbb C}.
$$
The central charge ${\cal Z}^0$ defines a stability condition on ${\mathscr D}_{{\cal X}}$. The stability condition that uses ${\cal Z}^0$ as central charge is due to Douglas \cite{Pi1, Pi2}, and is known as the $\Pi$-stability condition. As in Douglas' work, when ${\cal X}$ has a mirror, one can obtain ${\cal Z}^0$, the central charge associated to the $\Pi$-stability condition, from mirror symmetry. This is the case for us, as we explain in \cite{A2}.

\subsubsection{}
The function ${\cal V}[{\cal F}]$ generalizes ${\cal Z}^0[{\cal F}]$ in two different ways: firstly, by being a vector, coming from insertions of classes in $H_{\rm T}^*({\cal X})$ at the origin of ${\rm D}$; and secondly, by its dependence on equivariant parameters. Undoing the first generalization, but not the second, we get the scalar vertex function ${\cal Z}[{\cal F}]$ from \eqref{svf}. Undoing both we get the physical central charge  
\beq\label{central}
{\cal Z}^0[{\cal F}] = {\cal Z}[{\cal F}]|_{\vec{\lambda} =0}.
\eeq
This is partition function of ordinary, non-equivariant, Gromov-Witten theory on ${\rm D}$, with target ${\cal X}$, and brane ${\cal F}$ defining the boundary condition at infinity. The underlying sigma model to ${\cal X}$ is conformal, since we have set the parameter $\lambda_0=1/\kappa$ defined in section \ref{ss:GW} to zero. Since ${\cal X}$ has hyper-Kahler structure, ${\cal Z}^0$ receives no quantum corrections, once written in terms of flat coordinates. For our purposes, it suffices to identify these as the coordinates used by the A-model, in terms of which the WDVV equations hold, and the quantum differential equation is written in. Correspondingly, in our setting, there is a simple exact formula for the central charge:
\beq\label{ccsym}
{\cal Z}^0[{\cal F}] = \int_{\cal X} {\rm ch}({\cal F}) e^{-(J+iB)} \sqrt{{\rm td}({\cal X})},
\eeq
where $J\in H^{1,1}({\cal X}, {\mathbb R})$ is the Kahler class of ${\cal X}$, $B$ is the NS B-field, and ${\rm td}({\cal X})$ is the Todd class of ${\cal X}$. The fact that the formula is given in terms of the Todd class, and not the Gamma class as in \cite{iritani1} is another simplification due to the fact since ${\cal X}$ is hyper-Kahler; on a hyper-Kahler manifold, all the odd Chern classes of the tangent bundle vanish, and the Gamma and Todd classes coincide, see e.g. \cite{HM}. 
\subsubsection{}
The expression \eqref{ccsym} for the central charge ${\cal Z}^0$ reflects several elementary string theory facts: the fact the  A-model does not depend on $J$ and $B$ separately, but only on the combination $J_{\mathbb C} =J+iB$, the complexified Kahler class, and the fact that amplitudes depend on ${\cal F}$ and $B$ only in the combination ${\cal F}-B$, which is needed for invariance under $B$-field gauge transformations. Thus, \eqref{ccsym} is essentially determined by the classical symmetries of the A-model on ${\cal X}$. Our normalization is that ${\rm ch}({\cal F})= {\rm Tr} \, e^{i {\cal  F}}$  for a vector bundle on ${\cal X}$ with curvature ${\cal F}$, to naturally match the dependence of the A-model on the complexified Kahler class. 
We will explain in section \ref{s-three}, that for branes supported on holomorphic Lagrangians in ${\cal X}$, which in fact generate ${\mathscr D}_{\cal X}$, the expression for the central charge simplifies even further. 
\subsubsection{}
The relevance of the cigar compactification of the ${\cal N}=2$ theory to the monodromy problem originates from \cite{CV}. The quantum differential equation is the equation for flat holomorphic sections of the $tt^*$-connection of \cite{CV}, written in terms of ``special coordinates" - the coordinates in terms of which the WDVV equation holds. The relation of flat sections of the $tt^*$-connection to D-brane central charges was explained in \cite{HIV, mirrorbook, CGV}.
Further physical aspects of central charges are studied, for example in \cite{Ooguri, Douglas, Hori}, and mathematical aspects in the works of \cite{Hosono, iritani1, iritani2, KKP}.
\subsubsection{}
The A $\leftrightarrow$ B switch is characteristic of central charges. 
The combination of the A-twist in the interior with B-twist near the boundary, in the physical realization of the vertex functions, may seem unusual.  However, any boundary condition on the circle at infinity is the compatible with A-type supersymmetry in interior, since propagation in infinite ``time" along the cigar projects the boundary state on the $S^1$ at infinity to a supersymmetric ground state (a state in the Ramond sector, with fermions periodic around the $S^1$), and all such ground states are compatible with the A-twist.   Correspondingly, any boundary state on the $S^1$ at infinity can be traded for a point insertion of an A-model observable, while compactifying ${\rm D}$ to a ${\mathbb P}^1$. 
\subsubsection{}

For ${\cal Z}^0[{\cal F}]$ which is formulated in the superconformal sigma model to ${\scX}$, we may take ${\rm D}$ to be a finite cigar or a disk \cite{Ooguri}. 
Then, the fact that A-type twist in the interior of ${\rm D}$ is compatible with B-type boundary condition at $S^1= \partial {\rm D}$ can then be understood as follows \cite{Douglas}. 

The identification of a boundary condition as a brane on ${\cal X}$ is made by taking the time of the 2d theory to run along the boundary. This choice is the one relevant for studying the category of boundary conditions. 

The brane central charge, on the other hand, is a closed string quantity. In the closed string, we want to take the time to run along the cigar, and transverse to the boundary. 

The exchange of space and time direction in the 2d SCFT has the effect of reversing the sign of the left moving R-symmetry, which is what distinguishes the A- and the B-type boundary conditions. Away from $\lambda_0=0$, the superconformal invariance is broken, and ${\rm D}$ needs to be an infinite cigar for a B-type boundary condition on $S^1 =\partial {\rm D}$ to be compatible with A-type supersymmetry in the interior.

\subsection{Stability conditions and derived categories}\label{s-phys}
With ${\rm T}$-action on ${\cal X}$ turned off, the sigma model is superconformal. The category of B-type boundary conditions in a  superconformal sigma model to ${\cal X}$ is a subcategory ${\mathscr P}$ of $D^bCoh({\cal X})$, the category of topological B-type branes. Objects of ${\mathscr P}$ are semi-stable coherent sheaves on ${\cal X}$, where the stability condition is with respect to the central charge function ${\cal Z}^0$, defined in the previous section.
%$$
%{\cal Z}^0 [{\cal F}]\sim \int_{\cal X} e^{(-(B+iJ))} ch[{\cal F}] \,\widehat{\Gamma}_{\cal X}
%$$

The discussion of stability conditions may not seem relevant for us, because we do wish to work with a category ${\mathscr D}_{\cal X}$ of topological, rather than physical branes, and equivariantly with respect to the ${\rm T}$-action on ${\cal X}$. The relevance of stability structures for us is that the action of braiding on ${\mathscr D}_{\cal X}$ comes from variations of it.
\subsubsection{}

Bridgeland in \cite{Bridgeland1}, following Douglas in \cite{Pi1, Pi2} provided a precise mathematical formulation of the structure imposed by the existence of the pair $({\mathscr P}, {\cal Z}^0)$ on the category, and abstracted it to a general derived category setting (see also \cite{H} for a review). 
%\textcolor{red}{A Bridgeland stability condition is an abstract central charge function, a map form $K({\cal X})$ to ${\mathbb C}$, and a collection of abelian subcategories of ${\mathscr P}_{\phi}\subset {\mathscr D}_{\cal X}$ for each $\phi \in {\mathbb R}$, satisfying a collection of conditions motivated by studies of decays of BPS branes. Objects of ${\mathscr P}_{\phi}$ model the physical branes in the conformal supersymmetric sigma model to ${\cal X}$ which preserve a supersymmetry determined by the phase $\pi \phi$ of the central charge. Here, we regard the phase $\phi$ of the central charge as valued in ${\mathbb R}$, compensate by defining ${\mathscr P}(\phi+1) = {\mathscr P}(\phi)[1]$, where $[k]$ denotes the shift in cohomological degree by $k\in {\mathbb Z}$.}
By construction, the pair $({\mathscr P}, {\cal Z}^0)$ coming from the physical sigma model to ${\cal X}$ is expected to provide an example of Bridgeland stability condition. We will assume, without proof, that this is the case. The equivariant mirror A-model setting, described in \cite{A2}, should make a proof of this feasible, along the lines of \cite{TS}. Mirror symmetry also makes it easy to spell out which branes belong to ${\mathscr P}$, as special Lagrangians on the mirror, with chosen orientation.

\subsubsection{}

For ${\cal X}$ that admits an ${\rm T}$-action, requiring branes to be ${\rm T}$-equivariant makes them more stable, rather than less, so we will assume that all the objects in ${\mathscr P}$, the category of all physical branes can be made ${\rm T}$-equivariant. Furthermore, we will assume that ${\cal Z}^0$ not only provides a stability condition $D^bCoh({\cal X})$, the category of all B-branes, but also provides a stability condition on ${\mathscr D}_{\cal X}$, the category of ${\rm T}$-equivariant branes. 

Our understanding of why these assumptions are justified comes from mirror symmetry, described in \cite{A2}. Mirror symmetry is for us a key tool for understanding stability conditions - this is after all how $\Pi$-stability was discovered.  Fortunately, restricting to objects of $D^bCoh({\cal X})$ that can be made ${\rm T}$-equivariant is natural from mirror symmetry perspective, since ${\rm T}$- is a torus action on ${\cal X}$.

\subsubsection{}

Given a stability condition on ${\mathscr D}_{\cal X}$, which is the pair $({\mathscr P}, {\cal Z}^0)$ satisfying Bridgeland's stability conditions, we get an abelian subcategory 
$${\mathscr A} \subset {\mathscr D}_{\cal X},$$ obtained by taking all possible extensions of stable branes, i.e. objects of ${\mathscr P}$, whose central charge ${\cal Z}^0$ is in the upper half of the complex plane, with phase $0\leq \phi <1$. An exact sequence of objects in ${\mathscr A}$, gives an exact triangle in ${\mathscr D}_{\cal X}$.
Bridgeland proves in \cite{Bridgeland1} that, if $({\mathscr P}, {\cal Z}^0)$ is a stability condition, then ${\mathscr A}$ is the heart of a bounded $t$-structure on ${\mathscr D}_{\cal X}$, 
which in particular means that any non-zero object of $ {\cal F} \in {\mathscr D}_{{\cal X}}$ can be obtained from objects in ${\mathscr A}$, by taking direct sums, degree shifts and iterated cones. 
%Since we are assuming that $({\mathscr P}, {\cal Z}^0)$ define a Bridgeland stability condition on ${\mathscr D}_{\cal X}$ \cite{Bridgeland1}, ${\mathscr P}(\geq 0)$ and ${\mathscr P}(<0)$ are orthogonal subcategories of ${\mathscr D}_{\cal X}$. Then, ${\mathscr P}(\geq 0)$  defines a ${\rm t}$-structure on ${\mathscr D}_{\cal X}$ with ${\mathscr A} = {\mathscr P}(\geq 0) \cap {\mathscr P}^{\perp}[1]$ as its heart.

\subsubsection{}

On a general Calabi-Yau manifold, dependence of ${\cal Z}^0$ on the Kahler moduli is very complicated, and so is the stability structure that it imposes on the derived category. In the present setting, ${\cal X}$ is holomorphic symplectic, so the theory becomes much simpler. The physical central charge ${\cal Z}^0: {K}_{\rm T}({\cal X}) \rightarrow {\mathbb C}$ has the simple polynomial dependence \eqref{ccsym} on the Kahler moduli.
Consequently, the choice of ${\rm t}$-structure becomes the same as the choice of a chamber ${\fC}$ in Kahler moduli and the choice of resolution ${\cal X}_{\fC}$ of symplectic singularities. 

Given the many special features, the stability structures on ${\mathscr D}_{\cal X}$ with central charge function ${\cal Z}^0$ should provide a tractable generalization of previous examples of stability conditions constructed in \cite{ST, BridgelandADE}. (In fact, as we will explain in \cite{A2}, we loose little and gain much further simplicity by working with ${\mathscr D}_X \subset {\mathscr D}_{\cal X}$, the ``core subcategory". Objects of ${\mathscr D}_X$ have support on the core submanifold $X$, which is the fixed locus of the ${\mathbb C}^\times_{\fq}$ action on ${\cal X}$.)

\subsection{Monodromy and derived equivalences}

Consider a path $B$ in Kahler moduli from ${\cal X}$ to ${\cal X}' $, as in the previous section.
Analytic continuation of the vertex function ${\cal V}[{\cal F}]$, defined on ${\cal X}$, along the path $B$ is given by the monodromy of the quantum differential equation. It
 takes ${\cal V}[ {\cal F}]$ to a vertex function ${\cal V}[{\cal F}']$, defined on ${\cal X}'$. The two are related by ${\cal V}[{\cal F}'] = {\mathfrak B} {\cal V}[  {\cal F}]$, where ${\mathfrak B}$ acts
on the equivariant K-theory by  \eqref{BK}. In the underlying physical sigma model, the same action must come from the action of monodromy on the brane ${\cal F} \in {\mathscr D}_{\cal X}$ itself. It cannot come merely from the action on the K-theory class $[{\cal F}] \in {\mathscr D}_{\cal X}$ because the sigma model needs the brane ${\cal F}$ itself as the boundary condition. 
 \subsubsection{}
There is a well known expectation of how monodromy in Kahler moduli acts on the category of B-type boundary conditions, see e.g \cite{Douglas, AspinwallL}. 
Since $B$ is a path in (complexified) Kahler moduli on which the B-model and its category of branes do not depend, monodromy in Kahler moduli is expected to act as a derived equivalence functor ${\mathscr B}$, 
\beq\label{BC}
{\mathscr B}: {\mathscr D}_{\cal X} \cong {\mathscr D}_{{\cal X}'},
\eeq
The functor maps a brane ${\cal F}$ on ${\cal X}$ to a brane $ {\mathscr B} \scF$ on ${\cal X}'$
\beq\label{dc}
{\mathscr B}: \scF \;\in \;{\mathscr D}_{\cal X} \;\;\rightarrow \;\;  \scF' = {\mathscr B} \scF \;\in \; {\mathscr D}_{{\cal X}'},
\eeq
such that the action on $K$-theory reduces to 
\beq\label{red}
[{\cal F}'] = [\mathscr{B} {\cal F}]={\mathfrak B}[ {\cal F}],
\eeq 
as in \eqref{KB}. Both ${\mathfrak B}$ and ${\mathscr B}$ depend on the homotopy type of $B$, since they are in general non-trivial around any non-contractible loop in complexified Kahler moduli.  Thus, monodromies in Kahler moduli are expected to serve as a source of derived equivalences \cite{BO}.

\subsubsection{}
In general, defining the derived equivalence ${\mathscr B}$ corresponding to the homotopy type of the path $B$ in Kahler moduli is difficult. In the setting of ${\cal X}$ which is a holomorphic symplectic manifold with equivariant action scaling the holomorphic symplectic form, there is a systematic construction of such functors, due to Bezrukavnikov and Kaledin whose theory makes use of quantizations of ${\cal X}$ in characteristic $p\gg0$  \cite{BK, kaledin}.
The fact that the functor ${\mathscr B}$ is compatible with monodromy ${\mathfrak B}$ of the quantum-differential equation is a difficult theorem by Bezrukavnikov and Okounkov \cite{RO}:
\begin{theorem}[Bezrukavnikov-Okounkov]\label{t:two} 
If ${\cal X}$ is an equivariant symplectic resolution, monodromy ${\mathfrak B}$ of its quantum differential equation along the path $B$ in complexified Kahler moduli lifts to a derived equivalence functor ${\mathscr B}$ of ${{\mathscr D}}_{\cal X}$.
\end{theorem}

The sigma model origin of vertex functions implies that the monodromy of quantum differential equation of ${\cal X}$ lifts to a very specific derived equivalence functor ${\mathscr B}$.
\begin{conjecture}\label{c:one}
The functor ${\mathscr B}$ comes from the variation of stability with respect to the central charge function ${\cal Z}^0:K({\cal X}) \rightarrow {\mathbb C}$ defined in \eqref{ccsym}. 
\end{conjecture}
The central charge function ${\cal Z}^0$ is the physical central charge function of a supersymmetric sigma model ${\cal X}$, and as we will see, a specialization of the scalar vertex function ${\cal Z}$. 
%The corresponding stability condition is the $\Pi$-stability condition of \cite{Pi1, Pi2}. 

\subsubsection{}
Variations of stability conditions are a source of derived equivalences.   
In each chamber ${\fC}$ in Kahler moduli, we get a resolution of singularities ${\cal X}_{\fC}$, an abelian subcategory ${\mathscr A}_{\fC} $ of  ${\mathscr D}_{{\cal X}_{\fC}}$ that depends on the chamber, such that all the cohomology sheaves of ${\mathscr D}_{{\cal X}_{\fC}}$ are in ${\mathscr A}_{\fC}$.  
The category of topological branes does not depend on the Kahler moduli at all, so the derived categories $
{\mathscr D}_{{\cal X}_{\fC}}$ we obtain in this way must all be derived equivalent, for any ${\fC}$. We will often denote the resulting category simply as ${\mathscr D}_{{\cal X}_{\fC}}\cong {\mathscr D}_{\cal X}$ to indicate independence on the choice of resolution. What changes from one chamber to another is how  ${\mathscr D}_{{\cal X}}$ is generated, as the derived category of complexes with cohomology in ${\mathscr A}_{{\fC}}$.

\subsubsection{}
Given a homotopy class of a path $B$ from a chamber ${\fC}$ to a chamber ${\fC}'$ in complexified Kahler moduli, the central charges ${\cal Z}$ change by monodromy ${\mathfrak B}:K_{\rm T}({\cal X}_{\fC}) \rightarrow K_{\rm T}({\cal X}_{\fC}') $ of the quantum differential equation along the path. Along the path, central charges of some objects in ${\mathscr P}$ enter or leave the upper half of the complex plane, and the ${\rm t}$-structure changes. The dependence of ${\cal Z}:K_{\rm T}({\cal X}_{\fC}) \rightarrow {\mathbb C}$ on the path, and not only on the chambers, means that the functor 
$$
{\mathscr B}: {\mathscr A}_{{\fC}} \rightarrow {\mathscr A}_{{\fC}'}
$$
also depends on the path, and with it
the derived equivalence 
$$
{\mathscr B}: {\mathscr D}_{{\cal X}_{\fC}} \cong {\mathscr D}_{{\cal X}_{{\fC}'}}.
$$
Note that the stability condition depends on ${\cal Z}^0$, and not on ${\cal Z}$ - however only ${\cal Z}$ keeps track of action of ${\mathscr B}$ on ${\rm T}$-degrees of objects in ${\mathscr D}_{\cal X}$. 

A pair of paths from  ${\fC}$ to ${\fC}'$ that are not homotopic to each other differ by a closed, non-contractible loop $B$. The functor ${\mathscr B}$ corresponding to a closed loop $B$ is a derived auto-equivalence functor.  While the categories  ${\mathscr A}_{\fC}$ are well defined and only depend on the chamber ${\fC}$, their embeddings in ${\mathscr D}_{\cal X}$ are not well defined, but differ by action of auto-equivalences \cite{AD, Bridgeland1}.

\subsubsection{}
Proof of theorem \ref{t:two} comes from geometric representation theory in positive characteristic \cite{RO}. It uses the equivalence of ${\mathscr D}_{\cal X}$ and ${\mathscr D}_{{\mathscr { A}}}$, the derived category $D^b {\mathscr {A}}{\textup{-mod}}$ of modules of an algebra ${\mathscr { A}}$ obtained by quantization of ${\cal X}$ in characteristic $p\gg0$. The equivalence of ${\mathscr D}_{\cal X}$ and $D^b {\mathscr {A}}{\textup{-mod}}$ is the result of Bezrukavnikov an Kaledin \cite{BK}, who also show that such equivalences can be used to study equivalences of ${\mathscr D}_{\cal X}$ and ${\mathscr D}_{{\cal X}'}$ by relating them both to ${\mathscr D}_{{\mathscr { A}}}$. 

\subsubsection{}
While any detailed aspects of construction using characteristic $p$ are far from physics of our paper, the essential aspects of the equivalence of ${\mathscr D}_{\cal X}$ and  ${\mathscr D}_{{\mathscr { A}}}$ should follow from the equivalence of the category of B-branes in two regimes of complexified Kahler moduli. The two regimes correspond to generic real Kahler moduli with small imaginary parts, relevant for ${\mathscr D}_{\cal X}$, and to nearly vanishing real Kahler moduli, but generic $B$-fields, relevant for ${\mathscr D}_{{\mathscr { A}}}$.  Working with imaginary Kahler moduli has the effect of compactification \cite{A2}, which is the basic feature of working in characteristic  $p\gg 0$.
%\footnote{
%Namely, for ${\cal X}$ relevant in our paper, the Kahler moduli are the positions of vertex operators on ${\cal A}$, so one gets the former case by spreading the vertex operators along the axis of the cylinder (taking $y_i = {\ln} |a_i|$'s to be generic), and the latter case by positioning the vertex operators along a fixed $S^1$ in ${\cal A}$ (taking $y_i = {\ln} |a_i|$'s to coincide). }

More evidence for the fact qualitative aspects of the two stories are the same comes polynomial dependence of ${\cal Z}^0$ on ${\mathscr D}_{\cal X}$ on complexified Kahler moduli in \eqref{ccsym}.  This is the key property of the central charge function used to define a stability condition on $D^b {\mathscr {A}}{\textup{-mod}}$ in \cite{BS}, for  ${\cal X} = T^*G/B$.

\subsection{Categorification of monodromy matrix elements}\label{s-MC}

We will now explain, from the perspective of the ${\cal N}=2$ sigma model on ${\cal X}$, why the theorem \ref{t:two} must hold. We will begin by restating it in a convenient way.

%\subsubsection{}
%Let  ${\cal X}_0 = {\cal X}_{{\fC}_0}$ and ${\cal X}_1 = {\cal X}_{{\fC}_1}$ be two symplectic resolutions.
% corresponding to a pair of chambers ${\fC}_0$ and ${\fC}_1$ in Kahler moduli.
Pick a path $B$ in complexified Kahler moduli, from ${\cal X}_0$ to ${\cal X}_1$, avoiding singularities.
Corresponding to $B$ is a derived equivalence functor,
${\mathscr B}: {\mathscr D}_{{\cal X}_0}\rightarrow {\mathscr D}_{{\cal X}_1}$, and ${\mathfrak B}: K_{\rm T}({\cal X}_0)\rightarrow K_{\rm T}({\cal X}_1)$ as the monodromy of the quantum differential equation. We will explain below how ${\mathscr B}$ and ${\mathfrak B}$ arize from the sigma model perspective.

Given a B-brane ${\cal F}_{0} \in {\mathscr D}_{{\cal X}_{0}}$, let ${\cal V}_{0}= {\cal V}[{\cal F}_{0}]$ be its vertex function.  The image of ${\cal V}_0$ under the monodromy ${\mathfrak B}$ of the quantum differential equation is  ${\mathfrak B} {\cal V}_0$, viewed as a vertex function on ${\cal X}_1$. We also get a B-brane  $ {\mathscr B} {\cal F}_0$  in ${\mathscr D}_{{\cal X}_{1}}$ as the image of ${\cal F}_{0}$ under ${\mathscr B}$. Then, for any other brane ${\cal F}_1\in {\mathscr D}_{{\cal X}_1}$ with vertex function ${\cal V}_1= {\cal V}[{\cal F}_{1}]$, the following holds.
\begin{theorem}\label{t:three}
The graded Euler characteristic of  
\beq\label{conn}
{\rm Hom}^{*,*}({\cal F}_1,  {\mathscr B} {\cal F}_0 ),
\eeq
computed in ${\mathscr D}_{{\cal X}_1}$ coincides with the matrix element
\beq\label{connect}
({\cal V}_1, {\mathfrak B} {\cal V}_0).
\eeq
\end{theorem}
While this merely restates theorem \ref{t:two}, the physics explanation for why it holds is very different from the proof in \cite{BO}, which uses representation theory in characteristic $p\gg 0$. 
%From this perspective, the theorem \ref{t:three} we will end up with is an independent statement.
\subsubsection{}

The vertex function ${\cal V}_0 = {\cal V}[{\cal F}_0]$ is the partition function of the supersymmetric sigma model with target ${\cal X}_0$ on ${\rm D}$, the infinitely long cigar with A-type twist in the interior, and with the boundary condition at infinity given by the brane ${\cal F}_0 \in {\mathscr D}_{{\cal X}_0}$.
We will choose the ``time" coordinate $s$ along the cigar ${\rm D}$ so that the boundary is at $s=0$ and the tip of the cigar at $s\rightarrow \infty$.  The vertex function can be viewed as an overlap ${\cal V}_0 = ({\cal V}_{{\cal X}_0}|{\cal F}_0)$, of a closed string vacuum state $({\cal V}_{{\cal X}_0}|$ obtained by computing the path integral over the cigar for $s\geq 0$, with a state $|{\cal F}_0)$, determined by the brane. 
To get the theory on the cigar to compute the vertex function  ${\mathfrak B}{\cal V}_0$, obtained from ${\cal V}_0$ by monodromy of the quantum differential equation, corresponding to the path $B$, we proceed as follows.

\subsubsection{}
Take the cigar ${\rm D}$ with the same brane ${\cal F}_0 \in {\mathscr D}_{{\cal X}_0}$ as the boundary condition at $s=0$, but now we modify the theory for $s>0$, by taking the Kahler moduli to vary near the boundary along the path $a=a(s)$, from $s=0$ to $s=1$, and to be constant in the interior, so that $a(s) = a_1$ for $s\geq 1$. Thus, in the interior of the cigar we get the sigma model on ${\cal X}_1$, and the moduli interpolate from ${\cal X}_1$ at $s=1$ to ${\cal X}_0$ at $s=0$ along the path $B$. We are using shortcut notation, where $a$ stands for $a = (a_1, \ldots , a_h)$ where $h$ is the rank of  $H^{1,1}({\cal X})$. The singularities, one recalls, are in complex co-dimension one and our path must avoid them. 

\begin{figure}[!hbtp]
  \centering
   \includegraphics[scale=0.48]{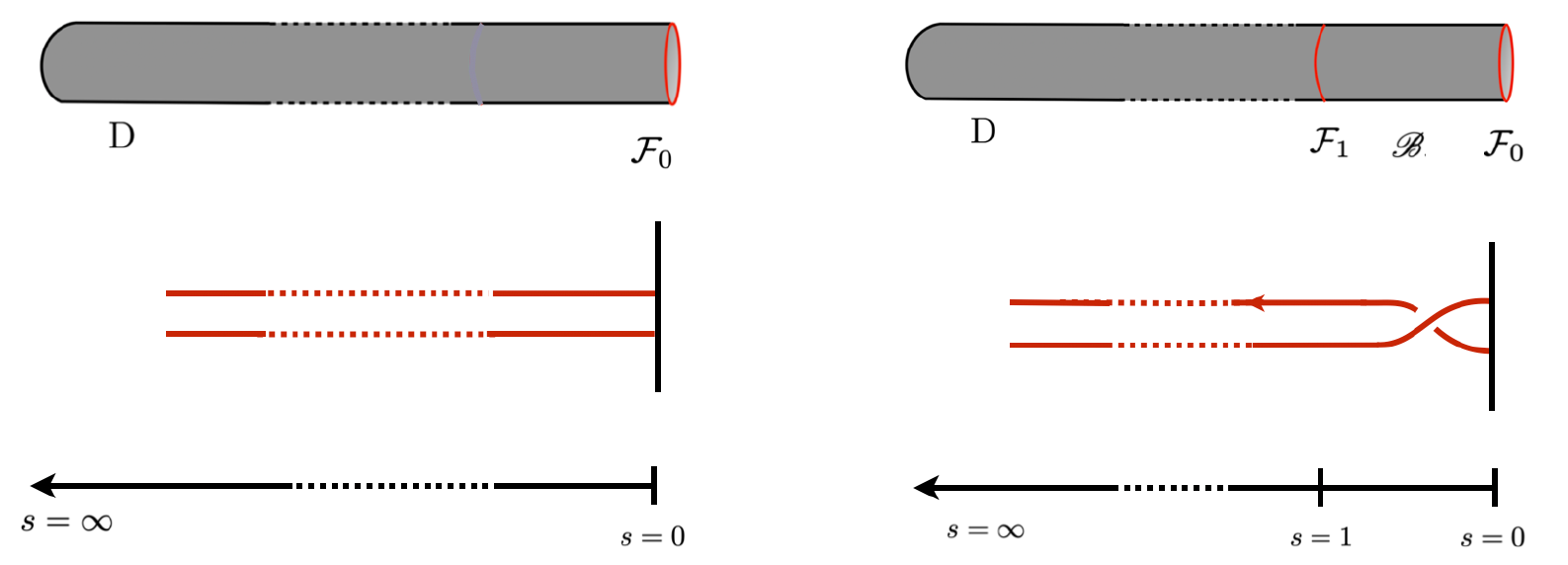}
 \caption{When ${\cal X}$ is the monopole moduli, the construction we just described is the sigma model realization of the action of braiding on conformal blocks from \cite{RCFT, MS}.}
  \label{f_gbraid}
\end{figure}
The path integral of the theory on ${\cal X}_1$ over the cigar for $s\geq 1$ produces the closed state vacuum state $({\cal V}_{{\cal X}_1}|$ at $s=1$. (The fact that the boundary of the cigar is at $s=1$ and not at $s=0$ is irrelevant, since both cigars are obtained from a hemisphere by adding an infinitely long neck.) Extending the path integral to include the interval from $s=1$ to $s=0$, the state we get at $s=0$ is obtained from $({\cal V}_{{\cal X}_1}|$ by evolution back in time with Kahler moduli which are taken to depend on time, as $a=a(s)$. %We will denote the state we get at ${s=0}$ by $({\cal V}_{{\cal X}_1}| {\mathscr B}$, where ${\mathscr B}$ denotes the time evolution operator, computed by the path integral of the theory on the annulus. 

To give an explicit characterization of the evolution of state $({\cal V}_{{\cal X}_1}|$ with parameters that vary with time is a Berry phase type-problem. The problem was solved for the class of theories at hand by \cite{CV, HIV, mirrorbook}. With the cigar of infinite length and A-twist in the interior, the relevant connection is the 
$tt^*$ connection of \cite{CV} on the space $H^*({\cal X})$ of supersymmetric ground states of the sigma model on ${\cal 
X}$, where one varies the complexified Kahler moduli of ${\cal X}$.  Written in the topological gauge, and in terms of flat coordinates used by the A-model, the $tt^*$ connection becomes the quantum connection of ${\cal X}_1$, defined solely in terms of quantum multiplication, with no reference to the $tt^*$ metric. 

It follows that, if path integral over the cigar with constant moduli and the brane ${\cal F}_0$ at the $s=0$ boundary computes the vertex function ${\cal V}_0 = {\cal V}[{\cal F}_0]$, the path integral with moduli that vary as we described computes  ${\mathfrak B}{\cal V}_0$, where ${\mathfrak B}$ is the monodromy of the quantum connection from $s=0$ to $s=1$ along the path $B$.
An alternative view of what we just did is to compute the overlap $({\cal V}_{{\cal X}_1}|{\mathscr B} {\cal F}_0)$
of the state $({\cal V}_{{\cal X}_1}|$, with the state $|{\mathscr B} {\cal F}_0)$ obtained from $|{\cal F}_0)$ by forward time evolution from $s=0$ to $s=1$, along the path $a=a(s)$. The state $|{\mathscr B} {\cal F}_0)$ corresponds to a B-type brane on ${\cal X}_1$ which we denote by ${\mathscr B} {\cal F}_0\in {\mathscr D}_{{\cal X}_1}$, and the overlap is its vertex function. In summary, we have shown that
${\mathfrak B}{\cal V}_0 = {\cal V}[{\mathscr B} {\cal F}_0].$

\subsubsection{}
As we will explain in more detail below, the path integral on the cigar depends on the homotopy type of the path $a=a(s)$. We call two paths $a_0(s)$ and $a_1(s)$ homotopic if there exists a smooth interpolation between them which avoids singularities of ${\cal X}$. This means we can take all the variation of $a=a(s)$ to happen in an arbitrarily small neighborhood of the $s=0$ boundary.  This way, the theory on the cigar that computes ${\cal V}[{\mathscr B} {\cal F}_0]$ can be taken to be on ${\cal X}_1$ for all times $s>0$, with the boundary condition at $s=0$ that defines the brane ${\mathscr B} {\cal F}_0 \in  {\mathscr D}_{{\cal X}_1}.$ 

Next, consider the same cigar which computed for us ${\cal V}[{\mathscr B} {\cal F}_0]$, except now we will cut the  cigar at $s=1$, by freezing the modes along the corresponding $S^1$ to lie on a subspace in field space corresponding to the brane ${\cal F}_1\subset {\mathscr D}_{{\cal X}_1}$, as in figure \ref{f_gbraid}. The path integral over the cigar with $s\geq 1$ with the boundary condition we just imposed computes ${\cal V}_1 = {\cal V}[{\cal F}_1]$. The path integral over the whole cigar, with $s\geq 0$ computes $  {\mathfrak B}{\cal V}_0={\cal V}[{\mathscr B} {\cal F}_0] $.
It follows that the path integral over the annulus with $s\in [0,1]$ with brane ${\cal F}_0$ as the boundary condition at $s=0$, brane ${\cal F}_1$ at $s=1$, and the moduli that vary over the interval according to the path $B$ computes the matrix element $
({\cal V}_1, {\mathfrak B}{\cal V}_0)$.

\subsubsection{}

\begin{figure}[!hbtp]
  \centering
   \includegraphics[scale=0.25]{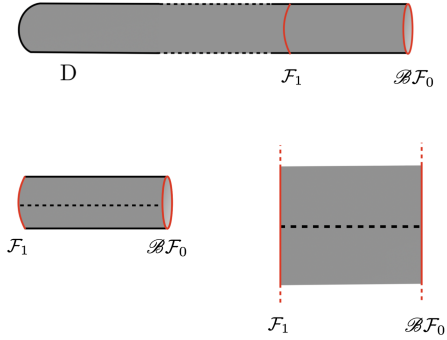}
 \caption{To extract matrix elements of ${\mathfrak B}$ from the sigma model, we cut the infinite cigar near the boundary, and insert a "complete set of branes". We recover the Homs which categorify the matrix elements by cutting the annulus open.}
  \label{f_Hom}
\end{figure}

The theory on the annulus which computes $
({\cal V}_{1}, {\mathfrak B} {\cal V}_{0})$ can be taken to be the supersymmetric sigma model with target ${\cal X}_1$, with branes ${\cal F}_{1} , \,{\mathscr B} {\cal F}_0 \in {\mathscr D}_{{\cal X}_1}$ imposing the boundary conditions at the two ends, and fermions which are periodic around the $S^1$, see figure \ref{f_Hom}.  (Fermions are periodic around the $S^1$ due to the A-type twist in cigar's interior  \cite{CV}. In the absence of the twist at the tip of the cigar, we would get fermions which are anti-periodic instead.)

In a theory such as ours, we have a choice of which direction we want to declare the Euclidian time. In computing the vertex functions via the topological A-model on ${\cal X}_0$ or ${\cal X}_1$, we took the time to run along the cigar, and hence along the annulus. Now, we will take the time to run around the $S^1$. Then, the path integral computes the supertrace, or the index of the supercharge $Q$-preserved by the branes at the two ends of the annulus. The fact one is computing an index means that the states on the interval $s\in [0,1]$ that contribute to the index are cohomology classes of a complex whose differential is the supercharge $Q$ preserved by the branes. Since the branes ${\mathscr B} {\cal F}_{0}$ and ${\cal F}_{1}$ are B-branes, objects of ${\mathscr D}_{{\cal X}_1}$, the cohomology of the supercharge $Q$ acting on the Hilbert space of the theory on the interval is
\beq\label{hoa}
{\rm Hom}^{*,*}({\cal F}_{1},  {\mathscr B}\, {\cal F}_{0} ),
\eeq
computed in ${\mathscr D}_{{\cal X}_1}$, by the usual connection between the category of B-branes on ${\cal X}_1$ and the derived category of coherent sheaves, see e.g. \cite{Hori} or \cite{AL, AspinwallL, Douglas, mirrorbook, Sharpe0, Sharpe}. 
The cohomology groups are bi-graded by cohomological and equivariant degrees corresponding to the ${\rm T}$-action on ${\cal X}_1$, with supercharge $Q$ which squares to zero, has cohomological degree $1$, and preserves ${\rm T}$-degree. 

Note that the theory on the annulus is not topologically twisted. The annulus being flat, twisting is not necessary.  This is fortunate, since the equivariant action breaks the $R$-symmetry needed to define the topological B-model on an arbitrary Riemann surface.

In formulating \eqref{hoa}, we have taken the moduli to be constant over the annulus, since this is the familiar setting. This is not necessary, as we will explain below in more detail. 

\subsubsection{}\label{s:conv}

Our notation, in Theorem \ref{t:three} and throughout, is such that ${\rm Hom}^{*,*}$ stands for
\beq\label{defhb}
{\rm Hom}^{*,*}({\cal F} , {\cal G}) = \bigoplus_{n \in {\mathbb Z}, {\vec k}\in {\mathbb Z}^{\rm{rk} T}} {Hom}_{{\mathscr D}_{\cal X}}({\cal F}, {\cal G}[n]\{\vec{k}\}),
\eeq
where $[n]$ denotes the homological grade shift, and $\{{\vec k}\}$ the ${\rm T}$-equivariant grade shift. (More precisely, ${\rm T}$-degrees live on a lattice which can be identified with ${\mathbb Z}^{\rm{rk} T}$ though not canonically. Sometimes fractional charges are more natural, as will be the case for $^LG = SU(n)$.) As is common, we will denote by ${\cal F}[n]$ a brane obtained from ${\cal F}$ by shifting its cohomological degree by $n\in {\mathbb Z}$. 
Objects of ${\mathscr D}_{\cal X}$ are also ${\rm T}$-graded, and all the morphisms in ${\mathscr D}_{\cal X}$ are ${\rm T}$-invariant.
Given a pair of ${\rm T}$-equivariant objects ${\cal F}$, ${\cal G}$, an equivariant degree $k$ morphism between them is becomes an element of ${Hom}_{{\mathscr D}_{\cal X}}({\cal F}, {\cal G}\{{\vec k}\})$.  Here ${\cal F}\{{\vec k}\}$ denotes the object ${\cal F}$ of $ {\mathscr D}_{\cal X} =D^bCoh_{\rm T}({\cal X})$, with equivariant grade shifted by ${\vec k} \in {\mathbb Z}^{\rm rk T}$. 

The elements of ${Hom}_{{\mathscr D}_{\cal X}}({\cal F}, {\cal G}\{{\vec k}\})$ are the same as morphisms from ${\cal F}$ to ${\cal G}$ in degree $\{{\vec k}\}$, so we write ${Hom}^*({\cal F}, {\cal G}\{\vec{k}\}) =  {Hom}^{*, \vec{k}}({\cal F}, {\cal G})$. 
Similarly, ${Hom}^{n, {\vec k}}({\cal F}, {\cal G}) = Hom_{{\mathscr D}_{\cal X}}({\cal F}, {\cal G}[n]\{\vec{k}\})$. This second viewpoint, where the it is the morphisms rather than branes which carry the equivariant degrees, is more familiar to physicists.

The Euler characteristic of \eqref{defhb}, where we keep track of equivariant grades, is
\beq\label{defhc}
 \chi({\cal F}, {\cal G})= \sum_{n \in {\mathbb Z}, {\vec k}\in {\mathbb Z}^{\rm{rk} T}}  (-1)^n {\fq}^{k_0-D/2} {\rm dim} \;{Hom}_{{\mathscr D}_{\cal X}}({\cal F}, {\cal G}[n]\{{\vec k}\}).
\eeq
It depends on equivariant K-theory classes of ${\cal F}$ and ${\cal G}$ only. Above, $D$ is half the complex dimension of ${\cal X}$, $D = {1\over 2 }dim_{\mathbb C}{\cal X}$. The shift in the ${\mathbb C}^{\times}_{\fq} \subset {\rm T}$ degree by $-D/2$ is introduced for the Euler characteristic to be invariant under Serre duality, $ \chi({\cal F}, {\cal G}) =  \chi({\cal G}, {\cal F})^*$.

\subsection{Invariance under cobordisms}
An important role in the preceding discussion is played by the fact that the theory depends only on the homotopy type of path $B$ in complexified Kahler moduli, and not on the choice of the path $a=a(s)$ itself. Recall that, by homotopy, we mean homotopy in complexified Kahler moduli with singularities in complex co-dimension one removed.  Now we will explain why this is the case.  

In the course of explaining why the homology group  ${\rm Hom}^{*,*}({\cal F}_{1},  {\mathscr B}\, {\cal F}_{0} )$  depends only on homotopy type of path $B$, we will also understand the relation between homology groups corresponding to a pair of paths $B_0$ and $B_1$ which are not homotopic.

For ${\cal X}$ which is the the main subject of our paper, a path $B$ in Kahler moduli with singularities removed is a colored braid in ${\cal A}\times [0,1]$. A pair of paths $B_0$ and $B_1$, defined up to homotopy, are related by a braid cobordism $S: B_0\rightarrow B_1$. We will discover that 
cobordisms of braids induce the corresponding maps on homology groups. The fact that homology groups are cobordism invariants is a key aspect of deeper structure one expects when working with homology as opposed to with Euler characteristics, as emphasized in \cite{DBN}. This emerges naturally for us, from the sigma model origin of geometric structures at hand. The fact that theory makes sense even if we let the moduli vary is due to \cite{GW, GMW1, GMW2}, in a setting related to our current one by mirror symmetry \cite{A2}.

\subsubsection{}

Cut open the annulus to a strip $\Sigma$, consisting of the interval $s\in [0,1]$ times the open string time $t$. The theory on $\Sigma$ can be viewed as supersymmetric quantum mechanics, which is a sigma model on an infinite dimensional space ${\mathscr  X}$ of maps $x:[0,1]\rightarrow {\cal X}$, constrained by the boundary conditions corresponding to the pair of B-branes ${\cal F}_0$ and ${\cal F}_1$ at $s=0$ and $s=1$. 
The metric on the target ${\mathscr X}$ is inherited from the metric on ${\cal X}$. For the time being, we will keep the moduli of ${\cal X}$ fixed, independent of the point on $\Sigma$.

For example, if ${\cal F}_{0}$ and ${\cal F}_1$ are vector bundles on ${\cal X}$, 
the target ${\mathscr X}$ of the supersymmetric quantum mechanics is the space of maps from $[0,1]$ to ${\cal X}$, obeying Neumman boundary conditions $\partial_sx =0$ at both boundaries. 
%
%By writing fermions of  the ${\cal N}=(2,2)$ theory: $\psi^{\overline b} ={\overline  {\psi_+^{ b}}}+{\overline  {\psi_-^{ b}}}$ and $\eta^{\overline b} ={\overline  {\psi_+^{ b}}}-{\overline  {\psi_-^{ b}}}$, the boundary condition sets $\partial_sx^a =0,$ at the boundary is supplemented by setting $\eta^a=0$ at the boundary. 
In quantum theory, the fermions of the ${\cal N}=2$ sigma model act as: 
$$\begin{aligned}
\psi^{\overline b}(s) \; \leftrightarrow\;  \delta\overline{x^{b}}(s)\wedge, \qquad  \eta^{\overline b}(s) \; \leftrightarrow\; g^{{\overline b} a} {\delta \over \delta {x^{a}(s)}}
\end{aligned}
$$
whose conjugates correspond to $\psi^{ b}(s)$ and $\eta^{b}(s)$. (This is worked out in \cite{HIV, mirrorbook}, for example. The boundary condition that sets $\partial_sx^a =0 $ at the boundary is paired by supersymmetry with one that sets $\eta^a=0$).  The Hilbert space ${\mathscr H}$ of the quantum mechanics is 
\beq\label{Hilbs}
{\mathscr H } = \Omega^{0, \star}({\mathscr X}, x_0^*{\cal F}_{0}^{\vee} \otimes x_1^*{\cal F}_{1} \otimes {\bigwedge}^\star T{{\mathscr X}}),
\eeq
the space of anti-holomorphic forms on ${\mathscr X}$, valued in a vector bundle.
Above, $ \bigwedge^\star T{{\mathscr X}} $ are exterior powers of tangent bundle to ${\mathscr X}$, spanned by vector fields $\delta/ \delta {x^{a}(s)}$, and $x_{0,1}: {\mathscr X}\rightarrow {\cal X}$ are holomorphic maps that send a path $x(s)$ to ${x(0)}$ and to $x(1)$. 
%The Hilbert space ${\mathscr H}$ is the space of anti-holomorphic forms, valued in a vector bundle over ${\mathscr X}$, on which $Q$ acts as a Dolbeault type-operator $Q = \overline{\delta_{A}}  + {\cal T}$, where
%
The supercharge $Q$ is a nilpotent operator, $Q^2=0$, which
% = \int_{0}^1 \Bigl(g_{a{\overline b}} \partial_t x^a \psi^{\overline b} + g_{a{\overline b}} \partial_s x^a \eta^{\overline b}\bigr) + A_{\overline a} \psi^{\overline a}|_{s=1} - A_{\overline a} \psi^{\overline a}|_{s=0}$
acts on ${\mathscr H}$ as a version of an equivariant Dolbeault operator  
\beq\label{Qop}
Q = {\overline \delta_{A}}  + {\cal T}\wedge.
\eeq
where
\beq\label{Ta}{\cal T} =  \int_{0}^1 ds \,  ( u\, g_{a {\overline b}}T^a(x) \delta{\overline x^{b}(s)}  +  \partial_s x^a(s){\delta \over \delta {x^{a}(s)}}),
\eeq
is the action of holomorphic vector fields on ${\mathscr X}$. The first term in ${\cal T}$ comes from the ${\rm T}$-action on ${\cal X}$. One recovers the usual description of the cohomology of $Q$ as $Hom^{*,*}({\cal F}_1, {\cal F}_0)$ by localization with respect to the ${\cal T}$-action - the fixed points of the second term in ${\cal T}$ are constant maps.

%
%The boundary conditions set $\eta^{\overline b}$ and $\partial_s x^a$ to zero at $s=0,1$. 
%
%where 
%$\delta_A$ is the Dolbeault operator on ${\mathscr X}$,
%$${\overline \delta_{A} =  \int_{0}^s ds\,\delta\overline{x^{b}}(s) {\delta \over \delta \overline {x^{a}}(s)} \; +  \;\delta\overline{x^{b}} A_{b}^{(1)}|_{s=1} -\delta\overline{x^{b}}A_{b}^{(0)}|_{s=0} 
%$$
%and ${\cal T}$ a holomorphic vector field acting on ${\mathscr X}$, 
%$${\cal T} =  \int_{0}^1 ds \,  \partial_s x^a(s){\delta \over \delta {x^{a}(s)}}.$$ 
%(We are suppressing the ${\rm T}$-action for the time being, since it is not important for the present discussion.)
%$${\cal T} =  \int_{0}^1 ds \,  \partial_s x^a(s){\delta \over \delta {x^{a}(s)}}.
%$$
%

The description of the theory on $\Sigma$ in terms of supersymmetric quantum mechanics lets us borrow many explanations from sections 10 and 11 of \cite{GMW1}. The key fact we will need is that 
the Hamiltonian of the theory on $\Sigma$ depends on the Kahler metric $g_{a{\overline b}}(x)$, complexified by the $B$-field, only through $Q$-exact terms -- terms of the form 
$\{Q, V\}$,  see for example  \cite{HIV, mirrorbook}. 
The theory is also insensitive to modifying $V$ by boundary terms, examples of which lead to holomorphic gauge transformations
which keep the holomorphic structure of ${\cal F}_0$ and ${\cal F}_1$ fixed, see for example \cite{Hori, GMW1}.

\subsubsection{}
Now let the moduli $a=a(s,t)$ vary smoothly over $\Sigma$, so that for every $s$ and $t$, the metric on ${\cal X}$ remains Kahler, $g_{a{\overline b}} = g_{a{\overline b}}(x; s,t)$, the $B$-field of type $(1,1)$. We keep the complex structure on ${\cal X}$ fixed, so that for every $s,t$, the ${\rm T}-$action on ${\cal X}$ remains a symmetry, as does the $U(1)_V$ R-symmetry that generates the fermion number, or cohomological grading. $Q$ has fermion number $1$ and ${\rm T}$-charge $0$.  

As a result of the explicit $s$- and $t$- dependence of the moduli, the Hamiltonian $H$ of our supersymmetric quantum mechanics becomes explicitly time dependent, $H=H(t)$. Despite the fact time translations are no longer a symmetry, all the relevant structure of the theory is preserved. We will start with the ${\rm T}$-action on ${\cal X}$ turned off, by setting the equivariant parameters $u$ in \eqref{Ta} to zero. Then, the B-model supercharge $Q$ (unlike ${\overline Q}$) generates a symmetry. Namely, it has no explicit time dependence, and commutes with the Hamiltonian which is $Q$-exact, of the form $H(t) = \{Q, V(t)\}$, so ${d\over dt} Q={\partial \over \partial t} Q + [H(t),Q]=0.$  
%Thus even if we let the complexified Kahler metric vary over $\Sigma$, and in this time dependent situation, the cohomology of $Q$ acting on the Hilbert space of the theory at any given time. 

The fact that $Q$ does not depend on time is no longer true after we turn the ${\rm T}$-action back on, since the term proportional to $u$ explicitly depends on the metric and hence on time.  $Q$ however remains nilpotent, it commutes with the Hamiltonian which remains $Q$-exact, and the cohomology of $Q$ remains time-independent even if the operator itself does depend on time. This follows since the supercharge $Q_t$ that acts on the Hilbert space ${\mathscr H}_t$ at time $t$ is related to the operator $Q_t|_{u=0}={\widehat Q}_t$  with no explicit time dependence by conjugation with an invertible operator: $Q_t = e^{u h(t)} {\widehat Q}_t e^{-uh(t)}$. Here $h$ is the potential on ${\mathscr X}$ which acts as the moment map for the $T$-action. The potential $h = h(t)$ becomes time dependent once the metric does.

Thus, at every time $t$,  we have an invertible map from a cohomology class of the time-independent operator ${\widehat Q}_t$ to a cohomology class of $Q_t$. The operator $e^{u h(t)}$ maps a ${\widehat Q}_t$-closed state to a $Q_t$-closed state, and a pair of states in the same cohomology class of ${\widehat Q}_t$ to a pair of states in the same cohomology class of $Q_t$ -- and vice versa. From now on, we will work with  the cohomology of the operator ${\widehat Q}_t$, which is technically simpler because the operator itself has no explicit time dependence.
(One subtlety is that, since ${\cal X}$ is non-compact, some states may become non-normalizable, by setting $u=0$. A way to deal with this is to view the moduli as smoothly varying, and split $h(t)$ into a time-independent potential and a slowly time varying part, which comes from varying moduli. In practice, of most interest to us will be branes with compact support on ${\cal X}$, for which the issue cannot arize anyhow.)
\subsubsection{}

Take now a pair of paths $B_0$ and $B_1$, for which $a= a_0(s)$ and $a= a_1(s)$, and which avoid singularities in complexified Kahler moduli. $B_0$ and $B_1$ are of the same homotopy type if we can find a family of paths $B_t$ for $t\in [0,1]$ that interpolate between them while avoiding
singularities. A path $B_t$ is described by  $a= a(s,t)$ where $a(s, 0 ) = a_0(s)$ and $a(s,1) = a_1(s)$. 
%
%Corresponding to the pair of paths $B_0$ and $B_1$, we get a pair of derived equivalence functors ${\mathscr B}_0$ and ${\mathscr B}_1$.

As in any quantum mechanical theory, the Hilbert spaces ${\cal H}_0$ and ${\cal H}_1$ are related by the time evolution operator which is the path ordered exponential ${\widehat U} =Pe^{-\int_{0}^1 {\widehat H}(t)dt}$ of the Hamiltonian ${\widehat H}(t) = H(t)|_{u=0}$, and which takes ${\widehat U}:{\cal H}_0 \rightarrow {\cal H}_1$. 
Since the Hamiltonian is ${\widehat Q}$-exact, and ${\widehat Q}$ is time-independent, the time evolution operator commutes with the supercharge ${\widehat Q}$. Denoting by ${\widehat Q}_0$ and ${\widehat Q}_1$, the supercharge ${\widehat Q}$ acting on the Hilbert space of the theory at $t=0$ and $t=1$, respectively, then
\beq\label{commute}
{\widehat U} {\widehat Q}_0 = { \widehat Q}_1 {\widehat U}.
\eeq
Thus, the operator $\widehat U$ corresponding to time evolution from path $a_0(s)$ to $a_1(s)$ 
gives us a map of the cohomology of the supercharge ${\widehat Q}_0$ acting on the Hilbert space ${\cal H}_0$ of the theory at $t=0$, to the cohomology of the supercharge ${\widehat Q}_1$ acting on the Hilbert space ${\cal H}_1$ of the theory at $t=1$.
This map turns out to depend only on the homotopy type of the path $a=a(s,t)$ interpolating between them. 
We will review the argument for this, which is a variant of the argument in \cite{GMW1}.

Suppose we choose a different path interpolating between $B_0$ and $B_1$. We will get a different time evolution operator ${\widehat U}' = Pe^{\int_{0}^1 {\widehat H}'(t)dt}$ which satisfies \eqref{commute} with ${\widehat U}$ replaced by ${\widehat U}'$. The two operators are not equal, since in general ${\widehat H}'(t)$ corresponding to the path $a=a'(s,t)$ is not the same as the Hamiltonian ${\widehat H}(t)$, corresponding to $a=a(s,t)$. However, since both Hamiltonians are $Q$-exact, the difference of two evolutions is of the form ${\widehat U}' - {\widehat U} = E {\widehat Q}_0 -{\widehat  Q}_1 E'$ where $ E, E'$ are operators that map ${\cal H}_0$ to ${\cal H}_1$  of cohomological degree $-1$ and vanishing ${\rm T}$-degree. While the time evolutions by ${\widehat U}$ and ${\widehat U}'$ are not the same, they induce equivalent maps of cohomology classes.

The cohomologies of ${{\widehat Q}}_0$ acting on ${\cal H}_0$ and ${{\widehat Q}}_1$ acting on ${\cal H}_1$ are equivalent if we can find a map going the other way which on cohomology acts as the inverse of $U$.  A direct way to do that is to note that if, for each $t$, the Hamiltonians $H(t)$ are well defined Hermitian operators acting on the Hilbert space ${\cal H}_t$ than map ${\widehat U}$ has an inverse ${\widehat U}'' = P e^{ -\int_0^1 {\widehat H}(t)dt}$, that simply propagates backwards. Thus for two homotopic paths $B_0$ and $B_1$, the cohomology groups $H_0 = {\rm Ker}({\widehat Q}_0)/{\rm Im}({\widehat Q}_0)$ and  $H_1={\rm Ker}({\widehat Q}_1)/{\rm Im}({\widehat Q}_1)$ are isomorphic.

There is a more elaborate explanation, which becomes essential in understanding what happens when $B_0$ and $B_1$ are {\it not} homotopic. 
Pick a path $a''(s,t)$ that interpolates from $B_1$ to $B_0$, so that $a''(s,0) = a_1(s)$, and $a''(s,1) = a_0(s)$. The corresponding Hamiltonian ${\widehat H}''(t)$ gives a map ${\widehat U}'' = P e^{i \int_0^1 {\widehat H}''(t)dt}$ from ${\cal H}_1$ to ${\cal H}_0$ that commutes with the supercharge ${\widehat Q}$,  ${\widehat U}''{\widehat Q}_1 = {\widehat Q}_0 {\widehat U}''$, and thus maps a cohomology class of the supercharge ${\widehat Q}_1$ acting on ${\cal H}_1$ to a cohomology class of the supercharge ${\widehat Q}_0$ acting on ${\cal H}_0$.  Now consider the compositions ${\widehat U} {\widehat U}''$ and ${\widehat U}''{\widehat U}$. If ${\widehat U}$ and ${\widehat U}''$ corresponded to a pair of paths from $B_0$ to $B_1$ which had the same homotopy type,
${\widehat U}{\widehat U}''$ represents a closed loop homotopic to identity, and correspondingly, it acts on the cohomology of ${\widehat Q}_0$ as the identity operator. Similarly, ${\widehat U}'' {\widehat U}$ acts on cohomology of ${\widehat Q}_1$ as the identity. Thus, ${\widehat U}''$ is a two sided inverse of ${\widehat U}$, and we have shown that the cohomology groups ${\rm Ker}({\widehat Q}_0)/{\rm Im}({\widehat Q}_0)$ and $ {\rm Ker}({\widehat Q}_1)/{\rm Im}({\widehat Q}_1)$ are the same. 

Having understood the fact that cohomology of the operator ${\widehat Q}$ depends only on the homotopy type of the path $B$, we get the same result for the cohomology of the supercharge $Q$ of the theory with $u\neq 0$, since the two are related by conjugation, as we explained earlier. 

\subsubsection{}
Now consider a pair of paths $B_0$ and $B_1$ which avoid the singularities in the moduli space, but which are not  homotopic to each other. 
In this case, it is not possible to find a smooth interpolation between $B_0$ and $B_1$, as a family of paths $B_t$ which avoid singularities of ${\cal X}$. 
We can still find an interpolation between $B_0$ and $B_1$ which is smooth except at some finite number of isolated points $P_1, \ldots P_k$ on $\Sigma$, where ${\cal X}$ must develop singularities. (As always, by a singularity of ${\cal X}$ we mean a singularity in complex co-dimension one of its complexified Kahler moduli, where the physical sigma model on ${\cal X}$ becomes singular - and not the singularity in the classical geometry of ${\cal X}$.)

As long as the sigma model on ${\cal X}$ develops singularities only at isolated points $P_i$ on $\Sigma$,  the path integral of the theory should still provide an operator ${\widehat U}: {\cal H}_0 \rightarrow {\cal H}_1$  that maps the Hilbert space of the theory at $t=0$ to that at $t=1$. As ${\widehat U}$ is the time evolution operator, it takes the supercharge ${\widehat Q}_0$ acting on ${\cal H}_0$ to supercharge ${\widehat Q}_1$ acting on ${\cal H}_1$ by ${\widehat U}{\widehat Q}_0 = {\widehat Q}_1 {\widehat U}$.  Correspondingly, the operator ${\widehat U}$ provided by the path integral of the theory on $\Sigma$, maps the cohomology of ${\widehat Q}_0$ at $t=0$ to cohomology of ${\widehat Q}_1$ at $t=1$. 

The price to pay for the fact $B_0$ and $B_1$ are not homotopic is that the operator ${\widehat U}$ may fail to be invertible. This is forced on us, since in general, the cohomology groups at $t=0$ and $t=1$ will have different dimensions. Picking a path $a''(s,t)$ that takes $B_1$ to $B_0$,  $a''(s,0)=a_1(s)$ and $a''(s,1) = a_0(s)$, we will get an analogous map ${\widehat U}''$ that similarly satisfies ${\widehat U}'' {\widehat Q}_1 = {\widehat Q}_0 {\widehat U}''$, and maps cohomology of ${\widehat Q}_1$ to cohomology of ${\widehat Q}_0$. Their compositions ${\widehat U} {\widehat U}'' $ and  ${\widehat U}'' {\widehat U}$ act as identity on cohomology only if the corresponding paths, from $B_0$ and $B_1$ back to themselves, are homotopic to a trivial, constant path.
\subsubsection{}

Specializing back to the setting of ${\cal X}$ which is the main subject of the paper, $B_0$ and $B_1$ become braids in ${\cal A} \times [0,1]$. The family of braids $B_t$ that interpolate between $B_0$ and $B_1$ describe a two-real dimensional surface in four real-dimensional space:
\beq\label{surf}
S\in {\cal A} \times \Sigma.
\eeq
The surface $S$ is called a braid cobordism if the projection of $S$ to $\Sigma =[0,1]^2$ contains simple branch-points only, and where in addition $z(0,t)$ and $z(1,t)$ are taken to be trivial braids, see for example \cite{KT}. Simple branch points correspond to a single pair of strands crossing, or equivalently, crossing a single wall in Kahler moduli of ${\cal X}$. More precisely, we are only interested in isotopy classes of surfaces $S$ (where we keep the boundary fixed), since any two surfaces $S$ and $S'$ that are homotopic modulo the boundary give equivalent maps of cohomology groups as we saw above.

\begin{figure}[!hbtp]
  \centering
   \includegraphics[scale=0.4]{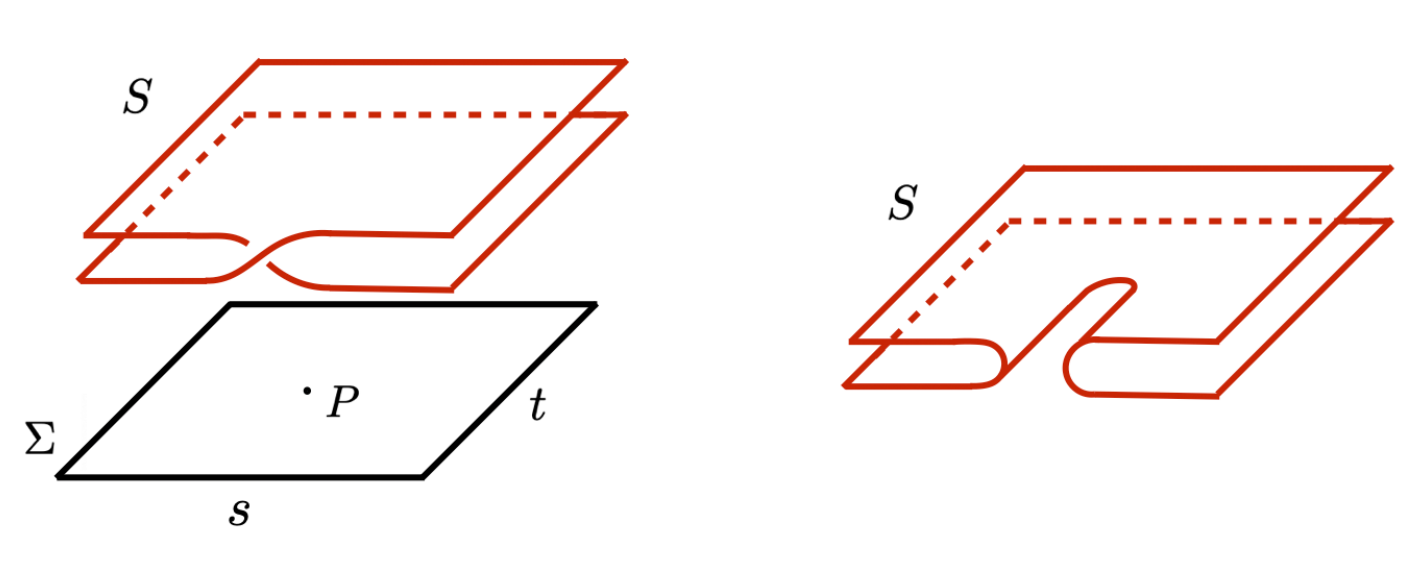}
 \caption{Two views of the same cobordism $S$ near its simple branch point. The map $S\rightarrow \Sigma$ is modeled locally by a Riemann surface $z^2 = u$, with $u=s+it$. Cobordisms are not holomorphic maps, in general. Time reversal of the one given here is modeled by $z^2 = {\overline u}$.}
  \label{f_cobor}
\end{figure}

The four manifold ${M}_4 ={\cal A}\times [0,1]^2 $ differs from the three manifold ${M}_3 ={\cal A}\times[0,1]$ where Chern-Simons theory "lives" by addition of the time direction, parameterized by $t\in [0,1]$. 
 As we will explain in \cite{A3}, one should view the surface $S$ as supporting a certain two dimensional defect in a six dimensional theory on $M_6 = M_4 \times {\mathbb C}$. The six dimensional theory is the $(0,2)$ theory of type $\fg$, which arizes in string theory. The two dimensional defects are what introduces knots and links in Chern-Simons theory, from string theory perspective. The theory on the defects has a description in terms of sigma model that describes maps $\Sigma \rightarrow {\cal X}$. The fact that when we chose the parameters of ${\cal X}$ to vary over $\Sigma$, the theory develops singularities at points $P_1, \ldots , P_k$ means that at those points, the description we chose breaks down. The quantum theory with defect supported on a smooth surface $S$ should make sense. This perspective also helps us understand why the time evolution map ${\widehat U}: {\cal H}_0\rightarrow {\cal H}_1$ fails to be invertible in presence of a branch point: The surface $S$ has nontrivial topology, see figure \ref{f_cobor}. In the example in the figure, if ${\widehat U}$ is the time evolution corresponding to the surface $S$ in the figure, and  ${\widehat U}''$ is its time reversal and the surfaces corresponding to ${\widehat U} {\widehat U}''$ and ${\widehat U}''  {\widehat U}$ are distinct and distinct from the surface corresponding to the identity evolution. 
Using mirror symmetry \cite{A2}, one can give a fairly explicit description of operator ${\widehat U}$ associated to the surface $S$, which in broad strokes follows the description in \cite{KT}, section 3.4.

\section{Categorification from ${\mathscr D}_{{\cal X}} = D^bCoh_{\rm T}({\cal X})$}\label{s-three}

We now return to study ${\cal X}$, the moduli of monopoles, whose vertex functions are conformal blocks of $\Lfgh$.
The complexified Kahler moduli of ${\cal X}$ is the configuration space of $n$ colored points on ${\cal A}$, modulo translations, and a path $B$ in the moduli space is a braid in ${\cal A}$ times ``time" parameterized by $s$. The action of monodromy ${\mathfrak B}$ on the vertex functions gives a geometric interpretation to action of braiding on conformal blocks. The sigma model origin of vertex functions implies
the action of braiding $B$ on the derived category is a derived equivalence functor 
\beq\label{Be}
{\mathscr B}: {\mathscr D}_{\cal X}  \cong {\mathscr D}_{{\cal X}'},
\eeq
on very general grounds, described in the previous section. In this section, we will describe the action of the functor ${\mathscr B}$ which one gets in our specific case. The specific geometry of ${\cal X}$ and the fact that its quantum differential equation is the Knizhnik-Zamolodchikov equation will now play an important role.

Consider a braid $B$ corresponding to reordering of a consecutive pair of vertex operators on ${\cal A}$
\beq\label{oporder}
\Phi_{V_i}(a_i) \otimes \Phi_{V_j}(a_j) \;\;\longrightarrow \;\;\Phi_{V_j}(a_j) \otimes \Phi_{V_i}(a_i),
\eeq
as in the figure below.
\begin{figure}[!hbtp]
  \centering
   \includegraphics[scale=0.4]{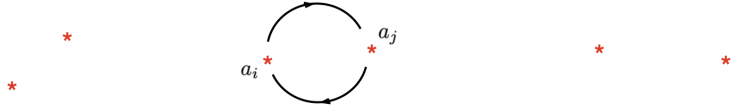}
 \caption{A path $B$ which is braiding a single pair of vertex operators.}
  \label{f_wall}
\end{figure}
Geometrically, the choice of ordering of vertex operators with respect to $y=\ln|a|$ is the choice of a chamber ${\fC} = {\fC}_{\vec \mu}$ in the real Kahler moduli of ${\cal X}$. The path $B$ goes from a chamber
${\fC} = {\fC}_{\vec \mu}$, in which $y_i < y_j$, to a chamber  ${\fC}' ={\fC}_{\vec \mu'}$, in which $y_j<y_i$. Going from one chamber to the other, changes the resolution of slices as:
$$
{\cal X} = {{\rm Gr}^{{\vec \mu}}}_{\nu}\;\; \longrightarrow \;\;{\cal X}' = {{\rm Gr}^{{\vec \mu'}}}_{\nu},
$$
where ${\vec \mu} =(\ldots, \mu_i, \mu_j, \ldots) $ and ${\vec \mu'} = (\ldots, \mu_j, \mu_i, \ldots)$. The chambers  ${\fC}$ and  ${\fC}'$ intersect in a real co-dimension 1-wall at $y_i = y_j$, although the singularity in the complexified Kahler moduli occurs only when the two vertex operators coincide, at $a_i=a_j$. The choice of the path $B$ is a choice of B-field on the wall separating the two chambers. Along the wall the real part of the complexified Kahler modulus vanishes. The imaginary part, given by the $B$-field, does not, and its sign distinguishes the two ways of going around the singularity.

To get a good description of the functor ${\mathscr B}$, we need to start by understanding the geometry of ${\cal X}$ and ${\cal X}'$ near the wall. 
As $a_i$ approaches $a_{j}$, ${\cal X}$ and ${\cal X}'$ develop singularities. We will see that the singularities that can occur reflect what happens in conformal field theory as $a_i$ approaches $a_j$, since conformal blocks are the generalized central charges of branes. This follows one of the lessons of mirror symmetry, which is that the behavior of central charges (or more precisely, of the $\Pi$-stability central charge ${\cal Z}^0$) near a singularity of ${\cal X}$, gives one a good picture of what happens to the geometry, and ultimately, to the derived category as well.

In conformal field theory, as $a_i$ approaches $a_j$, one gets a new natural basis of conformal blocks, corresponding to fusing $\Phi_{V_i}(a_i)$ and $\Phi_{V_j}(a_j)$, to $\Phi_{V_k}(a_j)$. The importance of this basis is that it diagonalizes the action of braiding ${\mathfrak B}$
on the space of conformal blocks. The eigenvalues of ${\mathfrak B}$ are labeled by representations $V_{k}$ in the tensor product $V_i\otimes V_j$.

While we can diagonalize the action of braiding by ${\mathfrak B}$ on conformal blocks, in general, eigenvectors of ${\mathfrak B}$ do not come from eigenvectors of ${\mathscr B}$, so the action of braiding on  ${\mathscr D}_{\cal X}$ and on ${\mathscr D}_{{\cal X}'}$ cannot be diagonalized. Instead, we will get a filtration whose terms are labeled by fusion products, and which is defined in terms of the behavior of the central charge function ${\cal Z}$ near the wall. The filtration is an example of filtrations considered by Cheung and Rouquier in \cite{CR}. Its existence implies that the functor ${\mathscr B}$ relating ${\mathscr D}_{\cal X}$ and ${\mathscr D}_{{\cal X}'}$ is given by degree shifts, computable in terms of eigenvalues of the braiding matrix acting on the fusion products, on certain quotient subcategories which we will describe. 

We learned about the role of filtrations for derived equivalences from Andrei Okounkov. The structure we will find here was anticipated in \cite{BS}, albeit in characteristic $p\gg0$.

\subsection{Fusion and vanishing cycles}

In conformal field theory, for widely separated vertex operators, the natural basis of conformal blocks is obtained by representing $\Phi_{V_i}(a_i)$ and $\Phi_{V_j}(a_j)$ as intertwiners of Verma module representations and sewing. As $a_i$ and $a_j$ approach each other, a different basis of solutions to the KZ equation is more natural.  

\subsubsection{}
\begin{figure}[!hbtp]
  \centering
   \includegraphics[scale=0.37]{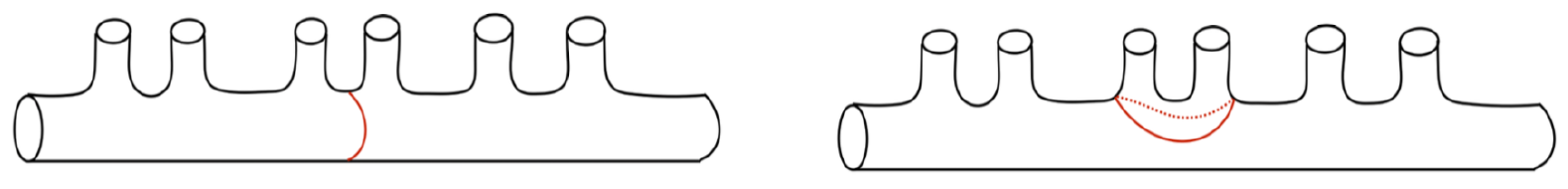}
 \caption{Two decompositions of the same Riemann surface.}
  \label{f_cut}
\end{figure}

\begin{figure}[!hbtp]
  \centering
   \includegraphics[scale=0.33]{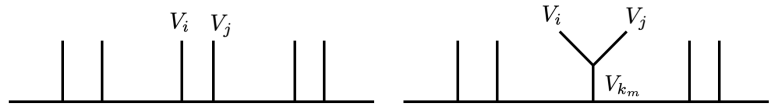}
 \caption{The second decomposition corresponds to fusing a pair of vertex operators.}
  \label{f_fuse}
\end{figure}

The two basis of confomal blocks correspond to two different sewing prescriptions of the underlying Riemann surface. The two sewings of the Riemann surface are described explicitly figure \ref{f_cut}, 
and more schematically in figure \ref{f_fuse}. 
The left sewing prescription corresponds to representing chiral vertex operators $\Phi_{V_i}(a_i)$ and $\Phi_{V_j}(a_j)$ as intertwiners of Verma module representations living on the horizontal legs of figure \ref{f_fuse} (see also figure \ref{f_blocks}). The right sewing corresponds to first bringing $\Phi_{V_i}(a_i)$ and $\Phi_{V_j}(a_j)$ together and taking their operator product expansion. 

Leading terms of the operator product expansion 
\beq\label{fuse}
\Phi_{V_i}(a_i) \otimes \Phi_{V_j}(a_j) \;\; \sim \;\; (a_i-a_j)^{h_k - h_i - h_j} \Phi_{V_k}(a_j),
\eeq
as $a_i$ tends to $a_j$, describe the Riemann surface which develops a long neck, corresponding to replacing $\Phi_{V_i}(a_i) \otimes \Phi_{V_j}(a_j)$ by a single vertex operator $\Phi_{V_k}(a_j)$. Here 
 $V_k$ can be any representation in the tensor product of $V_i$ and $V_j$, 
\beq\label{tensor}
{V}_i \otimes {V}_j =\bigotimes_{m=0}^{m_{max}} {V}_{k_m},
\eeq
since we are taking $\kappa \in {\mathbb C}$.  We will order the representations so that the highest weights $\mu_{k_m} $ of ${V}_{k_m}$ satisfy 
\beq\label{orderh}
\mu_{k_m} \leq \mu_{k_{m+1}},
\eeq for all $m$,
which means that $\mu_{k_{m+1}}-\mu_{k_m}$ is a sum of positive roots, as before. Above, $h_j$ is the conformal dimension of vertex operator $\Phi_{V_j}$.
%In \eqref{tensor}, we ordered the representations so that the highest weights $\mu_{k_m} $ of ${V}_{k_m}$ satisfy $\mu_{k_m} \leq \mu_{k_{m+1}}$ for all $m$. 
Unlike $V_{i}$ and $V_j$, the representations $ {V}_{k_m}$ in their tensor product are generically not minuscule. The subleading terms in the operator product expansion which we omitted in writing \eqref{fuse}, are fixed by conformal invariance, and correspond to working with the Riemann surface whose neck is finite, as in the figures above.

Solutions of the KZ equation where $\Phi_{V_i}(a_i)$ and $\Phi_{V_j}(a_j)$ fuse to $\Phi_{V_k}(a_j)$ can be identified by their asymptotic behavior as $a_i\rightarrow a_j$, 
 \beq\label{vanishing}
{\cal V}_k \;   \;\; = \;\; 
 (a_i-a_j)^{h_k-h_i-h_j} \times\textup{finite},
\eeq
where ``finite'' stands for terms that are single valued and non-vanishing in the limit. This follows from operator product expansion, the leading term of which is \eqref{fuse}. The subleading terms come from descendants of $\Phi_{V_k}(a_j)$ and have the same behavior under braiding $a_i$ and $a_j$ as the leading term. 
\subsubsection{}

Any two basis of solutions of KZ equation are related by a linear map. The linear map that relates the basis of conformal blocks associated to the left and and the right hand sides of figure \ref{f_fuse} is called ``fusion". A basic result of rational conformal field theory is that fusion diagonalizes braiding \cite{MS, RCFT}. 

To recall why this is the case, consider the action of braiding $\Phi_{V_i}(a_i)$ and $\Phi_{V_j}(a_j) $ along the path $B$ in figure \ref{f_wall}. In the basis of conformal blocks that uses the sewing prescription on the left side of figure \ref{f_fuse}, braiding acts by R-matrices of $U_{\fq}(^L\fg)$, as we reviewed.
The sewing prescription on the right hand side, in which $\Phi_{V_i}(a_i)$ and $\Phi_{V_j}(a_j) $ fuse to $\Phi_{V_k}(a_j) $, leads to the conformal block whose behavior as $a_i$ tends to $a_j$ is given in \eqref{vanishing}. On this $B$ acts by a phase
\beq\label{eigena}
e^{- \pi i ({h_k-h_i-h_j})} ={\fq}^{{1\over 2}(c_i+c_j-c_k)},
\eeq
where $h_i =c_i/\kappa$.
Thus, conformal blocks in which $\Phi_{V_i}(a_i)$ and $\Phi_{V_j}(a_j) $ fuse to $\Phi_{V_k}(a_j)$ are eigenvectors of braiding with eigenvalue in \eqref{eigena}. 

\subsubsection{}
It is not difficult to show that scalar conformal blocks have similar asymptotics to \eqref{vanishing}.  Corresponding to the conformal block in which $\Phi_{V_i}(a_i)$ and $\Phi_{V_j}(a_j)$ fuse
to $\Phi_{V_k}(a_j)$, we get a scalar conformal block
 ${\cal Z}_k$ with the following asymptotic behavior \cite{AFO}:
\beq\label{vana}
{\cal Z}_k =   (a_i- a_j)^{\Delta_{i} + \Delta_j -\Delta_k}  \times \textup{finite},
\eeq
%beq\label{vana}
%{\cal Z}_k(a) =   (a_i - a_j)^{\Delta_i+\Delta_j -\Delta_k} \times a_i^{\Lambda_i+\Lambda_j-\Lambda_k} \times \textup{single valued},
%\eeq
as $a_i$ tends to $a_j$. The asymptotic behavior of ${\cal Z}_k$ in \eqref{vana} differs from that of conformal blocks in \eqref{vanishing}, by applying the co-vector in \eqref{VS}.  
%As before, "regular`` stands for terms that are single valued and non-vanishing in the limit and  
Above,
\beq\label{ddf}
\Delta_i =  d_i-  c_i/\kappa,
\eeq
where $c_i$ is proportional to the conformal dimension of the vertex operator $\Phi_{V_i}$
$$h_i= c_i/\kappa,
$$ 
and $d_i$ may be thought of as its ``classical dimension". We will explain momentarily in which sense is this terminology natural. 
%The correction term
%$$
%h_{ij} = \beta c_{ij}.
%$$
%is related to the overall choice of normalization of conformal blocks. 
The parameter ${\lambda_0}$ has the geometric interpretation as the parameter that scales the holomorphic symplectic form of ${\cal X}$. 
The values of $d_i, c_i$ are
\beq\label{dimensions}
d_i \equiv \langle {\mu}_i, \rho\rangle, \qquad c_i \equiv \,  {1\over 2} \langle {\mu}_i,  {\mu}_i+2\, {^L\!\rho}\rangle,
\eeq
where $^L\rho$ is the Weyl vector and $\rho$ the Weyl co-vector of $^L\fg$. 

Finally, a note of caution. The formulas \eqref{vanishing} and \eqref{vana} are written assuming the conventional normalization of conformal blocks, where each conformal block is multiplied by an overall factor that vanishes as $(a_i-a_j)^{c_{ij}/\kappa}$
where
\beq\label{correction}
c_{ij} = \langle ^Lw_i, ^Lw_j\rangle.
\eeq
This factor has no geometric interpretation in terms of ${\cal X}$, however it affects the $U_{\fq}(^L\fg)$ braiding matrices. Were we to omit it, as may be natural from geometric perspective, some of the Reidermeister moves needed to prove we are getting link invariants would hold only up to grading shifts. Since it is link invariants we want, we will stick to the conventions familiar from knot theory. As the result some of our formulas may not look as natural geometrically. 

%Note that we are not spelling out the finite terms explicitly, and we used that to write the formula in the way that is easiest to remember and most resembles \eqref{vanishing}. 

\subsubsection{}
The scalar conformal blocks have a geometric interpretation as central charges of B-type branes in ${\mathscr D}_{\cal X}$, so they should reflect the geometry of ${\cal X}$ near the singularity where $a_i \rightarrow a_j$. 

We will show below that near the wall, ${\cal X}$ develops a collection of vanishing cycles
\beq\label{geomrel}
F_{k} \;\; \leftrightarrow \;\; {V}_{k},
\eeq
which are labeled by representations ${V}_{k}$ in the tensor product \eqref{tensor}, whose dimension is
\beq\label{diml}
{\rm dim}_{\mathbb C}\, F_{k} = d_i+d_j -d_{k} \equiv D_k,
\eeq
with $d_i$ given in \eqref{dimensions}. The fact "classical dimensions'' $d_i$ of vertex operators enter the classical dimension of the cycle $F_k$ is what motivates their name. 

Corresponding to a conformal block in which $\Phi_{V_i}$ and $\Phi_{V_j}$ fuse to $\Phi_{V_{k_m}}$ are
branes %
$${\cal F}_{k} \in {{\mathscr D}}_{\cal X},
$$
which have support on the vanishing cycle $F_{k}$ in ${\cal X}$. The generalized central charge of such a brane ${\cal F}_{k}\in {\mathscr D}_{\cal X}$ is a solution to scalar KZ equation which
has the same 
behavior, as $a_i \rightarrow a_j$ as the conformal block in \eqref{vana},
\beq\label{Zkg}
{\cal Z}[{\cal F}_k] \sim {\cal Z}_k.
\eeq
Above, ``$\sim$'' stands for terms of the same, or faster order of vanishing corresponding to possible mixing with conformal blocks ${\cal Z}_{k'}$ with $D_{k'}\geq D_k$.  The reason why we do not have an ``$=$'' sign in \eqref{Zkg} is that, while any brane in ${\mathscr D}_{\cal X}$ leads to a solution of the KZ equation, not every solution of the KZ equation arizes as the central charge of a brane. 

Setting $\lambda_0=0$, specializes to the physical central charge, which vanishes as 
\beq\label{Zk}
 {\cal Z}^0[{\cal F}_{k}]  \sim   {\cal Z}^0_{k} = \log(a_i/a_j)^{{\rm dim}_{\mathbb C}\, F_{k} }  \times  \textup{finite}.
 \eeq
This says that the physical central charge of the brane ${\cal F}_{k}\in {\mathscr D}_{\cal X}$ vanishes at least as fast as the complexified volume of $F_k$,
\beq\label{Zk0}
 {\cal Z}^0[{\cal F}_{k}]  \sim \bigl(\int_{F_k} \, (-J_{\mathbb C})^{\,{\rm dim}_{\mathbb C}F_{k} } \bigr) \times  \textup{finite}.
\eeq
and moreover, the volume of $F_k$ is controlled by a single Kahler modulus
$$\log a_j/a_i = \int_{C_{ij}} J_{\mathbb C},
$$ 
associated to a curve class $C_{ij} \in H_2({\cal X}, {\mathbb Z})$, so that in fact $h^{1,1}(F_k) = 1$.  

Below, we will show that ${\cal X}$ really has branes ${\cal F}_k$ whose central charge has this behavior, for each representation $V_k$ in the tensor product $V_i\otimes V_j$.

\subsection{Central charge of a holomorphic Lagrangian}

We will pause to compute the exact central charge ${\cal Z}^0$ of a holomorphic Lagrangian in ${\cal X}$. 
 This is a key link between the behavior of conformal blocks and geometry vanishing cycles of ${\cal X}$, as holomorphic Lagrangians in ${\cal X}$ (with partial support on $F_k$) are primary examples of branes whose central charge has the behavior in \eqref{Zk0}.

The central charge ${\cal Z}^0$ of branes in ${\cal X}$ takes the simple exact form in equation \eqref{ccsym}, due to the fact ${\cal X}$ is holomorphic symplectic. For branes which are supported on holomorphic Lagrangians in ${\cal X}$, that formula simplifies yet further.  
This can be viewed as one of the reasons why holomorphic Lagrangians play a special role in our story. (Among others is the fact that ${\mathscr D}_{\cal X}$ becomes generated by such branes after tensoring with the structure sheaf of its core $X$. The core $X$ is a fixed locus of the ${\mathbb C}^\times_{\fq}$ action on ${\cal X}$,  and is itself a union of holomorphic Lagrangians. It plays an important role in \cite{A2}.)

Later, we will study the geometry of ${\cal X}$ in some detail and find the holomorphic Lagrangians whose existence is predicted by the behavior of conformal blocks in \eqref{Zk}.

\subsubsection{}

We will begin by recalling a subtle, but well known fact. Let $F$ be a submanifold of ${\cal X}$ and $f:F \rightarrow {\cal X}$ its embedding. A B-type brane on $F$ supporting a vector bundle $ {\cal E}$  determines a sheaf $ f_*{\cal E}$ on ${\cal X}$. Naively, this sheaf is the object ${\cal F} \in {\mathscr D}_{\cal X}$ of the derived category corresponding to the B-brane. This is not quite correct, instead,
\beq\label{trueid}
{\cal F} = f_*({\cal E} \otimes K_F^{-1/2})\;\;\in\;\; {\mathscr D}_{\cal X}.
\eeq
The tensor product with $K_F^{-1/2}$, the fractional power of the canonical line bundle of $F$, is crucial whenever $F$ is not a spin manifold. If $F$ is a spin manifold (its second Steifel-Whitney class vanishes), $K_F^{-1/2}$ is an honest line bundle with integral first Chern class $+{1 \over 2} c_1(F)$. What happens when $F$ is not spin was explained by Freed and Witten in \cite{FW}. When $F$ is not spin, the vector bundle ${\cal E}$ on the B-brane cannot be an ordinary bundle, but itself is a twisted bundle, in the sense defined there. It is twisted in precisely such a way so that, while neither ${\cal E}$ nor $K_F^{-1/2}$ are ordinary bundles, their tensor product, ${\cal E} \otimes K_F^{-1/2}$ is an ordinary bundle on $F$ \cite{KatzS, Sharpe, AspinwallL}. This ordinary bundle enters the object ${\cal F}$ corresponding to the B-brane.
\subsubsection{}

For an object ${\cal F}$ of the derived category, supported on $F$ which is a holomorphic Lagrangian with a bundle $ {\cal E}' = {\cal E}\otimes K_F^{1/2}$, the central charge ${\cal Z}^0[{\cal F}]$ given by \eqref{ccsym}, simplifies further to
\beq\label{ccLa}
 {\cal Z}^0[{\cal F}] =  \int_{F} {\rm ch}({\cal E}) e^{-f^*\!J_{\mathbb C}}.
\eeq
In showing this, one uses the Grothendieck-Riemann-Roch theorem to rewrite the central charge in \eqref{ccsym} as
\beq\label{ccLb}
\int_{{\cal X}} {\rm ch}({\cal E}') e^{-f^*J_{\mathbb C}}\sqrt{{\rm td}({\cal X}) }
=\int_{F} {\rm ch}({\cal E}' ) e^{-f^*J_{\mathbb C}}\sqrt{{\rm td}(F)\over {\rm td}(N) },
\eeq
where ${\rm td}(F)$ is the Todd class of the tangent bundle to $F$, and ${\rm td}(N)$ the Todd class of the normal bundle to $F$ in ${\cal X}$. 
Then, for $F$ which is a holomorphic Lagrangian in ${\cal X}$, ${\rm td}(F)$ and ${\rm td}(N)$ are related by
$${\rm td}(F) = \prod_i {x_i \over 1-e^{-x_i}} =  \prod_i {-x_i \over 1-e^{x_i}} e^{x_i} ={\rm ch}(K_F^{-1}) {\rm td}(N).
$$
since the normal bundle is its cotangent bundle, the dual of the tangent bundle.
Above, $x_i$ are the Chern roots of the tangent bundle, and $-x_i$ of its dual. This takes \eqref{ccLb} to \eqref{ccLa}.
\subsubsection{}
As an example, take ${\cal F}= {\cal O}_F$ which is the structure sheaf of a holomorphic Lagrangian $F$ in ${\cal X}$. By \eqref{ccLa}, the central charge computes the holomorphic volume of $F$:
\beq\label{ccL}
 {\cal Z}^0[{\cal F}] =  \int_{F} (-f^*\!J_{\mathbb C})^{{\rm dim}_{\mathbb C}F}.
\eeq
There is in fact another way to obtain the same result in this case. A hyper-Kahler rotation takes ${\cal F}= {\cal O}_F$ to an ordinary Lagrangian on ${\cal X}$, and its central charge becomes the integral of the top holomorphic form over $F$, which computes its classical volume. Undoing the hyper-Kahler rotation gives us our result.

\subsection{Near a singularity of ${\cal X}$}\label{s_nearX}

We will now study the geometry of the monopole moduli space ${\cal X} = {{\rm Gr}^{\vec \mu}}_{\nu}$ as the pair of singular monopoles come together. This is the geometric counterpart of bringing together operators $\Phi_{V_i}(a_i)$ and $\Phi_{V_j}(a_j)$, where 
the monopoles are at $y_i = \log|a_i|$ and $y_j = \log|a_j|$, with charges given by the highest weights $\mu_i$ and $\mu_j$ of representations $V_i$ and $V_j$. 
From the geometric interpretation of conformal blocks, we deduced that in this regime ${\cal X}$ must develop a collection of vanishing cycles, that contract to a point in ${\cal X}$ as $y_i$ approaches $y_j$, 
$$
F_k \subset {\cal X},
$$
and which are labeled by representations $V_k$ in the tensor product $V_i\otimes V_j$, of dimension $D_k$ given in \eqref{diml}. We will now show this is indeed is the case.

\subsubsection{}
 At $y_i=y_j$,
 we get a single singular monopole of  charge $
\mu_{ij} = \mu_i + \mu_j$. Since $\mu_{ij}$ is not a minuscule weight, presence of such a singular monopole
introduces singularities in the monopole moduli space.  The point $y_i=y_j$ is a wall in the Kahler moduli, along which ${\cal X}$ becomes a singular manifold ${\cal X}^{\times}$. The resulting monopole moduli space ${\cal X}^{\times}$ can also be described in terms of the affine Grassmannian:
\beq\label{singX}
 {\cal X}^{\times} = {{\rm Gr}^{{\vec \mu_{ij}}^{\times}}}_{\nu}  = \cup_{\mu_k \leq \mu_{ij}} {{\rm Gr}^{{\vec \mu}_{k}}}_{\nu},
\eeq
where ${\vec \mu}_{k}$ is obtained from ${\vec \mu} = (\ldots, \mu_i, \mu_j, \ldots)$ by replacing $(\mu_i, \mu_j)$ by $\mu_{k}$, ${\vec \mu}_{k} = (\ldots, \mu_{k}, \ldots)$.
%similarly when $\mu_k$ is taken to equal $\mu_{ij}$.
${\cal X}^{\times}$ is singular, with singularities which are due to monopole bubbling phenomena. 

Monopole bubbling occurs when smooth monopoles concentrate at a location of a singular monopole of charge $\mu_{ij}$ to leave behind a singular monopole of lower charge $\mu_k$. The types of monopole bubbling that can occur \cite{KW}, correspond to dominant weights $\mu_k$ such that $\mu_{k} \leq \mu_{ij}$.  Alternatively, possible monopole bubblings are labeled by representations $V_{k_m}$ in the tensor product of $V_i\otimes V_j$, and $\mu_k = \mu_{k_m}$ are their highest weights. 

${\cal X}^{\times}$ is a union of $T_{k_{max}} ={{\rm Gr}^{{\vec \mu}_{ij}}}_{\nu}$, where no monopole bubbling occurs, but which is open, with lower dimensional loci  $T_{k_m} ={{\rm Gr}^{{\vec \mu_{k_m}}}}_{\nu}$, for $m<m_{max}$, where exactly $\mu_{ij}-\mu_{k_m}$ monopoles bubble off, and which provide its partial closures. Each $T_{k_m}$ is itself open for $0<m$, of dimension which decreases with $m$.  We have ordered the representations as in \eqref{tensor}, so that $\mu_{k_{max}}=\mu_{ij}$ corresponds to no bubbling at all, and that $\mu_{k_m} \leq\mu_{k_{m+1}}$ for $m=0,\ldots,m_{max}$.

\subsubsection{}

The locus in  ${\cal X}^{\times}$ where smooth monopoles have concentrated at the singular monopole of charge $\mu_{ij}$ to leave behind a singular monopole of charge $\mu_{k_m}$ is 
\beq\label{Tbase}
{ T}^{\times}_{k_m}= {{\rm Gr}^{{\vec \mu_{k_m}}^{\times}}}_{\nu} =  \cup_{\mu_k \leq \mu_{k_m}} {{\rm Gr}^{{\vec \mu}_{k}}}_{\nu} = \cup_{s=0}^m \,T_{k_s}.
\eeq
This is simply the moduli space of monopoles with the singular monopole of charge  $\mu_{ij}$ replaced by a singular monopole of charge $\mu_{k_m}$. It is obtained as the closure of the locus $T_{k_m}$, where exactly $\mu_{ij}-\mu_{k_m}$ monopoles bubble off, by including lower dimensional loci where additional monopole bubble off. ${ T}^{\times}_{k_m}$ should give the geometric interpretation to conformal blocks where $\Phi_{V_i}(a_i) \Phi_{V_j}(a_j)$ get replaced by insertion of a single vertex operator $\Phi_{V_{k_m}}(a_j)$ at $a_i=a_j$, and  the Riemann surface in figure \ref{f_cut} develops an infinitely long neck. (Since $T^{\times}_{k_m}$ is singular, making this precise requires work, but this is exactly what is needed to understand the generalization of our story to links colored by arbitrary, as opposed to only minuscule representations.)

\subsubsection{}
Consider now the transverse slice to ${T}^{\times}_{k_m}$ in ${\cal X}^{\times}$. 
This is the moduli space ${W}_{k_m}^{\times}$ of smooth monopoles whose positions we need to tune to equal that of the singular monopole so that the bubbling of type $\mu_{k_m}$ can occur:
\beq\label{dimW}
{W}_{k_m}^{\times}= {{\rm Gr}^{{ \mu_{ij}^{\times}}}}_{\mu_{k_m}}.
\eeq
${W}_{k_m}^{\times}$ is the transverse slice, in the sense that the tangent space to ${\cal X}^{\times}$ at any point on $T^{\times}_{k_m}$, splits into the tangent space to $T^{\times}_{k_m}$ at that point, and the tangent space to a ${W}_{k_m}^{\times} $ at the point $z^{-\mu_{k_m}} \in {W}_{k_m}^{\times}$.  (As we recall in the appendix in more detail, 
${{\rm Gr}^{ \mu_{ij}^{\times}}}_{\mu_{k_m}}$
is per definition the transverse slice to the ${{\rm Gr}^{\mu_{k_m}}
} = G[[z]] z^{-\mu_{k_m}}$ orbit inside ${{\rm Gr}^{{ \mu_{ij}^{\times}}}} = \cup_{\mu_k \leq \mu_{ij}} {{\rm Gr}^{{ \mu_{k}}}} $ at $z^{-\mu_{k_m}}$.  This orbit is identified with the top-dimensional component of $T^{\times}_{k_m}$.)

\subsubsection{}
We would now like to find the vanishing cycles in ${\cal X}$. We obtained ${\cal X}^{\times}$ from ${\cal X}$ by bringing two monopoles, originally at for $y_i<y_j$, together at $y_i=y_j$. Consider the map
\beq\label{down}
m_{ij} :  {\cal X} = {{\rm Gr}^{{\vec \mu}}}_{\nu}  \rightarrow   {\cal X}^{\times} = {{\rm Gr}^{{\vec \mu^{\times}}}}_{\nu} 
\eeq
that contracts ${\cal X}$. By a vanishing cycle in ${\cal X}$ we mean a cycle that 
gets contracted by $m_{ij}$ to a point.

Like ${\cal X}^{\times}$, ${W}_k^{\times}$ corresponding to any $V_k$ in the tensor product of $V_i\otimes V_j$, has a singularity that gets resolved as we take $y_i<y_j$.
The preimage of $W_k^{\times}$ of this map is smooth manifold  ${W}_{k}$ given by
\beq\label{cota0}
 {W}_{k}={\rm Gr}^{ ({\mu}_i, {\mu}_j)}_{{\mu}_k} = m_{ij}^{-1}(W_k^{\times}).
\eeq 
$ {W}_{k}$ is a symplectic resolution of singularities of $W_k^{\times}$: it is smooth, since $\mu_{i,j}$ are minuscule weights, and its holomorphic symplectic form is a pull back of the holomoprhic symplectic form on $W_k^{\times}$. 

As $y_i\rightarrow y_j$, $ {W}_{k}$ develops a singularity since a cycle inside it contracts to a point.
The cycle is the pre-image of the fixed point  $z^{-\mu_k} \in W_k^{\times}$ of the ${\rm T}$-action on $W_k^{\times}$:
\beq\label{vanishingc}
F_k :=m_{ij}^{-1}(z^{-\mu_k}).
\eeq
The vanishing cycle $F_k$ is in fact a holomorphic Lagrangian in $ {W}_{k}$ -- the holomorphic symplectic form of $ {W}_{k}$ vanishes when restricted to $F_k$ since it is a pullback of a holomorphic symplectic form on $W_k^{\times}$ restricted to a point. Since in addition it has a single Kahler modulus, $W_k$ is the cotangent bundle $W_k = T^* F_k.$

 In ${\cal X}'$, for $y_j>y_i$, we get an analogous resolution, 
 \beq\label{cotb}
{W}'_{k} = {\rm Gr}^{ ({\mu}_j, {{\mu}}_i)}_{{\mu}_k}
\eeq
with roles of $i$ and $j$ reversed. At $y_i=y_j,$ this becomes 
 the same singular cone as in \eqref{down}
\beq\label{downb}
m_{ji} : \qquad {W}'_{k} \;\; \longrightarrow \;\;   {W}^{\times}_{k}={{\rm Gr}^{{\mu_{ij}^{\times}}}}_{\nu},
\eeq
with the vanishing cycle
\beq\label{vanishingc}
F'_k =m_{ji}^{-1}(z^{-\mu_k}),
\eeq
of the same dimension as $F_k$. While they have the same dimension, in general $F_k \neq F_k'$ as symplectic manifolds.

\subsubsection{}

The cycles $F_k$ or $F_k'$ are the cycles whose existence and dimension are predicted by the behavior of scalar conformal blocks as the pair of vertex operators $\Phi_{V_i}(a_i)$ and $\Phi_{V_j}(a_j)$ come together: $F_k$ is a vanishing cycle in $W_k$, and hence also in ${\cal X}$. In ${\cal X}$, $F_k$ comes in a family fibered over $T_k^{\times}$, where the total space of this fibration is
$
m_{ij}^{-1}(T_k^{\times}) \subset {\cal X}.
$

To compute dimension of $F_k$, note that as a holomorphic Lagrangian inside $W_k$, its the complex dimension is half that of $W_k$ -- which is in turn the same as the dimension of $W_{k}^{\times}$, and the co-dimension of $T^{\times}_k$ in ${\cal X}^{\times}$. It equals the number of smooth monopoles we need to tune to get ${T}^{\times}_k$.
% namely ${\rm dim}_{\mathbb C} W^{ij}_{k} =2 \langle \mu_{i} + \mu_j - \mu_{k}, \rho \rangle$. 
Thus, the dimension of the vanishing cycle $F_k$ is
$${\rm dim}_{\mathbb C} F_{k} = \langle \mu_{i} + \mu_j - \mu_{k}, \rho \rangle =  D_k.
$$
This is the dimension $D_k$ of the vanishing cycle we expected to find from the behavior of conformal blocks \eqref{diml},
illustrating their geometric origin.
\begin{comment}
\subsubsection{}
A way to get objects ${\cal F}_k \in {\mathscr D}_{\cal X}$ whose central charge vanishes at least as fast as \eqref{vana} is as follows.
Take any holomorphic Lagrangian in $T_k^{\times}$, and denote it by $G_k$. Since $T_{k}^{\times}$ is singular Its pre-image in ${\cal X}$ is $H_k = m_{ij}^{-1}(G_k)$ is a holomorphic Lagrangian in ${\cal X}$ which fibers over $G_k$ with fibers $F_k$. Take ${\cal F}_k \in {\mathscr D}_{\cal X}$ be the structure sheaf of $H_k$, 
${\cal F}_k = {\cal O}_{H_k}$. The physical central charge ${\cal Z}^0[{\cal F}_k]$ is the holomorphic volume of $H_k$, by \eqref{ccL}. 
Since the fibers $F_k$ contract as one approaches the singularity, one reproduces the vanishing behavior in \eqref{Zk0}, as $a_i$ approaches $a_j$.
Other objects whose central charge vanishes in the same way, or faster, by replacing $G_k$ by an arbitrary B-brane on $T_k^{\times}$. In section \ref{s:example} we will give explicit examples.
\end{comment}

\subsection{Diagonalization vs. filtration}\label{integrable}

We explained that the action of braiding 
%$\Phi_{V_i}(a_i)$ and ${\Phi}_{V_j}(a_j)$ 
on the space of (scalar) conformal blocks is implemented by a  $U_{\fq}(^L\fg)$ matrix ${\mathfrak B}$, and that this action can always be diagonalized. The same is not true for the action of braiding by ${\mathscr B}$ on ${\mathscr D}_{\cal X}$. Only some special eigenvectors of ${\mathfrak B}$ come from eigensheaves of ${\mathscr D}_{\cal X}$.

\subsubsection{}

The eigenvectors of ${\mathfrak B}$ corresponding to braiding $\Phi_{V_i}(a_i)$ and ${\Phi}_{V_j}(a_j)$,
are conformal blocks obtained by fusing $\Phi_{V_i}(a_i)$ and ${\Phi}_{V_j}(a_j)$ to ${\Phi}_{V_{k}} (a_j)$, where $V_k$ is a representation in $V_i\otimes V_j$.
The corresponding solution of scalar KZ equation has the form
 \beq\label{Zka}
{\cal Z}_{k_m} =  (a_i-a_j)^{\Delta_i + \Delta_j  - \Delta_{k_m}}  \times  \textup{finite},
\eeq
where ``finite'' stands for terms that are both non-vanishing and regular in the neighborhood of $a_i=a_j.$
The eigenvalue corresponding to braiding along the path $B$ in figure \ref{f_wall}
can be read off from this. Using \eqref{ddf}, it is given by
\beq\label{eigenval}
e^{-i \pi({\Delta_i + \Delta_j - \Delta_{k_m}})} = (-1)^{D_m} \, {\fq}^{{1\over 2}C_m},
\eeq
where $D_k$ is the dimension of the vanishing cycle $F_k$
\beq\label{shifta}
D_m = d_i+d_j - d_{k_m},
\eeq
and $C_k$ takes value
\beq\label{shiftb}
 C_m =  c_i+c_j  - c_{k_m}.
\eeq
${\mathfrak B}$ has an eigenvector with eigenvalue in \eqref{eigenval} for every weight of a representation $V_{k} \subset V_i \otimes V_j$, which contributes to the weight $\nu$ subspace of representation $V$ in \eqref{representation}.

\subsubsection{}
In general, conformal blocks of the form \eqref{Zka} do not come from actual objects of ${\mathscr D}_{\cal X}$. 
While we can always diagonalize the action of braiding ${\mathfrak B}$ on the space of conformal blocks, there is no reason to expect that this lifts to the derived category, in any generality. So, the action of braiding on ${\mathscr D}_{\cal X}$ is not diagonalizable. 
The fact that the action of braiding ${\mathscr B}$ on the derived category is not diagonalizable means that ${\mathscr B}$ induces a filtration on the derived category instead, as we will describe in the next subsection.

The derived category ${\mathscr D}_{\cal X}$ can contain some special objects which are "eigensheaves" of the action of braiding by ${\mathscr B}$ on the derived category. 
For us, an eigensheaf of the action of braiding by ${\mathscr B}$ is an object ${\cal E}_{k_m}\subset {\mathscr D}_{\cal X}$, which is mapped to itself by the action of ${\mathscr B}$ up to degree shifts: 
\beq\label{eigensheaf}
{\mathscr B}{\cal E}_{k_m} = {\cal E}_{k_m}[-D_m]\{C_m\},
\eeq
with $D_m$ and $C_m$ as in \eqref{shifta} and \eqref{shiftb}. The central charge of such an object is ${\cal Z}_{k_m} = {\cal Z}[{\cal E}_{k_m}]$ in \eqref{Zka}.
While the eigenvectors of ${\mathfrak B}$ with eigenvalue in \eqref{eigenval} correspond to weights of representation $V_{k_m}$ which contribute the space of solutions of the KZ equation but are otherwise arbitrary, only the highest weight vectors of $V_{k_m}$ should lead to eigensheaves of the action of
${\mathscr B}$ on the derived category, as we explain in the appendix.

Recall that the central charge function in \eqref{Zka} contains the contribution from overall normalization of conformal blocks. Working in the normalization of ${\cal Z}$
that would be natural from geometry, but not from representation theory perspective, would result in 
\beq\label{shiftc}
 C_m^{\cal X}  =  c_i+c_j  - c_{k_m} + c_{ij}.
\eeq
as the degree shift in \eqref{eigensheaf}.

\subsection{Central charge filtration of ${\mathscr D}_{\cal X}$}\label{s_filtration}

In this subsection, we will show that near a wall in Kahler moduli where $y_i = \log|a_i|$ and $y_j = \log |a_j|$ coincide,
the central charge ${\cal Z}^0$ gives rise to a filtration on ${{\mathscr D}}_{\cal X}$,
\beq\label{filtration0a}
 {{\mathscr D}}_{k_{0}} \subset {{\mathscr D}}_{k_{1}} \ldots \subset {{\mathscr D}}_{k_{max}} = {{\mathscr D}}_{\cal X}.
\eeq
The different terms of the filtration are labeled by representations $V_{k_m}$ in the tensor product,
$$
{V}_i \otimes {V}_j = \bigoplus_{m=0}^{\rm{max}} {V}_{k_m},
$$
ordered so that
\beq\label{reporder}
\mu_{k_m} \leq \mu_{k_{m+1}},
\eeq
where $\mu_{k_m}$ are the highest weights of ${V}_{k_m}$, for all $m$.
As we will explain, the filtration is by the order of vanishing of ${\cal Z}^0$ as we approach the wall. One gets the same kind of filtration on the other side of the wall, on ${{\mathscr D}}_{{\cal X}'}$,
\beq\label{filtration0b}
 {{\mathscr D}}_{k_{0}}' \subset {{\mathscr D}}_{k_{1}}' \ldots \subset {{\mathscr D}}_{k_{max}}' = {{\mathscr D}}_{{\cal X}'},
\eeq
since the central charge function is analytic in complexified Kahler moduli, and ${\cal X}$ and ${\cal X}'$ are separated by a wall in codimension one in real Kahler moduli.

 \subsubsection{}

The central charge filtration of ${\mathscr D}_{\cal X}$ in \eqref{filtA} comes from a central charge filtration of its heart ${\mathscr A}\subset {\mathscr D}_{\cal X}$. The heart ${\mathscr A}$ of ${\mathscr D}_{\cal X}$ is an abelian category whose objects are branes in the physical sigma model to ${\cal X}$, preserving B-type supersymmetry, as we reviewed in section \ref{s-s}.  More precisely, the branes in ${\mathscr A}$ are semi-stable coherent sheaves ${\cal F}$, whose central charge ${\cal Z}^0[{\cal F}]$ is in the upper half of the complex plane, with phase $\phi \in [0, 1)$.  The central charge being in the upper half plane distinguishes what we call branes from anti-branes. (Given a brane ${\cal F}$, its anti-brane differs from it by a cohomological degree shift ${\cal F}[1]$, and has central charge which is equal and opposite.)
By taking direct sums, degree shifts, and iterated cones, ${\mathscr A}$ generates ${\mathscr D}_{\cal X}$.

Every non-zero object of ${\cal F} \in {\mathscr D}_{\cal X}$ has central charge ${\cal Z}[{\cal F}]$ whose asymptotic behavior near the wall is ${\cal Z}[{\cal F}] \sim {\cal 
Z}_{k_m}$ for some $V_{k_m}$ in $V_i \otimes V_j$. 
This follows since, on one hand, the central charge ${\cal Z}[{\cal F}]$ of every non-zero brane is a non-zero solution to the scalar KZ equation by theorem \ref{t:two}, and on the other hand,
solutions ${\cal 
Z}_{k_m}$ where ${\Phi}_{V_i}(a_i) \otimes {\Phi}_{V_j}(a_j)$ fuse to ${\Phi}_{V_{k_m}}(a_j)$ provide a basis of all solutions to the scalar KZ equation near the wall. 

Let ${\mathscr A}_{k_m} $ be the subcategory of ${\mathscr A}$ generated by objects whose central charge vanishes near the wall at least as fast as ${\cal Z}_{k_m}^0$ in \eqref{Zk}, or equivalently, at least as fast as
\beq\label{vanac}
{\cal Z}^0_{k_m} = (a_i-a_j)^{D_{{m}}} \times {\rm finite},
\eeq
where $D_m$ is the dimension of the vanishing cycle corresponding to the representation $V_{k_m}$. Because the objects of ${\mathscr A}_{k_m}$ are all branes (as opposed to branes and anti-branes), there are no cancelations in the central charge formula. It follows ${\mathscr A}_{k_\ell}$ is a subcategory of ${\mathscr A}_{k_m}$ whenever $\ell \leq m$. Consequently,  ${\mathscr A}$ is filtered by subcategories ${\mathscr A}_{k_m} $:
\beq\label{filtA}
{{\mathscr A}}_{k_{0}} \subset {{\mathscr A}}_{k_{1}} \ldots \subset {{\mathscr A}}_{k_{max}} = {{\mathscr A}},
\eeq
where the filtration is by the order of vanishing of central charge ${\cal Z}^0$. Since %
\beq\label{order}
D_m \geq D_{{m+1}},
\eeq
the lower the order in the filtration, the faster central charge vanishes, and since $D_m$ is also the dimension of the vanishing cycle as we showed in section \ref{s_nearX}, the higher is the dimension of the vanishing cycle. The ordering in \eqref{order} is also the same as the ordering of the representations
\beq\label{order1}
{\mu}_{k_m}  \leq {\mu}_{k_{m+1}},
\eeq
since
the dimension is given by $D_m = d_i+d_j - d_{k_m}$, with $d_{k_m} = \langle {\mu}_{k_m}, \rho\rangle$. 
%The ordering $\nu \leq \mu$  for a pair of weights  $\mu$, $\nu$  of $^L\fg$ means, as before, that $\mu-\nu$ is a non-negative integral combination of simple positive roots of $^L{\fg}$. 

%\subsubsection{}
The wall where the filtration is defined separates a pair of chambers ${\fC}$, where $y_i<y_j$, and which corresponds to ${\cal X}$, from ${\fC}'$, where $y_j<y_i$ which corresponds to ${\cal X}'$. If we get a filtration \eqref{filtA} in the chamber ${\fC}$, on one side of the wall, we get an identical looking filtration on the other side of the wall,
\beq\label{filtAb}
{{\mathscr A}'}_{k_{0}} \subset {{\mathscr A}'}_{k_{1}} \ldots \subset {{\mathscr A}'}_{k_{max}} = {{\mathscr A}'},
\eeq
defined in the same way. This follows 
since the central charge 
in \eqref{vana} is analytic in $a_i$ and $a_j$ away from the singularity at $a_i=a_j$.

%\subsubsection{}

 The filtration of the heart ${\mathscr A}$ of ${\mathscr D}_{\cal X}$, by subcategories ${\mathscr A}_{k_m} $, induces the filtration of ${\mathscr D}_{\cal X}$, as follows.
Define ${\mathscr D}_{k_m}$ to be the subcategory of ${\mathscr D}_{\cal X}$ whose heart is ${\mathscr A}_{k_m}$. 
If ${\mathscr A}_{k_\ell}  $ is a subcategory of ${\mathscr A}_{k_m} $, then ${\mathscr D}_{k_\ell} $ is a subcategory of ${\mathscr D}_{k_m} $; this is the case whenever
 $\ell\leq m$. Alltogether, the subcategories ${\mathscr D}_{k_m}$ give a central charge filtration of ${\mathscr D}_{\cal X}$, in \eqref{filtration0a}.
 Similarly, on the other side of the wall, the filtration of ${\mathscr A}'$ in \eqref{filtAb} generates the filtration of ${\mathscr D}_{{\cal X}'}$ in \eqref{filtration0b}.
\subsection{Perverse equivalence and central charge filtration}

A key aspect of the filtrations is that the derived equivalence ${\mathscr B}$ preserves them. As we cross the wall, objects at any given order $m$ in the filtration can get mixed up with those at lower order, with faster vanishing of the central charge, but not with those from above. For this reason, the filtration is the key tool for describing the equivalence relating  ${{\mathscr D}}_{\cal X}$ and ${{\mathscr D}}_{{\cal X}'}$.  

The derived equivalence ${\mathscr B}$ not only preserves the filtrations, but in addition, it acts on the quotient subcategories
$$
{\mathscr B}: \; {\mathscr D}_{k_m}/{\mathscr D}_{k_{m-1}} \cong {\mathscr D}_{k_m}'/{\mathscr D}_{k_{m-1}}'
$$
by shifts in cohomological and equivariant degrees that depend only on the order $m$ in the filtration.  We will be able to predict exactly what these degree shifts are. Derived equivalences with these properties are called a ``perverse equivalences" by Chuang and Rouquier \cite{CR}.   
 
\subsubsection{}

Pick a path $B$ from ${\fC}$ to ${\fC}'$, around the singularity at $a_i = a_j$ clockwise, as in figure \ref{f_wall},
Along such a path B-branes whose central charge vanishes to any given order in the filtration, can get mixed up with lighter branes, whose central charge vanishes faster near the wall, but not the other way around. This follows from theorem \ref{t:two} and the interpretation of solutions of KZ equation as generalized central charges of ${\mathscr D}_{\cal X}$. It also follows from mirror symmetry \cite{A2}, where it is a 
direct consequence Picard-Lefshetz monodromy of vanishing cycles.
Without theorem \ref{t:two} or mirror symmetry, the one can understand this by restricting to a subcategory of ${\mathscr D}_{\cal X}$ generated by holomorphic Lagrangians. After a hyper-Kahler rotation, objects in this subcategory become ordinary Lagrangians on ${\cal X}$, and the statement again becomes a consequence of Picard-Lefshetz theory.

\subsubsection{}
Consider the quotient category, 
\beq\label{mgR}
{\rm gr}_{m}({\mathscr A}) = {\mathscr A}_{k_m}/{\mathscr A}_{k_{m-1}},
\eeq
obtained from $ {\mathscr A}_{k_m}$ by treating all objects that come from ${\mathscr A}_{k_{m-1}}$ as zero. The statement that ${\mathscr B}$ preserves the filtration means that it acts as an equivalence of the quotient categories
$$
{\mathscr B}: \;\;{\rm gr}_{m}({\mathscr A}) \;\;  { \xrightarrow{\sim}}  \;\; {\rm gr}_{m}({\mathscr A}'),
$$
even though it does not preserve $ {\mathscr A}_{k_m}$ or  $ {\mathscr A}_{k_m}'$ themselves.
The central charge any object ${\cal F}$ of either ${\rm gr}_{m}({\mathscr A})$ or of ${\rm gr}_{m}({\mathscr A}')$ vanishes exactly like ${\cal Z}_{k_m}$, which is an eigenvector of the action of braiding on conformal blocks in \eqref{Zka}: 
\beq\label{vanaDC}
{\cal Z}_{k_m} = (a_i-a_j)^{D_{{m}}-\lambda_0 C_m} \times {\rm finite}.
\eeq
This is because any contribution to the central charge that vanishes faster comes from objects that are treated as zero in the quotient categories $ {\mathscr A}_{k_m}/{\mathscr A}_{k_{m-1}}$ and $ {\mathscr A}'_{k_m}/{\mathscr A}'_{k_{m-1}}$, and there are no contributions from objects whose central charge vanishes slower to either ${\mathscr A}_{k_m}$ or $ {\mathscr A}_{k_m}'$, per their definitions.

It follows from \eqref{vanaDC} that, along the path $B$, the central charge ${\cal Z}[{\cal F}]$ of an arbitrary object ${\cal F}\in {\rm gr}_{m}({\mathscr A})$
changes by the following phase: 
\beq\label{zsa}
 {\cal Z}[{\cal F}]\; \longrightarrow e^{ - \pi i D_m} \,{\fq}^{ {1\over 2} C_m}  \,  {\cal Z}[{\cal F}].
\eeq
The phase does not depend on ${\cal F}$ at all, but only on the order $m$ in the filtration and the path $B$ we chose. It reflects the cohomological and equivariant degree shifts.

The functor ${\mathscr B}$, therefore,
acts on ${\rm gr}_{m}({\mathscr A})$ by a cohomological and equivariant degree shifts
 \beq\label{gs}
 {\mathscr B}: \;{\rm gr}_{m}({\mathscr A})\; \xrightarrow{} \; {\rm gr}_{m}({\mathscr A}') \; \cong \;{\rm gr}_{m}({\mathscr A})[-D_m]\{C_m\}.
 \eeq
which depend only on the order $m$ in the filtration.
Equivalently, the functor ${\mathscr B}[D_m]\{-C_m\}$ corresponding to composing ${\mathscr B}$ with the functors $[D_m]$ and $\{-C_m\}$ that act by cohomological and equivariant degree shifts, is an equivalence of categories ${\rm gr}_{m}({\mathscr A})$ and ${\rm gr}_{m}({\mathscr A}')$.

The shift of cohomological degree is by $[-D_m]$, and comes from the ${ -i \pi} D_m$ change of phase. This is the change of phase of the physical central charge ${\cal Z}^0[{\cal F}]$. There, $D_m$ is an integer, 
\beq\label{shifta}
D_m = d_i+d_j - d_{k_m},
\eeq
 equal to the dimension $D_m$ of the vanishing cycles $F_{k_m}$ and $F'_{k_m}$.  
 The shift of equivariant grade, corresponding to the ${\mathbb C}^{\times}_{\fq}$ symmetry that scales the holomorphic symplectic form, is by 
 \beq\label{shiftb}
 C_m =  c_i+c_j - c_{k_m},
 \eeq
 with $d$'s and $c$'s given in \eqref{dimensions}. 

The ${\rm T}$-grading defined by \eqref{shiftb} way may end up fractional, rather than integral. This reflects our overall normalization of the central charge function ${\cal Z}$ which, as we explained near equation \eqref{correction}, is natural from representation theory perspective, but not from perspective of ${\cal X}$. 
 
 Using the central charge function that naturally comes from geometry of ${\cal X}$, 
 and which differs from ${\cal Z}$ by overall normalization, 
 would lead to equivariant degree shifts given by
 \beq\label{shiftca}
C_m^{\cal X} = C_m + c_{ij}
\eeq
which are always integral \cite{A2}. The two versions of the braiding functor ${\mathscr B}$ which use $\{C_m\}$ and $\{C_m^{\cal X}\}$  in \eqref{gs} differ by a equivariant shift functor $\{c_{ij}\}$ which is trivially an auto-equivalence of ${\mathscr D}_{\cal X}$. We will see this effect in the example below. 

\subsubsection{}

If instead we 
loop around the singular locus, ${\cal X}$ comes back to itself. We get an auto-equivalence of ${\mathscr D}_{\cal X}$ which extends to an auto-equivalence of its subcategories  ${{\mathscr D}}_{k_{m}}$, but acts on the heart ${\mathscr A}$ by a shift of gradings  
$${\mathscr B}: \; {\rm gr}_{m}({\mathscr A})\;   \rightarrow \; {\rm gr}_{m}({\mathscr A})[-2D_m]\{2C_m\},$$
individual objects in ${\rm gr}_{m}({\mathscr A})$ come back to themselves only up to a degree shift. This is a basic example of monodromies that leave ${\mathscr D}_{\cal X}$ unchanged, but act non-trivially on
${\mathscr A} \subset {\mathscr D}_{\cal X}$.

\subsection{Derived equivalences from generalized flops}\label{s:example}

Derived equivalences ${\mathscr B}$ relating the derived categories of ${\cal X}$ and ${\cal X}'$ come from generalized flops.
The facts that flops generate derived equivalences is well known \cite{ST, HT}. In an ordinary flop, a single  ${\mathbb P}^1$ shrinks to a point. A generalization of that is a single ${\mathbb P}^n$ shrinking in a local geometry $T^*{\mathbb P}^n$, for $n>1$. The corresponding derived equivalence was constructed in \cite{HT}. 

Ours is a further generalization. Corresponding to a pair of vertex operators $\Phi_{V_i}(a_i)$ and $\Phi_{V_j}(a_j)$ coming together, either ${\cal X}$ or ${\cal X}'$ develops a singularity, depending on whether $\log|a_j/a_i|>0$ or $\log|a_j/a_i|<0$.  In both ${\cal X}$ and ${\cal X}'$, one gets vanishing cycles labeled by representations $V_{k_m}$ in the tensor product $V_i\otimes V_j$. 
%(The $V_{k_{max}} = V_i\otimes V_j$ is the trivial case, for which the vanishing cycle is a point.)
On the wall, at $|a_i|=|a_j|$ we get a singular manifold ${\cal X}^{\times}$ in which all the vanishing cycles contract to points. The case corresponding to an ordinary flop corresponds to taking $V_i=V_j$ to be the defining representation of $^L\fg=\mathfrak{sl}_2$.

As an illustration of the derived equivalence ${\mathscr B}$, we will work out an $^L\fg=\mathfrak{sl}_n$ example, where the answer is already known by other means, from \cite{KamnitzerD}. 

\subsubsection{}
Take ${\cal X}$ and ${\cal X}'$ to be 
cotangent bundles to Grassmannians
$$
{\cal X} =T^*{\mathbb G}(\ell,n), \qquad {\cal X}' =T^*{\mathbb G}(n-\ell,n).
$$
In the language of our paper, they can be represented as slices in the affine Grassmannian
$${\cal X} = {\rm Gr}^{\mu_i, \mu_j}_0, \qquad {\cal X}' = {\rm Gr}^{\mu_j, \mu_i}_0,
$$
where the ${^LG}=SU(n)$,  $V_i$ and $V_j$ are conjugate representations whose highest weights are the fundamental weights $\mu_i = ^L\!w_{\ell}$ and $\mu_j = ^L\!w_{n-\ell}$, and we take the zero weight subspace of $V_i\otimes V_j$. 
\begin{figure}[!hbtp]
  \centering
   \includegraphics[scale=0.22]{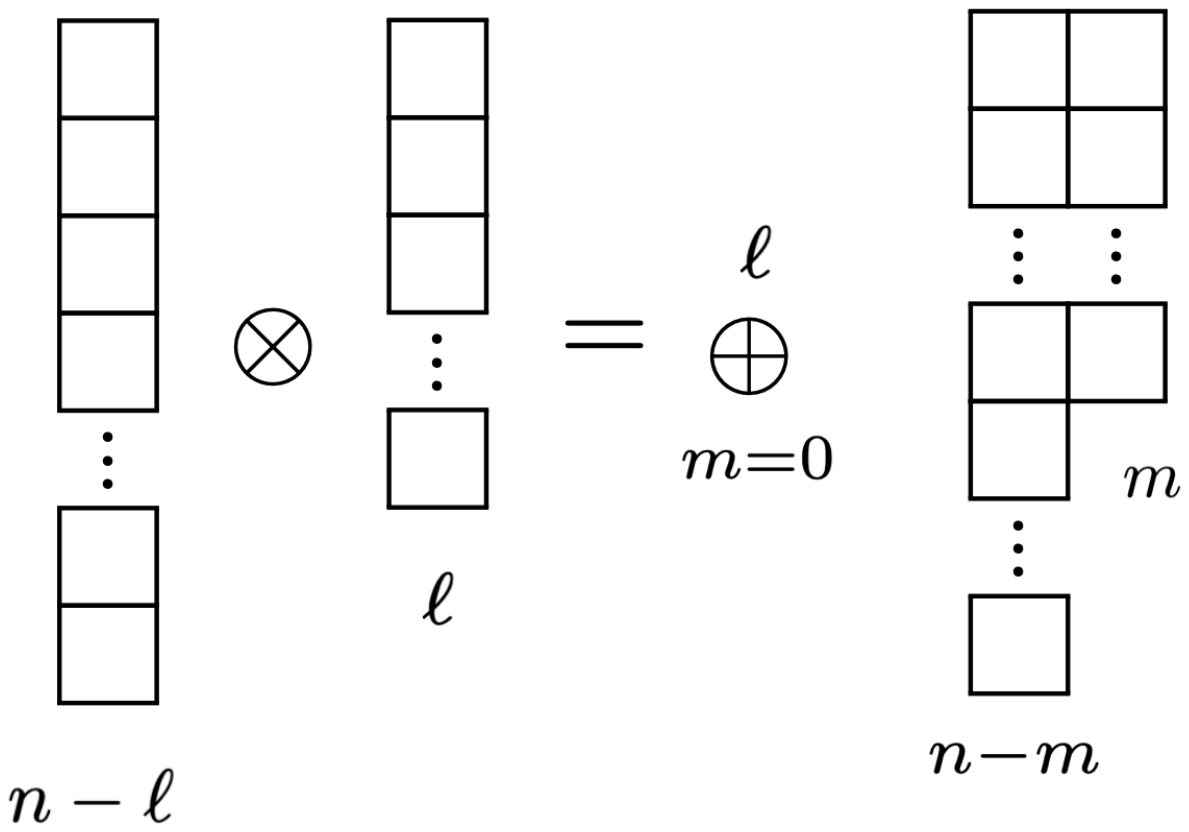}
 \caption{Tensor product $V_i\otimes V_j = \oplus_{m=0}^\ell V_{k_m}$ in this example.}
  \label{f_tensor}
\end{figure}
This means $V_i$ and $V_j$ are the 
 $\ell$-th and $n-\ell$'th antisymmetric representation of $^LG$, respectively, and the representations $V_{k_m}$  in their tensor product are have   highest weights 
$\mu_{k_m} = {^Lw}_{m} + {^Lw}_{n- m}$, with $m = 0, \ldots,  \ell$,  assuming $n\geq  2\ell$. 

The vanishing cycles in ${\cal X}$ and ${\cal X}'$ associated to the representation $V_{k_m}$ are the Grassmannians
$$
 F_{k_m} = {\mathbb G}(\ell-m,n-2m)  , \qquad  F'_{k_m} = {\mathbb G}(n-m-\ell,n-2m), 
$$ 
with dimensions $D_m = (\ell - m)(n - m-\ell)$ that increase as $m$ decreases (dimension of ${\mathbb G}(k,n)$ is $k(n-k)$). 
Their cotangent bundles
$$W_{k_m} = T^*F_{k_m}, \qquad W'_{k_m} = T^*F_{k_m}',$$ 
are
$
W_{k_m} =  {\rm Gr}^{\mu_i, \mu_j}_{\mu_{k_m}}$ and $W_{k_m} '=  {\rm Gr}^{\mu_j, \mu_i}_{\mu_{k_m}}$. 

At the wall in Kahler moduli, ${\cal X}$ and ${\cal X}'$ contract to the same singular manifold ${\cal X}^{\times}$ and all the vanishing cycles contract to a point, as described in section \ref{s_nearX}. Monopole bubbling makes ${\cal X}^{\times}$ into a singular union
$
{ X}^{\times} = \cup_{m=0}^{max} \; T_{k_m}.
$
Each $T_{k_m}$ is describes a locus where some fixed number of smooth monopoles have bubbled off; 
%the number of monopoles that have bubbled off is zero when $m$ is maximal and increases as $m$ decreases. 
its closure ${ T}^{\times}_{{k_m}}= \cup_{p=0}^m \; T_{k_p},$ may be obtained by starting with either $T^*{\mathbb G}(m,n)$ or $T^*{\mathbb G}(n-m,n)$, and contracting their bases. 

At the wall, 
$W_{k_m}$ and $W_{k_m}'$ contract to
$W_{k_m}^{\times}$ which is the transverse slice to $T_{k_m}^\times$ in ${\cal X}^{\times}$, and which parameterizes the moduli space of smooth monopoles, whose positions we need to tune to restrict ${\cal X}^{\times}$ to $T^{\times}_{k_m}$.

\subsubsection{}

The derived category ${\mathscr D}_{\cal X}$ of ${\rm T}$-equivariant coherent sheaves on ${\cal X}$ has an  $\ell+1$-term filtration, 
%${{\mathscr D}}_{k_{0}} \subset {{\mathscr D}}_{k_{1}} \ldots \subset {{\mathscr D}}_{k_\ell} = {{\mathscr D}}_{\cal X}$ and ${{\mathscr D}'}_{k_{0}} \subset {{\mathscr D}'}_{k_{1}} \ldots \subset {{\mathscr D}'}_{k_\ell} = {{\mathscr D}}_{{\cal X}'}$
%\begin{comment}
\beq\label{filtration0a}
{{\mathscr D}}_{k_{0}} \subset {{\mathscr D}}_{k_{1}} \ldots \subset {{\mathscr D}}_{k_\ell} = {{\mathscr D}_{\cal X}}
\eeq
%and
%\beq\label{filtration0b}
%{\mathscr D}'_{k_{0}} \subset {\mathscr D}'_{k_{1}} \ldots \subset {\mathscr D}'_{k_\ell} = {{\mathscr D}}_{{\cal X}'}
%\eeq
%end{comment}
corresponding to $\ell+1$ representations $V_{k_m}$ in the tensor product of $V_i\otimes V_j$. 
%Objects in
%${{\mathscr D}}_{k_{m}}$ are roughly of the form 
%$m^{*}({\cal T}_{k_m})$
%where ${\cal T}_{k_m}$ is any object of ${\mathscr D}_{T_{k_m}^{\times}}$, the ${\rm T}$-equivariant derived category of coherent sheaves on $T^{\times}_{k_m}$ or its suitable replacement since $T^{\times}_{k_m}$ is singular, and $m:{\cal X} \rightarrow {\cal X}^{\times}$ is the contraction of ${\cal X}$. The same holds for ${\cal X}'$. $m^*$

The functor ${\mathscr B}: {\mathscr D}_{\cal X} \cong {\mathscr D}_{{\cal X}'}$ preserves the filtrations, identifies
$
{\mathscr D}_{k_{m}}/{\mathscr D}_{k_{m+1}} \cong {\mathscr D}'_{k_{m}}/{\mathscr D}'_{k_{m+1}}$
and acts by degree shifts
\beq\label{BE}
{\mathscr B}:\;  {\mathscr A}_{k_{m}}/{\mathscr A}_{k_{m+1}}\rightarrow {\mathscr A}'_{k_{m}}/{\mathscr A}'_{k_{m+1}} \cong {\mathscr A}_{k_{m}}/{\mathscr A}_{k_{m+1}}\ [-D_m]\{C_m\} .
\eeq
which can be calculated from \eqref{shifta} and \eqref{shiftb} to be 
\beq\label{ourp}
D_m = (\ell -m)(n - \ell - m), \qquad C_m = D_m+\ell - m - {\ell^2 \over n},
\eeq
where $D_m$ is just the dimension of the Grassmannians $F_{k_m}$ and $F_{k_m}'$.  

The derived equivalence functor that comes naturally from the geometry of ${\cal X}$ is obtained by composing ${\mathscr B}$ with the equivariant shift functor  $\{c_{ij}\}$,
where $c_{ij}$ is given in \eqref{correction} and equals $\langle ^Lw_\ell, ^Lw_{n-\ell}\rangle  = {\ell^2 \over n}$ in our case. The functor ${\mathscr B} \circ \{c_{ij}\}$  acts as in \eqref{BE} with $C_m$ replaced with
\beq\label{ourpa}
C_m^{\cal X} =  D_m+\ell - m.
\eeq

The derived equivalence relating ${{\mathscr D}}_{\cal X}$ and $ {{\mathscr D}}_{{\cal X}'}$ was constructed in \cite{KamnitzerD}, by completely different means, using stratified Mukai flops, with stratification that has $\ell+1$ terms. The formulas for $D_m$ and $C_m^{\cal X}$ we just gave reproduce the result of theorem 2.8 of \cite{KamnitzerD}, obtained through a much more complex calculation.
% More precisely, the formula in \cite{KamnitzerD} is missing the  $- {\ell^2\over n}$ factor in \eqref{ourp}, which is however necessary\footnote{Working with $^LG = U(n)$ instead of $^LG=SU(n)$ one gets almost the same answer, as the $U(1)$ factor essentially decouples. The cohomological degree shift $D_m$ stays the same, but the equivariant degree shift becomes $C_m = D_m - m$. The difference between $U(n)$ and $SU(n)$ was important in a related context, in \cite{framed}.}
% for decategorification to give quantum invariants of links colored by the representations $V_i$ and $V_i^*$ of $^LG =SU(n)$.

\section{Categorified link invariants from ${\cal X}$}\label{s-four}
 
Quantum link invariants are certain very special matrix elements of the $U_{\fq}(^L\fg)$ braiding matrix, acting on the space of conformal blocks. As we reviewed in section \ref{s-one}, any link $K$ can be represented as a closure of a braid. We can take the braid $B$ to be a path in configuration space of $2m$ points on the Riemann surface ${\cal A}$, such that the closures at the top and at the bottom
represent collections of $m$ consecutive non-intersecting cups or caps, as in figure \ref{f_inout}. Then, the link invariant is given by 
\beq\label{Kb}
{\cal J}_{K}({\fq}) = ({\mathfrak U}_1, {\mathfrak B} \,{\mathfrak U}_0), 
\eeq
where ${\mathfrak U}_0$ and ${\mathfrak U}_1$ are the conformal blocks that correspond to the chosen closures, and where ${\mathfrak B}$ is the $U_{\fq}(^L\fg)$ braiding matrix corresponding to $B$. 

To categorify the link invariant in \eqref{Kb}, the derived categories must contain very special branes ${\cal U}_0 \in {\mathscr D}_{{\cal X}_{0}}$ and ${\cal U}_1 \in {\mathscr D}_{{\cal X}_{1}}$, one of whose properties is that their vertex functions 
\beq\label{Uvertex}
 {\mathfrak U}_0 = {\cal V}[{\cal U}_0]\quad \textup{ and }\quad {\mathfrak U}_1 = {\cal V}[{\cal U}_1]
 \eeq 
are the conformal blocks from \eqref{Kb}. These conformal blocks, one recalls, describe vertex operators colored by pairs of complex conjugate representations, which fuse together to identity. We will give an explicit geometric construction of ${\cal U}_{0}$ and ${\cal U}_1$, as structure sheaves of cycles which vanish as we approach the corresponding intersection of walls in Kahler moduli of ${\cal X}$.
Their second special property is:
\begin{theorem*}\label{t:four}
Bigraded homology groups
\beq\label{Bhom}{ Hom}^{*,*}({\cal U}_1,  {\mathscr B} \,{\cal U}_0 ),\eeq
are invariants of framed links.
\end{theorem*}
We call this a theorem with a $^*$, since its proof relies on an assumption that perverse filtrations of ${\mathscr D}_{\cal X}$ described in previous section exist at walls in Kahler moduli of ${\cal X}$.  The assumption can be shown to follow from theorem \ref{t:two}, with some work.\footnote{Note added: The theorem was proven in \cite{webster2} after this work appeared.}

By theorems \ref{t:two} or \ref{t:three}, ${\mathfrak B}$ lifts to a derived equivalence functor ${\mathscr B}:{\mathscr D}_{{\cal X}_0}\cong {\mathscr D}_{{\cal X}_1}$. Given two objects ${\cal U}_0 \in {\mathscr D}_{{\cal X}_{0}}$ and ${\cal U}_1 \in {\mathscr D}_{{\cal X}_{1}}$ such that \eqref{Uvertex} holds, the Euler characteristic of the homology group in \eqref{Bhom}
\beq\label{chiJ}
 \chi({\cal U}_1,  {\mathscr B} \,{\cal U}_0) = \sum_{n, k \in {\mathbb Z}} (-1)^n {\fq}^{k-D/2} \; {\rm dim} \,{Hom}({\cal U}_1,  {\mathscr B} \,{\cal U}_0[n]\{k\} )
\eeq
is the $U_{\fq}(^L\fg)$ link invariant which coincides 
with the matrix element of ${\mathfrak B}$ in \eqref{Kb}
\beq\label{J2}
\chi({\cal U}_1,  {\mathscr B}\, {\cal U}_0)= ({\mathfrak U}_1, {\mathfrak B}{\mathfrak U}_0) = {\cal J}_{K}({\fq}).
\eeq

Theorem \ref{t:four}, that homology groups ${ Hom}^{*,*}({\cal U}_1,  {\mathscr B} \,{\cal U}_0 )$ which categorify $ \chi({\cal U}_1,  {\mathscr B} \,{\cal U}_0)$ are themselves invariants of the link is a stronger statement.

\subsection{Mirror symmetry from Serre duality}
A well-known property of $U_{\fq}(^L\fg)$ link invariants
is that a link $K$ and the link $K^*$ which is its mirror reflection, have the same invariants 
\beq\label{mirr}
{\cal J}_{K}({\fq}) = {\cal J}_{{K^*}}({\fq}^{-1}).
\eeq
up to exchanging ${\fq}$ to ${\fq}^{-1}$. This is referred to as mirror symmetry of quantum link invariants. 
We will give a geometric explanation for it, based on ${\mathscr D}_{\cal X}$.

From Chern-Simons perspective, \eqref{mirr} is a consequence of the following elementary fact, explained in \cite{Jones}. Taking $K$ to $K^*$ is the same changing the orientation of the three manifold, which in turn is the same as changing the sign of ${\kappa}$, the effective Chern-Simons level ($\kappa$ multiplies the classical Chern-Simons action). This takes ${\fq}$ to ${\fq}^{-1}$, since ${\fq} = e^{2\pi i/\kappa}$. Viewing the link invariant as the matrix element of the braiding matrix
acting on conformal blocks, this says:
\beq\label{mirr2}
%$$
({\mathfrak U}_1,{\mathfrak B}  {\mathfrak U}_0)({\fq}) = ({\mathfrak B}{\mathfrak U}_0, {\mathfrak U}_1)({\fq}^{-1}),
%$$
\eeq
which is the same as \eqref{mirr} since ${\cal J}_{K}({\fq}) = ({\mathfrak U}_1,{\mathfrak B}  {\mathfrak U}_0)({\fq})$ and ${\cal J}_{K^*}({\fq}) = ({\mathfrak B}  {\mathfrak U}_0, {\mathfrak U}_1)({\fq})$.

From geometric perspective, mirror symmetry in \eqref{mirr} comes from an equally basic property of derived categories and the B-model, which is Serre duality. Serre duality is an isomorphism
\beq\label{mirr3}
{Hom}_{{\mathscr D}_{\cal X}}({\cal F}, {\cal G}[n]\{k\}) ={Hom}_{{\mathscr D}_{\cal X}}({\cal G}, {\cal F}[2D-n]\{D-k\}),
\eeq
of $Q$-cohomology with branes ${\cal F}, {\cal G}$ as boundary conditions at the two ends of the interval $s\in [0,1]$, and $Q$-cohomology obtained by reflection on $s$, which exchanges ${\cal G}$ and ${\cal F}$.
%
%
%and 
%$$
%{Hom}_{{\mathscr D}_{\cal X}}({\cal G}\{k\}, {\cal F}[2d-n]) = {Hom}_{{\mathscr D}_{\cal X}}({\cal G}, {\cal F}[2d-n]\{\textcolor{red}{d}-k\})
%$$
In writing \eqref{mirr3}, we used the fact that the canonical line bundle $K_{\cal X}$ of ${\cal X}$ is trivial, since ${\cal X}$ is holomorphic symplectic, and $2D={\rm dim}_{\mathbb C} {\cal X}$ is its complex dimension. The shift in equivariant  ${\mathbb C}^{\times}_{\fq}$ degree by $\{D\}$ on the right hand side of \eqref{mirr3} comes from the fact that, while $K_{\cal X}$ of ${\cal X}$ is trivial, its unique holomorphic section is not invariant under the ${\rm T}$-action on ${\cal X}$, but transforms with character ${\fq}^{D}$, coming from the action of ${\rm T}$ on the holomorphic symplectic form.

The two actions in \eqref{mirr2} and in \eqref{mirr3} have to be the same in our case because the three manifold where Chern-Simons theory lives is ${\cal A} \times {\mathbb R}$, with ${\mathbb R}$ parameterized by $s$: taking $s$ to $-s$ leads to both \eqref{mirr2} and \eqref{mirr3}. 

It follows from \eqref{mirr3} that the Euler characteristic of our homology theory $H^{*,*}({\cal U}_1, {\mathscr B} {\cal U}_0)$, as defined in \eqref{defhc}, satisfies
\beq\label{mirr4}
%$$
\chi({\cal U}_1, {\mathscr B} {\cal U}_0)({\fq}) = \chi({\mathscr B} {\cal U}_0, {\cal U}_1) ({\fq}^{-1}).
%$$
\eeq
The left and the right hand sides of this equations exactly coincide with the left and the right hand sides of \eqref{mirr} and \eqref{mirr2}.

\subsection{Cups and caps as objects}\label{s-uk}

Take ${\cal X}$ to correspond to a sequence of $2m$ vertex operators ${\Phi}_{V_i}(a_{2i-1})$ and ${\Phi}_{V_i^*}(a_{2i})$, colored by conjugate representations, and order the vertex operators so that $y_i= \ln |a_i|$ increases with $i$. We choose
the weight $\nu$ which vanishes. Then, 
$${\cal X} = {{\rm Gr}^{{\vec \mu}_{2m}}_{0}},
$$ 
where ${\vec \mu}_{2m}$ is a vector with $2m$ entries which a sequence of pairs  $({\mu}_i , {\mu}_{i}^*)$, for $i=1, \ldots, m$.

\subsubsection{}

\begin{figure}[!hbtp]
  \centering
   \includegraphics[scale=0.4]{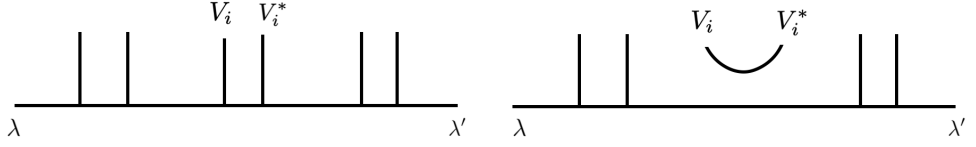}
 \caption{Fusing a pair of vertex operators to the identity.}
  \label{f_cap}
\end{figure}

As $a_{ 2i-1}$ approaches $a_{2i}$, the vertex operators  ${\Phi}_{V_i}(a_{2i-1})$ and ${\Phi}_{V_i^*}(a_{2i})$ can fuse to the identity operator, corresponding to the trivial representation in the tensor product of ${V_i}$ and ${V_i^{*}}$, and disappear.  A conformal block corresponding to this process contains a cup colored by representation ${V}_i$, as figure \ref{f_cap}. 

As $a_{2i-1}$ and $a_{2i}$ approach each other with $y_{2i-1}<y_{2i}$, 
${\cal X}$ develops singularities. The possible singularities are labeled by representations ${V}_{k_m}$ in the tensor product ${{V}_i} \otimes{{V}_i^{*}}$, as explained in section \ref{s-three}.  We will denote by $W^{(i)}$ the singularity corresponding to taking ${V}_{k_m}$ to be the trivial representation
 \beq\label{W0}
W^{(i)} = {\rm Gr}^{(\mu_i, \mu_i^*)}_0 = T^* U_i.
\eeq
It is not difficult to show that the vanishing cycle $U_i $ is the homogenous space 
\beq\label{cap}
 U_i = {\rm Gr}^{{\mu}_i}=G/P_i,
\eeq
(see appendix for derivation). As in \eqref{basic}, $P_i$ is a maximal parabolic subgroup of $G$ corresponding to $\mu_i$. 
\subsubsection{}
Let now all the $2m$ vertex operators come together in pairs, ${\Phi}_{V_i}(a_{2i-1})$ and ${\Phi}_{V_i^*}(a_{2i})$  approaching each other for all $i=1, \ldots, m$, keeping their ordering fixed. Then, ${\cal X}$ develops a vanishing cycle 
$U$ which is the product 
\beq\label{basic}
U =U_1\times \ldots \times U_m=  G/P_1\times \ldots  \times G/P_m,
\eeq
of vanishing cycles one obtains by colliding vertex  operators pairwise.

${\cal X} $ itself has a local neighborhood where it factors as
$$
{\cal X}  \;\sim \; W^{(1)}\, {\times} \,W^{(2)}\,  {\times} \,\ldots  \, {\times}\, W^{(m)}  = T^*U.
$$
This is just the description of the geometry of a holomorphic symplectic manifold in the neighborhood of its vanishing cycle.
\subsubsection{}
We claim that the special object we need, 
$${\cal U} \in {\mathscr D}_{{\cal X}},$$ 
corresponding to the conformal block ${\mathfrak U}$ where the vertex operators fuse to the identity is
the structure sheaf of the vanishing cycle $U$, 
\beq\label{cupU}
 {\cal U}= {\cal O}_{U}.
 \eeq
 The sheaf ${\cal U
 }$ generates the bottom part of the filtration \eqref{filtration0a} near the intersection of $m$ walls in Kahler moduli, where $y_{2i}$ and $y_{2i+1}$ approach each other, remaining otherwise distinct. Its generalized central charge
 $$
 {\mathfrak U} = {\cal V}[{\cal U}]
 $$ 
 is the conformal block in figure \ref{f_inout}.

\subsubsection{}
Globally, the vanishing cycle $U$ is the set of all points in ${\cal X} = {\rm Gr}^{\vec{\mu}_{2m}}_0$ of the form
$$
U = \{(L_1, \ldots ,L_{2m} ) \in {\cal X} | \, L_{2j} = z^0, \textup{for all }j\}.
$$
As $a_{2i-1}$ and $a_{2i}$ approach each other pairwise, ${\cal X}$ develops singularities  
$$m: \; {\cal X} =  {\rm Gr}^{{\vec \mu}_{2m}}_0\; \rightarrow  \;{\cal X}^{\times} = {\rm Gr}^{{\vec \mu}^{\times}_{m}}_0,$$
where ${\vec {\mu}}^{\times}_{m} = ({\mu}_1+{\mu}_1^*, \ldots , {\mu}_m+{\mu}_m^*)$. This gives us an alternative description of the vanishing cycle, as  the locus $U=  m^{-1} (z^{0}, \ldots , z^{0} )$  in ${\cal X}$.

\subsection{Unknot homology}
As a simple check, consider
\beq\label{defh}
{\rm Hom}^{*,*}_{{\mathscr D}_{\cal X}}({\cal U},  {\cal U})
\eeq
for 
${\cal U} = {\mathscr O}_{U},$ with $U$ as in  \eqref{basic}. Since $U$ is embedded in ${\cal X}$ as the zero section of its local $W = T^*U$ neighborhood, the Hom's in \eqref{defh} can be computed in ${\mathscr D}_{W} = D^bCoh_{\rm T}(W)$.  The non-vanishing Hom's are (see \cite{CK2}, remark 5.11)
\beq\label{key}
Hom_{{\mathscr D}_{\cal X}}({\cal U},  {\cal U}[2 j ]\{j\}) =Hom_{{\mathscr D}_{W}}({\cal O}_{U},  {\cal O}_U[2j]\{j\}) =H^{j}(U),
\eeq
Above, $H^j(U)$ denotes $H^{j,j}(U)$.
\subsubsection{}

It follows that the Euler character
\beq\label{Ea}
{\chi}({\cal U}, {\cal U}) = \sum_{j, k} (-1)^j {\fq}^{k-D/2} Hom_{{\mathscr D}_{\cal X}}({\cal U},  {\cal U}[j ]\{k\}), 
\eeq
where $D = {1\over 2} dim_{\mathbb C} {\cal X} = dim_{\mathbb C} {U}  $,  computes the Poincare polynomial of $U$, up to a pre-factor
\beq\label{UP}
{\chi}({\cal U}, {\cal U})  = {\fq}^{-D/2} P_U({\fq}),
\eeq
where 
$$P_U({\fq})= \sum_{j=0}^D {\fq}^{j} \, {\rm dim} \,H^{j}(U).
$$
The cohomology of $U$ is the product
$$
 H^{*}(U ) = \bigotimes_{i=1}^m\, H^*(G/P_i),
$$
and dimensions add:
$$D = {\rm dim}\, U= \sum_{i=1}^{m} {\rm dim} \,U_i,
$$
so the Euler character of the sheaves on $W=T^*U$ is the product of Euler characters of sheaves coming from its $T^*(G/P_i)$ factors.

\subsubsection{}
The Poincare polynomial of
$$
U_i = {\rm Gr}^{\mu_i} = G/P_i
$$
can be computed by Morse theory.  

Recall that the torus ${\rm T}$ acts on $U_i$ with isolated fixed points. Pick a suitably generic 1-parameter subgroup, ${\mathbb C}^{\times} \subset {\rm T}$. The Hamiltonian that generates it is a real Morse function on $U_i$, whose critical points are the fixed points of the ${\mathbb C}^{\times}$-action. Given a fixed point, the number of attracting directions of its gradient flow, or equivalently, the number of negative eigenvalues of the Hessian of the Morse function, is the degree of the class which the fixed point contributes to $H^*(U_i)$. (In our case, only even cohomology shows up in this calculation, so there are no corrections due to instantons.)

As we review in the appendix, the fixed points of the ${\rm T}$-action on $U_i$ are in one to one correspondence with the weights $^Lw$ in the representation ${V_i}$ of $^L\fg$. In turn, the tangent space to $U_i$ at the fixed point labeled by the weight $^Lw$ is spanned by vectors which correspond to roots $\alpha$ of $\fg$ with the property that $\langle \alpha, {^Lw}\rangle = 1$ \cite{V, Danilenko}. 

For a convenient ${\mathbb C}^{\times}$-action, the split of the tangent space into the attracting and repelling directions 
$$
T_{
^Lw}U_i = T^+_{
^Lw}U_i \oplus T^-_{
^Lw}U_i,
$$
is the split of {\it positive} roots ${\alpha}$ into those with $\langle \alpha, ^Lw \rangle = +1$, corresponding to attracting directions, and those with $\langle \alpha, ^Lw \rangle = -1$, corresponding to repelling directions.

We can count the number of attracting directions at $^Lw$, and therefore compute the degree of the cohomology class in $H^*(U_i)$ that corresponds to it, as follows.  Take the Weyl vector ${\rho}$ of ${\fg}$, which is half the sum of its positive roots (more precisely, $\rho$ is the Weyl co-vector of $^L\fg$), and consider 
\beq\label{number}
\langle \rho, {^Lw}\rangle = {1\over 2}\sum_{\alpha>0}  \langle \alpha, {^Lw}\rangle = {1\over 2 } (n_+ - n_-) = n_+ - {\rm dim}\, U_i.
\eeq
The first equality is the definition of the Weyl vector. Since $^Lw$ is a minuscule weight, its inner product with any root can only be $0$ or $\pm 1$. A given positive root $\alpha$ contributes to the right hand side only if $\pm \alpha$ is in the tangent space. It contributes $+1/2$ if it is an attracting, and $-1/2$ if it is a repelling direction, which gives us the second equality. Altogether, $\langle \rho, {^Lw}\rangle$ equals the number $n_+$ of attracting directions, up to a constant shift. For $n_+=j$, the fixed point corresponding to $^Lw$ contributes to $H^j({U_i})
$.

The Poincare polynomial is therefore given by 
\beq\label{PUf}
P_{U_i} (\fq) = \sum_{j} {\fq}^j {\rm dim} \,H^j({U_i}) = {\fq}^{{\rm dim}\, U_i} \sum_{^Lw} {\fq}^{{\langle \rho, ^Lw\rangle}}, 
\eeq
which is a sum over all weights $^Lw$ of the minuscule representation $V_i$ of $^L\fg$. 
%The weight $^Lw$ corresponds to the cohomology class in $H^j({U_i}) $, for $j$ equal to $j = n_+(^Lw) = {1\over 2} d_i + \langle \rho, {^Lw}\rangle$.
The sum on the right hand side is just the character of $^{L}{\fg}$ in representation $V$, so we can rewrite the equation as 
\beq\label{charan}
\chi({\cal U}_i, {\cal U}_i)={\fq}^{-{\rm dim}\, U_i} P_{U_i}(\fq) ={{\rm tr}}_{\,{V}} \, {\fq}^{ \rho}.
\eeq
This is the specialization of the Euler characteristic to having a single cap, instead of $m$ of them. In other words, one regards here ${\cal U}_i $ as the structure sheaf of $U_i$ inside $T^*U_i$, ${\cal U}_i= {\cal O}_{U_i}$.

\subsubsection{}

When $^L\fg = {\fg}$ is simply laced, the roots and co-roots, and weights and co-weights get identified, and in particular, the Weyl vectors of $^L{\fg}$ and of ${\fg}$ coincide 
\beq\label{Weyls}
^L\rho = \rho.
\eeq
(When the Lie algebra is not simply laced, this is not the case, not even up to a rescaling: the vectors $^L\rho$ and $\rho$ are then truly different.)
Then, the Euler characteristic of \eqref{charan} coincides with the $U_{\fq}(^L\fg)$ invariant of the unknot, 
\beq\label{unknot prediction}
\chi({\cal U}_i, {\cal U}_i) = {{\rm tr}}_{\,{V_i}} \, {\fq}^{ ^L\!\rho} = J_{{\bigcirc, {V_i}}}({\fq})
\eeq
colored by representation $V_i$ of $^L\fg$ in "vertical framing". The relation of the quantum link invariant to the trace in representation $V_i$ of $^L\fg$
 comes from Verlinde algebra, which says that 
\beq\label{unknot prediction}
J_{{\bigcirc, {V_i}}}({\fq}) =  S_{V_i 0}/S_{00} = {{\rm tr}}_{\,{V_i}} \, {\fq}^{ ^L\!\rho},
\eeq
where $S_{V_i V_j}$ is the $S$-matrix element of the affine $\Lfgh$ Lie algebra. The character $ {{\rm tr}}_{\,{V_i}} \, {\fq}^{ ^L\!\rho}$ is sometimes referred to as the quantum dimension of the representation $V_i$ of $^L\fg$.
 Thus, the Hom in \eqref{defh} categorifies the $U_{\fq}(^L\fg)$ invariant of the collection of $m$ unknots colored by minuscule representations $V_i$.

\subsubsection{}
When $^L\fg$ is not simply laced, ${^L\fg}$ and ${\fg}$ are distinct, and so are their Weyl vectors $^L\rho \neq \rho$. (Since we are restricting to minuscule representations, the only non-simply laced Lie algebras that have them are of $B_n$ and $C_n$ types, and they are exchanged by  Langlands duality.) While the unknot invariant is given by ${{\rm tr}}_{\,{V_i}} \, {\fq}^{ ^L\!\rho}$, the Euler characteristic is given by ${{\rm tr}}_{\,{V_i}} \, {\fq}^{\rho}$, and the two are distinct. 

Correspondingly, while one might have guessed that to extend our story to non-simply laced Lie algebras, one should simply working with slices in affine Grassmannian of the Lie group $G$, this cannot be the case. Indeed, the naive extension of the theorem \ref{t:one} does not hold either \cite{Danilenko}.

This is all expected from string theory perspective, where to obtain non-simply laced Lie algebras do not appear directly. Rather, to obtain them requires an extra step, which is to start with a theory based on a simply laced Lie algebra and use a version of folding. 
We will describe this in \cite{A2} and \cite{A3}.

\subsection{Cap and cup functors}\label{s-cf}

Fusing a pair of vertex operators to the identity
\beq\label{vi}
{\mathfrak C}_i^{\vee} : \;\;\Phi_{{V}_i}(a_{2i-1}) \otimes \Phi_{{V}_i^{*}}(a_{2i}) \; \;\; \rightarrow \;\;\ \mathbbm{1}
\eeq
and the inverse process of pair creation
\beq\label{iv}
{\mathfrak C}_i :\;\;  \mathbbm{1}\;\; \rightarrow \;\; \Phi_{{V}_i}(a_{2i-1}) \otimes \Phi_{{V}_i^{*}}(a_{2i}),
\eeq
gives a pair of maps, the first of which takes a conformal block of the form 
\beq\label{electrica}
\langle \lambda|\Phi_{V_1}(a_1)\cdots \, \Phi_{{V}_i}(a_{2i-1}) \otimes \Phi_{{V}_i^{*}}(a_{2i})\,\cdots  \Phi_{V_{n}}(a_{n})| \lambda'\rangle,
\eeq
which comes from ${\cal X}_{n} = {\rm Gr}^{{\vec \mu}_{n}}_{\nu}$,
  to a conformal block of the form
\beq\label{electric'}
\langle \lambda|\Phi_{V_1}(a_1)\cdots \, \mathbbm{1}\, \cdots  \Phi_{V_{n}}(a_{n})| \lambda'\rangle,
\eeq
which comes from 
 ${\cal X}_{n-2} = {\rm Gr}^{{\vec \mu}_{n-2}}_{\nu}$. 
 Here, ${\vec \mu}_{n-2}$ is a vector with $n-2$ entries is obtained from ${\vec \mu}_{n}$ by deleting the pair  ${\mu}_i , {\mu}_{i}^*$  in the $2i-1$'th and $2i$'th slots. The second map, ${\mathfrak C}_i$ goes the other way.
Thus, the caps and cups define maps between spaces of conformal blocks. 

%The fact that the matrix elements of ${\mathfrak B}$ are invariants of links implies that these maps have to satisfy certain moves which we will recall below. 
\subsubsection{}\label{s-cap}
  
The pair of maps ${\mathfrak C}_i$ and ${\mathfrak C}_i^{\vee}$ in \eqref{vi} and \eqref{iv}, acting on spaces of conformal blocks lift to a pair of functors acting on the derived categories ${\mathscr D}_{{\cal X}_{n-2}}$ and  ${\mathscr D}_{{\cal X}_{n}}$,
\beq\label{cupcap}
{\mathscr  C}_i : {\mathscr D}_{{\cal X}_{n-2}} \longrightarrow {\mathscr D}_{{\cal X}_{n}}, 
 \qquad {\mathscr  C}^{\vee}_i : {\mathscr D}_{{\cal X}_{n}} \longrightarrow {\mathscr D}_{{\cal X}_{n-2}}, 
\eeq

To construct them, following \cite{CK1, CK2}, we start with 
a holomorphic Lagrangian $C_i$ on the product ${\cal X}_{n-2} \times {\cal X}_n$. $C_i$ embeds into ${\cal X}_{n}$ as a subspace of codimension ${\rm dim}( U_i)$, and fibers over ${\cal X}_{n-2}$ with fiber $U_i$.  $U_i$ is the vanishing cycle in ${\cal X}_{n}$ that leads to the cap colored by representation $V_i$. It is easy to see that the dimension count works since ${\rm dim} \,C_i = {\rm dim}({\cal X}_{n-2}) + {\rm dim}(U_i) = {1\over 2}{\rm dim}({\cal X}_{n-2} + {\cal X}_n)$. Explicitly,
$C_i$  is obtained by identifying points 
$$ \{(L_1, \ldots ,L_{2i-1} = L_{2i+1}, \ldots L_n) \in {\cal X}_{n-2} \},$$
which is all of ${\cal X}_{n-2}$,
with points 
$$
 \{(L_1, \ldots L_{2i-1},  L_{2i}, L_{2i+1}, \ldots L_n) \in {\cal X}_{n}|L_{2i-1} = L_{2i+1}\},
$$
in ${\cal X}_n$. 

The pair of functors in \eqref{cupcap} are Fourier-Mukai transforms with kernel
${\cal C}_i \in {\mathscr D}({\cal X}_{n-2} \times {\cal X}_n)$ which is the structure sheaf of $C_i$
$${\cal C}_i = {\cal O}_{C_i}.
$$
The Fourier-Mukai transform 
${\mathscr C}_i $ takes
$$
{\mathscr C}_i : \; {\cal F}_{n-2} \in {\mathscr D}_{{\cal X}_{n-2}} \rightarrow \pi_{2*}(\pi_1^*({\cal F}_{n-2}) \otimes {\cal C}_i)
$$
where $\pi_1$ and $\pi_2$ are projections to the first and the second factor of ${\cal X}_{n-2} \times {\cal X}_n$: $\pi_1^*({\cal F}_{n-2})$ interprets a sheaf ${\cal F}_{n-2}$ on ${\cal X}_{n-2}$ as a sheaf on the product, and $ \pi_{2*}$ takes a sheaf on the product to a sheaf on ${\cal X}_{n}$. Since ${\cal C}_i$ is the structure sheaf of $C_i$, $\pi_{2*}(\pi_1^*({\cal F}_{n-2}) \otimes {\cal C}_i)$ is the same as $i_*(f^*({\cal F}_{n-2}))$ where $f$ projects to ${\cal X}_{n-2}$ by forgetting the $U_i$ fiber of $C_i$:
$$f: C_i  \rightarrow\, {\cal X}_{n-2},$$
and $i$ describes inclusion of $C_i$ into ${\cal X}_n$ as a divisor
$$i: C_i \,\rightarrow \,{\cal X}_{n},$$ 
see \cite{CK1, CK2}.
More generally, it follows that, given any two objects ${\cal F}_{n-2} \in {\mathscr D}_{{\cal X}_{n-2}}$ and  ${\cal G}_n \in {\mathscr D}_{{\cal X}_{n}}$  the functors act as
$${\mathscr C}_i ({\cal F}_{n-2})= i_*(f^*({\cal F}_{n-2})), \qquad {\mathscr C}_i^{\vee} ({\cal G}_{n})= f^*(i_*({\cal G}_{n})).
$$
\subsubsection{}
An important property of the functor ${\mathscr C}_i$ is that sends every object of ${\cal F}_{n-2} \in  {\mathscr D}_{{\cal X}_{n-2}}$ to a unique object 
 \beq\label{imv}
 {\cal F}_{n}  = {\mathscr C}_i({\cal F}_{n-2})\in  {\mathscr  D}_{{\cal X}_{n}}
 \eeq
 which sits at the bottom part of the filtration associated to bringing $a_{2i-1}$ and $a_{2i}$ together.

If ${\cal F}_{n} \in  {\mathscr  D}_{{\cal X}_{n}}$ comes from some ${\cal F}_{n-2}\in  {\mathscr  D}_{{\cal X}_{n-2}}$ via the functor ${\mathscr C}_i$, its central charge ${\cal Z}[{\cal F}_{n}]$ corresponds to a scalar conformal block where 
$\Phi_{{V}_i}(a_{2i-1})$ and $ \Phi_{{V}_i^{*}}(a_{2i}) $ fuse to the identity as $a_{2i-1}$ and $a_{2i}$ approach each other. More precisely its  asymptotic behavior as $a_{2i-1}$ approaches $a_{2i}$ is
$${\cal Z}[{\cal F}_{n}]\;\; = \;\; {\cal Z}[{\cal F}_{n-2}] (a_{2i-1} - a_{2i})^{2\Delta_i}\times{\textup{finite}},$$
where the finite terms are regular and equal $1$ in the limit. The KZ equation and these asymptotics uniquely fix ${\cal Z}[{\cal F}_{n}]$ given  ${\cal Z}[{\cal F}_{n-2}]$; this is nothing but the imprint of the functor ${\mathscr C}_i$ on $K_{{\rm T}}({\cal X})$.

\subsection{Isotopy invariance}

To prove the theorem \ref{t:four}, stating that the homology groups in  \eqref{Bhom} are invariants of framed links we need to show they are invariant under Reidermeister I, II and III moves, and the ``pitchfork'' and the ``$S$-move'', see for example \cite{webster}. 

Invariance Reidermeister II and III moves is the statement the homology group is invariant under ambient braid isotopies. We explained why this is the case in section \ref{s-two}. Below, we will explain why it is invariant under the remaining three moves as well.

\subsubsection{}

The pitchfork identity states that

\begin{figure}[!hbtp]
  \centering
   \includegraphics[scale=0.2]{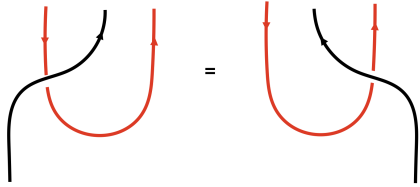}
 \caption{Pitchfork identity...}
  \label{f_rel2}
\end{figure}
By further braiding, we get an equivalent identity:

\begin{figure}[!hbtp]
  \centering
   \includegraphics[scale=0.2]{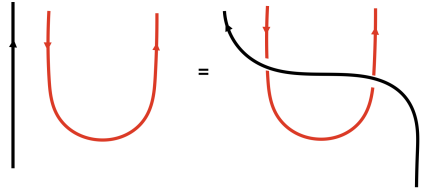}
 \caption{....and its equivalent.}
  \label{f_rel1}
\end{figure}
At the level of categories, the figure \ref{f_rel1} relates the cap functor ${\mathscr  C}_i : {\mathscr D}_{{\cal X}_{n-2}} \longrightarrow {\mathscr D}_{{\cal X}_{n}}$, to ${\mathscr  C}''_i : {\mathscr D}_{{\cal X}_{n-2}} \longrightarrow {\mathscr D}_{{\cal X}''_{n}}$ by crossing two consecutive walls. It expresses an equivalence of functors 
\beq\label{BC}
{\mathscr B}\circ {\mathscr C}_i \cong  {\mathscr C}_i'',
\eeq 
where the functor ${\mathscr B}$ is an equivalence of categories ${\mathscr B}: {\mathscr D}_{{\cal X}_n} \cong {\mathscr D}_{{\cal X}''_n}$, which comes from braiding
 $\Phi_{{V}_{k}}(a_{k})$ with $(\Phi_{V_{i}}(a_{i}) \otimes \Phi_{{V}_{i}^*}(a_{j}))$.  We will now explain why \eqref{BC} holds, using results of previous sections.
 
 \subsubsection{}
Consider the more general equivalence of categories ${\mathscr B}: {\mathscr D}_{\cal X} \cong {\mathscr D}_{{\cal X}''}$ which comes from braiding three arbitrary vertex operators  $\Phi_{V_{i}}(a_{i})$, $\Phi_{V_{j}}(a_{j})$ and  $\Phi_{V_{k}}(a_{k})$, from $y_i<y_j<y_k$ in ${\cal X}$ to $y_k<y_i<y_j$ in ${\cal X}''$.
This requires crossing a pair of intersecting walls, the first at $y_i<y_j=y_k$, the second at $y_k = y_i<y_j$.

In ${\cal X}$, we can assume that $y_j-y_i$ and $y_k - y_j$ both go to zero, in such a way that $0<y_j-y_i \ll y_k- y_j$.   
In this regime, the KZ equation has a basis of solutions labeled by a pair of representations occurring in the tensor product
\beq\label{start}
(\Phi_{{V}_i}(a_i) \otimes \Phi_{{V}_j}(a_j) ) \otimes \Phi_{{V}_k}(a_k).
\eeq
They correspond to conformal blocks in which  $\Phi_{V_{i}}(a_{i})$ and $\Phi_{V_{j}}(a_{j})$ first fuse to $\Phi_{V_{\ell}}(a_{j})$, and then $\Phi_{V_{\ell}}(a_{j})$ and $\Phi_{V_{k}}(a_{k})$ fuse to $\Phi_{V_{m}}(a_{k})$. 
The corresponding scalar conformal block ${\cal Z}_{\ell, m}$ labeled by a pair of representations $(V_{\ell}, V_{m})$ has the following asymptotics 
\beq\label{double}
{\cal Z}_{\ell, m} = (a_i-a_j)^{\Delta_i+\Delta_j - \Delta_\ell} \;  (a_j-a_k)^{\Delta_\ell+\Delta_k - \Delta_m} \times {\rm finite},
\eeq
where $\Delta_i = d_i -  c_i/\kappa$, as in \eqref{ddf}. The representations $V_{\ell}$, ${V}_{m}$ are not of course minuscule, in general.

The filtration of ${\mathscr D}_{\cal X}$ by orders of vanishing of ${\cal Z}^0$ now looks as follows.  Let ${\mathscr A}_{\ell_s}$ be the subcategory corresponding to objects whose central charge vanishes at least as fast as $ (a_i-a_j)^{D_{\ell_s}^{ij}}$, where $D_{\ell_s}^{ij} = d_i+d_j - d_{\ell_s}$ is the dimension of the vanishing cycle corresponding to representation $V_{\ell_s}$ in the tensor product $V_i\otimes V_j$. All the vanishing cycles we need are found in section \ref{s-three}. Let ${\mathscr D}_{\ell_s}$ be the subcategory of ${\mathscr D}_{{\cal X}}$ whose objects are generated from objects in ${\mathscr A}_{\ell_s}$ by grade shifts, direct sums, and taking iterated cones. This gives a filtration of the derived category ${\mathscr D}_{\cal X}$, as in the previous section:
\beq\label{firstfil}
{\mathscr D}_{\ell_0} \subset\ldots \subset {\mathscr D}_{\ell_{s_{max}}}  = {\mathscr D}_{{\cal X}}.
\eeq
The quotient category  
$${\mathscr D}^{(1)}_{\ell_s} = {\mathscr D}_{\ell_s}/{\mathscr D}_{\ell_{s-1}}$$ 
has cohomology objects in $
{\mathscr A}^{(1)}_{\ell_s} = {\rm gr}_{\ell_s}({\mathscr A}) ={\mathscr A}_{\ell_{s}}/{\mathscr A}_{\ell_{s-1}}
  $
Next, each  $ {\mathscr D}^{(1)}_{\ell_{s}}$ itself gets a second filtration by orders of vanishing of ${\cal Z}^0$ near $y_j=y_k$
\beq\label{secondfil}
{\mathscr D}^{(1)}_{\ell_s, m_{t_0}} \subset\ldots \subset {\mathscr D}^{(1)}_{\ell_s, m_{t_{max}}}  = {\mathscr D}^{(1)}_{\ell_{s}}.
\eeq
The second filtration has as many terms as there are representations $V_{m_t}$ in the tensor product of ${V}_{\ell} \otimes {V}_k$. It comes from vanishing cycles that develop within $T_{\ell_s}^{\times}$, in notation of section \ref{s-three}, as the pair of singular monopoles, corresponding to vertex operators $\Phi_{{V}_{{\ell}_s}}(a_j)\otimes\Phi_{{V}_{k} }(a_k)$ approach each other. The vanishing cycle corresponding to $V_{m_t}$ has dimension $D^{k \ell_s}_{m_t} = d_k+d_{\ell_s} - d_{m_t}$, and shrinks as $y_j$ approaches $y_k$. At $y_j=y_k$, monopole bubbling can occur which leaves behind a single singular monopole of charge $\mu_{m_t}$, which is the highest weight of representation $V_{m_t}$.
Letting ${\mathscr A}^{(1)}_{\ell_s,m_t} = {\mathscr A} \cap {\mathscr D}^{(1)}_{\ell_s, m_t}$, 
 $$
 {\rm gr}_{\ell_s, m_{t}}({\mathscr A}) ={\mathscr A}^{(1)}_{\ell_s, m_t}/{\mathscr A}^{(1)}_{\ell_s, m_{t-1}}
  $$
 consists of objects supported on the vanishing cycle, and whose central charge vanishes exactly as fast as $(a_i - a_j)^{D^{ij}_{{\ell_s}}}(a_j - a_k)^{D^{k \ell_s}_{m_t}}$.
 
Consider now braiding $\Phi_{V_{i}}(a_{i})$, $\Phi_{V_{j}}(a_{j})$ and  $\Phi_{V_{k}}(a_{k})$ to get to ${\cal X}''$. We end up with fusion products in the following order:
\beq\label{finish}
 \Phi_{{V}_k}(a_k) \otimes (\Phi_{{V}_i}(a_i) \otimes \Phi_{{V}_j}(a_j) ).
\eeq
The order of vertex operators has changed to $y_k < y_i < y_j$, where $y_j - y_i\ll y_j- y_k$. This corresponds to a different chamber ${\fC}''$  in Kahler moduli, giving a different resolution ${\cal X}''$ of the singularity, different stability structure, and different abelian subcategory ${\mathscr A}''$ of its derived category ${\mathscr D}_{{\cal X}''}$. 
What does not change from chamber ${\fC}$ to ${\fC}''$ is the fact that the KZ equations has solutions which correspond to fusing vertex operators in the order \eqref{finish}. Correspondingly, in both chambers one gets a basis of solutions to the scalar KZ equation with the asymptotics given by \eqref{double}. 

The derived equivalence ${\mathscr D}_{\cal X} \cong {\mathscr D}_{{\cal X}''}$ preserves the filtrations in \eqref{firstfil} and \eqref{secondfil}, but the identification of their abelian hearts is non-trivial. Any half monodromy that takes one from chamber ${\fC}$ to ${\fC}'$ changes the phase of central charges, due to the second factor in \eqref{double}. This preserves the double filtration, and acts by a shift of gradings:
\beq\label{g3}
 {\rm gr}_{\ell_s, m_{t}}({\mathscr A}'') =  {\rm gr}_{\ell_s, m_{t}}({\mathscr A})[-D_{s,t}]\{C_{s,t}\},
\eeq
where $D_{s,t} = d_k + d_{\ell_s} - d_{m_t} ={D^{k\; \ell_s}_{m_t}} $ is the dimension of the vanishing cycle, and $C_{s,t} = c_k + c_{\ell_s} - c_{m_t}$.
On an object in a graded subcategory ${\mathscr D}_{\ell_s, m_t}^{(1)}$, the braiding functor acts by a degree shift, plus some contribution of objects from lower orders in the filtration. However, on any object in the bottom term of the double filtration, which is ${\mathscr D}^{(1)}_{\ell_0, m_{t_0}}$, it acts only by degree shifts,
\beq\label{g3}
{\mathscr B}{\mathscr D}^{(1)}_{\ell_0, m_{t_0}} \cong  {{\mathscr D}''}^{(1)}_{\ell_0, m_{t_0}}   [-D_{0,0}]\{C_{0,0}\}.
\eeq
This will be important below.

\subsubsection{}

Now we can go back to the identity in figure \ref{f_rel1}. With the conjugate pair of representations, ${V}_{j} = {V}_{i}^*$, the bottom part of the first filtration ${\mathscr D}_{\ell_0} \subset {\mathscr D}_{{\cal X}_n}$ corresponds to $\Phi_{V_{i}}(a_{i}) \otimes \Phi_{{V}_{i}^*}(a_{j})$ fusing to the identity. Then second filtration then has only one term, because in the tensor product of $V_k$ with identity only one representation occurs. Thus, the bottom part of the double filtration is ${\mathscr D}_{\ell_0}$ itself,
$
{\mathscr D}^{(1)}_{\ell_0, m_{t_0}} \cong {\mathscr D}_{\ell_0}.
$
This is the subcategory of  ${\mathscr D}_{{\cal X}_n}$ which one gets as the image of the functor ${\mathscr C}_i$ from a category ${\mathscr D}_{{\cal X}_{n-2}}$ corresponding to omitting $\Phi_{V_{i}}(a_{i}) \otimes \Phi_{{V}_{i}^*}(a_{j})$ all together:
\beq\label{gra}
{\mathscr D}_{\ell_0} ={\mathscr C}_i {\mathscr D}_{{\cal X}_{n-2}} \subset {\mathscr D}_{{\cal X}_n}.
\eeq
By the same reasoning, on ${\cal X}_n''$, we have 
\beq\label{grb}
{\mathscr D}''_{\ell_0}={\mathscr C}''_i {\mathscr D}_{{\cal X}_{n-2}} \subset {\mathscr D}_{{\cal X}_n''}.
\eeq

Now consider the action of ${\mathscr B}$ corresponding to braiding $\Phi_{V_{k}}(a_{k}) $ with $\Phi_{V_{i}}(a_{i}) \otimes \Phi_{{V}_{i}^*}(a_{j})$
on ${\mathscr C}_i {\mathscr D}_{{\cal X}_{n-2}}$. Since  ${\mathscr C}_i {\mathscr D}_{{\cal X}_{n-2}} = {\mathscr D}_{\ell_0}$ is the bottom part of a double filtration the functor ${\mathscr B}$ acts at most by a degree shift from \eqref{g3}. In our case, the degree shift is trivial, since the representation ${V}_{\ell}$ is trivial and the only representation in the tensor product of the trivial representation ${V}_\ell$ with ${V}_k$ is ${V}_m = {V}_k$. Consequently, the contributions of ${V}_m$ and ${V}_k$ to the shifts in \eqref{g3} cancel, and ${V}_{\ell}$ contributes zero, as it is a trivial representation. A braiding functor that acts by a trivial degree shift is identity, so we get 
${\mathscr B}\circ {\mathscr C}_i \cong  {\mathscr C}_i''$, which is what we wanted to show.

\subsubsection{}

The $S$-move, or Reidermeister $0$, corresponds to the following diagram:

\begin{figure}[!hbtp]
  \centering
   \includegraphics[scale=0.2]{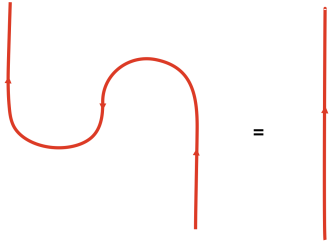}
 \caption{The S-move or Reidermeister $0$}
  \label{f_rel3}
\end{figure}

Start with a double filtration above, but now we assume that ${V}_j^* ={V}_i = {V}_k$. In the chamber ${\fC}$ corresponding to $y_i<y_j<y_k$, there are two asymptotic regimes: one where $y_i-y_j \ll y_j - y_k$ and the other where $y_j-y_k \ll y_i - y_j$. While we remain in a single chamber ${\fC}$, the two asymptotic regions give two different filtrations, the first corresponds to being near the wall where $y_i=y_j$, the second to being near the wall where $y_j = y_k$. The two filtrations come from different asymptotics of conformal blocks:
\beq\label{doublea}
{\cal Z}_{\ell, m} \sim (a_i-a_j)^{\Delta_i+\Delta_j - \Delta_\ell} \;  (a_j-a_k)^{\Delta_\ell+\Delta_k - \Delta_m} \times {\rm finite}
\eeq
in the regime with $y_i-y_j \ll y_j - y_k$  and 
\beq\label{doubleb}
{\cal Z}''_{m'', \ell''} \sim (a_j-a_k)^{\Delta_j+\Delta_k - \Delta_{m''}} \;  (a_i-a_j)^{\Delta_{m''}+\Delta_i - \Delta_{\ell''}} \times {\rm finite}
\eeq
in the regime with $y_j-y_k \ll y_i - y_j$.

Relating the asymptotic solutions of the KZ equation in the two regimes
%$y_i-y_j \ll y_j - y_k$, corresponding to \eqref{doublea}, and $y_j-y_k \ll y_i - y_j$ corresponding to \eqref{doubleb}, 
is a connection matrix which may be obtained as a product of two fusion matrices of the $\Lfgh_{\kappa}$ WZW model. One fusion matrix, which we will denote by $F_{ij}$, maps the space of conformal blocks with $y_i \ll y_j \ll y_k$ to $y_i-y_j \ll y_j - y_k$, and corresponding to fusing $\Phi_{V_{i}}(a_{i})$ to $\Phi_{{V}_{i}^*}(a_{j})$. The second fusion matrix gives the map $F_{jk}$ from the space of conformal blocks with $y_i \ll y_j \ll y_k$ to  that with $y_j-y_k \ll y_i - y_j$, and corresponds to fusing $\Phi_{{V}_{i}^*}(a_{j})$ to $\Phi_{{V}_{i}}(a_{k})$. The connection matrix, relating solutions in the first regime, the one with $y_i-y_j \ll y_j - y_k$, to the second, with $y_j-y_k \ll y_i - y_j$, is given by $F_{ij}^{-1}F_{jk}$.

The $S$ diagram on the left hand side of figure \ref{f_rel3} is a special matrix element of this matrix, the one that maps the space of conformal blocks where $\Phi_{{V}_i}(a_i)$ and $\Phi_{{V}_i^*}(a_j)$ fuse to the identity, for $y_i-y_j \ll y_j - y_k$, to the space of conformal blocks
where  $\Phi_{{V}_i^*}(a_j)$ and $\Phi_{{V}_i}(a_k)$ fuse to the identity, for  $y_j-y_k \ll y_i - y_j$. It is an elementary exercise to show that this matrix element of the product of two fusion matrices is identity - this is just the affine Lie algebra expression of the relation in the figure \ref{f_rel3}, as explained in \cite{MS, RCFT}.

The affine Lie algebra identity may be understood as consequence of the following categorical statement, interpreting the conformal blocks as brane central charges. Let ${\cal X}_{n-2}$ be obtained from ${\cal X}_n$ by omitting  $\Phi_{{V}_i}(a_{i})$ and $\Phi_{{V}_i*}(a_{i+1})$, and ${\cal X}'_{n-2}$ by omitting $\Phi_{{V}_i^*}(a_{i+1})$ and $\Phi_{{V}_i}(a_{i+2})$, where $a_{i+1} = a_j$ and $a_{i+2} = a_k$. On ${\cal X}_{n-2}\times {\cal X}_n$ and on ${\cal X}_n \times {\cal X}'_{n-2}$ we get a pair of sheaves ${\cal C}_{i}$ and ${\cal C}_{i}'$, constructed as in section \ref{s-cap}: We start with a pair of holomorphic Lagrangians $C_i$ and $C_i'$:
$$C_{i} \subset {\cal X}_{n-2} \times {\cal X}_n, \qquad C_{i}' \subset {\cal X}_n \times  {\cal X}'_{n-2}.
$$  
$C_{i}$
is given by identifying points 
$$ \{(L_1, \ldots, L_{i-1} ,L_{i} = L_{i+2}, L_{i+3},  \ldots ,L_n) \in {\cal X}_{n-2} \},$$
which is all of  ${\cal X}_{n-2}$,
with points 
$$
\{( L_1, \ldots L_{i} , L_{i+1},  L_{i+2}, L_{i+3}, \ldots, L_n)\in {\cal X}_n|L_{i} = L_{i+2}\},
$$
in ${\cal X}_n$. 
 $C_{i}'$
is given by identifying points 
$$ \{(L_1, \ldots ,L_{i} , L_{i+1} = L_{i+3},  \ldots ,L_n) \in {\cal X}'_{n-2} \},$$
which is all of  ${\cal X}'_{n-2}$,
with points 
$$
\{(L_1,  \ldots L_{i} , L_{i+1},  L_{i+2}, L_{i+3}, \ldots ,L_n)\in {\cal X}_n|L_{i+1} = L_{i+3}\},
$$
in ${\cal X}_n$. Each of these gives a pair of corresponding functors ${\mathscr C}_i$ and ${\mathscr C}_i^{\vee}$ and ${\mathscr C}_i'$ and ${\mathscr C}_i'^{\vee}$, as in section \ref{s-cf}.

The composition  of corresponding functors ${\mathscr C}_i^{\vee'} \circ {\mathscr C}_{i}$ gives a functor from ${\mathscr D}_{{\cal X}_{n-2}}$ to
${\mathscr D}_{{\cal X}'_{n-2}}$,
\beq\label{ccomp}
{\mathscr C}_i^{\vee'} \circ {\mathscr C}_{i}: {\mathscr D}_{{\cal X}_{n-2}} \rightarrow {\mathscr D}_{{\cal X}'_{n-2}}
\eeq  
The corresponding Fourier Mukai kernel is an object of  ${\mathscr D}_{{\cal X}_{n-2} \times {\cal X}'_{n-2} }$ given as follows (this is analogous to constructions in \cite{CK1, CK2}). Let $\pi_{12}$ be a projection of ${{\cal X}_{n-2}} \times {{\cal X}_n} \times  {{\cal X}'_{n-2}}$ to the first two factors, i.e. to ${{\cal X}_{n-2}} \times {{\cal X}_n} $, $\pi_{23}$ the projections to the last two factors, and $\pi_{13}$ the projection to the first and the last factor. Then, the Fourier Mukai kernel of the functor  ${{\mathscr C}'}_i^{\vee}\circ {\mathscr C}_{i}$ is
$$
\pi_{13 *}(\pi_{12}^*({\cal C}_{i}) \otimes  \pi_{23}^*( {\cal C}_{i}'))
$$
The pullback $\pi_{12}^*$ reinterprets the sheaf on the first two factors as the sheaf on the triple product. So $\pi_{12}^*({\cal C}_{i} )= {\cal O}_{C_{i}}$ is the structure sheaf of $C_{i}$ on the triple product, and similarly  $\pi_{23}^*({\cal C}_{i}')= {\cal O}_{C_{i}'}$. Since $C_{i}$ and $C_{i}'$ are smooth subspaces of ${{\cal X}_{n-2}} \times {{\cal X}_n} \times  {{\cal X}'_{n-2}}$ intersecting transversely, 
$$
{\cal O}_{C_{i}} \otimes {\cal O}_{C_{i}'} = {\cal O}_{C_{i} \cap C_{i}'}
$$
by Lemma 5.5 in \cite{CK1}. It is not difficult to see that the projections of $\pi_{1}$ and $\pi_3$ map $C_i \cap C_i'$ identically back to  ${{\cal X}_{n-2}}$ and to ${\cal X}_{n-2}'$, which are moreover isomorphic, ${\cal X}_{n-2} = {\cal X}_{n-2}'$. Therefore 
$\pi_{13*}( {\cal O}_{C_{i} \cap C_{i}'})
$ is the structure sheaf of the diagonal $\Delta \subset {\cal X}_{n-2}  \times {\cal X}_{n-2}'$,
$$
\pi_{13*}( {\cal O}_{C_{i} \cap C_{i}'}) = {\cal O}_{{\Delta}}.
$$
It follows that the composition of the two functors is the identity,
$${{\mathscr C}'}_i^{\vee} \circ {\mathscr C}_{i} \cong  id,
$$
which is what the S-move states.
\subsubsection{}

\begin{figure}[!hbtp]
  \centering
   \includegraphics[scale=0.33]{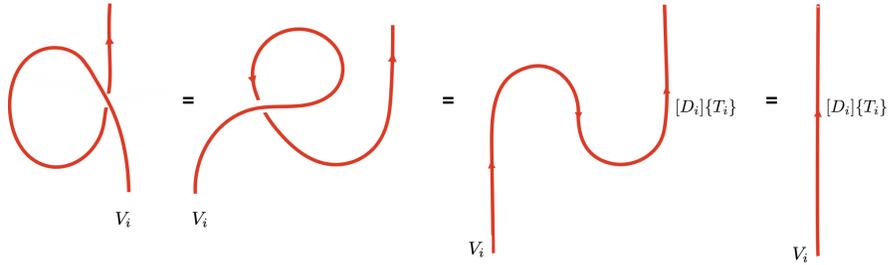}
 \caption{Reidermeister I move relates the initial and the final tangles.}
  \label{f_movie}
\end{figure}

The final move we need to prove theorem \ref{t:four} is the framed Reidermeister I move, relating the utmost left to the utmost right of the figure \ref{f_movie}.

Chern-Simons link invariants are invariants of framed links, and for a framed tangle, Reidermeister I move is not identity.  Recall that a framed knot is a ribbon, obtained by considering a knot together with a normal vector field. Implicitly, all of our diagrams are in "vertical" framing, where at each point on the knot the vector field points out of the plane of the paper.  The leftmost and the rightmost tangles in figure \ref{f_movie}, regarded as ribbons in vertical framing, differ by one full twist, which corresponds to one unit of framing. 

The Reidermeister I move relating the utmost left to the utmost right of the figure \ref{f_movie}, can be broken up into a sequence of the three moves in between. The first move is the pitch-fork move, the last is the S-move and they both preserve the framing. The change of framing occurs in the second move.

\subsubsection{}
The second move, which we isolated in the figure \ref{f_frame} below, corresponds to exchanging a pair of vertex operators
$\Phi_{{V}_i}(a_{2i-1})$ and $\Phi_{{V}_i^*}(a_{2i})$.  Geometrically, doing so corresponds to crossing a wall in the Kahler moduli which takes ${\cal X} = {\cal X}_{{\vec \mu}}$ to ${\cal X}' = {\cal X}_{{\vec \mu}'}$,  where ${\vec \mu}$ and ${\vec \mu}'$ differ by exchanging a pair of neighboring entries corresponding to $a_{2i-1}$ and $a_{2i}$. 
\begin{figure}[!hbtp]
  \centering
   \includegraphics[scale=0.37]{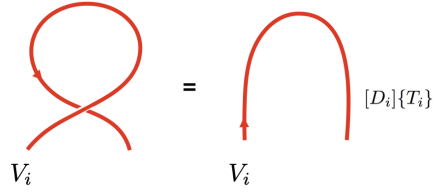}
 \caption{Removing a twist from a ribbon changes the framing.}
  \label{f_frame}
\end{figure}

The derived categories ${\mathscr D}_{{\cal X}}$ and ${\mathscr D}_{{\cal X}'}$ of ${\cal X}$ and ${\cal X}'$ each have a filtration by the order of vanishing of central charge at the wall, of the form \eqref{filtration0a} and \eqref{filtration0b}, where the different terms in the filtration correspond to representations in the tensor product $
{V}_i \otimes {V}_i^*$.  The derived equivalence functor
$${\mathscr B}_{i} : {\mathscr D}_{{\cal X}} \cong {\mathscr D}_{{\cal X}'}
$$
corresponding to braiding $a_{2i}$ and $a_{2i+1}$ clockwise preserves the filtrations and acts by a shift of grading in \eqref{gs}. For the second move in figure \ref{f_movie}, or equivalently for the move in figure \ref{f_frame}, we need to know how the functor ${\mathscr B}_i$ acts on the bottom part of the filtration corresponding to bringing $a_{2i}$ and $a_{2i+1}$ together. The functor maps bottom part ${\mathscr D}_0$ of the filtration on ${\mathscr D}_{{\cal X}}$,  to bottom part of the filtration ${\mathscr D}_0'$ on ${\mathscr D}_{{\cal X}}'$, mapping an object ${\cal F} \in {\mathscr D}_0$ to the object $ {\cal F}' \in {\mathscr D}_0'$, where
\beq\label{twist}
 {\cal F}' = {\mathscr B}_i{\cal F}  = {\cal F}[-D_i]\{C_i\}.
\eeq
The degree shifts in \eqref{twist} can be read off from \eqref{vana}, as in 
section \ref{s-three}.
The shift $D_i$ in cohomological grade is the dimension of the vanishing cycle $U_i = G/P_i$
\beq\label{Ds} D_i = {\rm dim}[U_i] = 2 \langle \mu_i, \rho\rangle .
\eeq
The equivariant shift is by
\beq\label{Ts}
C_i =  c_i+c_{i*} =  \langle \mu_i, \mu_i +2 \,{^L\!\rho} \rangle.
\eeq
corresponding to the ${\mathbb C}^{\times}$ action that scales the symplectic form; the other ${\rm T}$-grading shifts are trivial.
Above, we used that, for conjugate representations $V_i$, $V_i^*$, their classical and conformal dimensions are equal, $d_i=d_i^*$ and $c_i=c_i^*$.
%\begin{figure}[!hbtp]
 % \centering
 %  \includegraphics[scale=0.1]{framing.pdf}
 %\caption{}
 % \label{f_frame}
%end{figure}
%\begin{figure}[!hbtp]
  %\centering
  % \includegraphics[scale=0.3]{twistb.pdf}
 %\caption{Applied to a ribbon, Reidermeister I the framing by one unit (drawing curtesy of \cite{integrable}.)}
  %\label{f_twist}
%\end{figure}
%\subsubsection{}

The geometric action on the objects of ${\mathscr D}_{\cal X}$ by \eqref{twist} implies that, once we take the Euler characteristic \eqref{chiJ} and use the equation \eqref{J2}, adding a twist to a cap colored by $V_i$ changes the corresponding quantum link invariant by
$$
J_K(\fq) \rightarrow (-1)^{D_i} {\fq}^{C_i\over 2}  J_K(\fq)
$$
This is the change of framing of a strand colored by $V_i$ by one unit, as we reviewed in section \ref{s-one}. In fact, the signs introduced due to \eqref{Ds} are fairly subtle to fix in the decategorified theory and for this reason, they are often neglected, see for example \cite{RCFT}. Here we have no choice, the homological degree shifts are fixed from geometry.

\newpage
\appendix

\section{Affine Grassmannians and Monopole Moduli Spaces}\label{s-B}
In this section we collect some conventions and results about affine Grassmannians, and their connection to monopole moduli spaces.

\subsection{Singular monopole moduli spaces}

We will start by briefly reviewing why slices in affine Grassmannian describe the holomorphic structure of the moduli space of singular monopoles on ${\mathbb R}^3$, following \cite{KW}.

\subsubsection{Monopole equations}
Moduli space of monopoles on  ${\mathbb R}^3$ is the space of solutions Bogomolny equations
$$
F=*D\Psi,
$$
where $D$ is the connection, $\Psi$ a real scalar field and $F$ is the curvature; all are ${\fg}$-valued.  

Choose a splitting of ${\mathbb R}^3 ={\mathbb R} \times {\mathbb C}$. For us, ${\mathbb R}$ is naturally identified with the radial direction of ${\cal A}$, and ${\mathbb C}$ with the complex plane rotated by ${\fq}$.  Let $z$ be the complex coordinate on ${\mathbb C}$, and $y$ a real coordinate on ${\mathbb R}$.
The Bogomolny equations
imply that 
\beq\label{gauge}
[{{\mathscr D}}_{\overline z}, {{\mathscr D}}_y] =0,
\eeq
where ${{\mathscr D}}_{\overline z} = \partial_{\overline z} + A_{\overline z}$, and
${{\mathscr D}}_{y} =\partial_y  + A_y + i \Psi$. 
We  view $A_{\overline z}(y)$ as a family of holomorphic connections on ${\mathbb C}$, parameterized by $y$.  The equation \eqref{gauge} says that ${{\mathscr D}}_y A_{\overline{z}}(y)$ is a gauge transformation generated by $\Psi$. So, as long as $\Psi$ is well defined (which means away from the points on the $y$ axis where the singular monopoles are located), the holomorphic type of the bundle is $y$-independent. 

\subsubsection{Singular monopoles}
A singular Dirac monopole at ${\vec y} = {\vec y}_i$ is a solution where $F=  *d \bigl( {i \over 2}{1\over |{\vec y} - {\vec y}_i |}\bigr) {t}_{\mu_i}$, $\Psi =\bigl( {1\over 2}{1\over |{\vec y} - {\vec y}_i |}\bigr) { t}_{\mu_i}$, where ${\vec y}$ is a coordinate on ${\mathbb R}^3$ and ${ t}_{\mu_i}: {\mathfrak u}(1) \rightarrow {\mathfrak h}$ comes from a homomorphism $T_{{\mu}_i}: U(1) \rightarrow H$, were $H$ is the maximal torus of $G$. Such homomorphisms are classified by dominant co-weights $\mu_i$ of $G$. Having chosen the split ${\mathbb R}^3 ={\mathbb R} \times {\mathbb C}$, this implies that at the location of the singular monopole
the holomorphic type of the bundle can jump. For us, all such monopoles are at the origin of ${\mathbb C}$, and we fix their positions on $y=y_i$ on ${\mathbb R}$.

For simplicity, take first all the singular monopoles to coincide, so we get a single singular monopole of charge $\mu = \sum_i \mu_i$ at $y=0$.  Let $L_{-}$ and $L_+$ be the holomorphic $G$-bundles on ${\mathbb C}$ for $y<0$ and $y>0$. Then, the connection corresponding to $L_+$ is obtained from the one corresponding to $L_-$ by a singular gauge transformation $g(z)$. 

A-priori, the possible choices for $g(z)$ are any Laurent polynomial in $z$, corresponding to gauge transformations which may have poles at $z=0$ and $z=\infty$, modulo gauge transformations that are trivial, i.e. holomorphic, on ${\mathbb C}$. 
This implies all possible gauge transformations from $L_-$ to $L_+$  are elements of the affine Grassmannian of $G$, 
\beq\label{Grg}
{\rm Gr}_G := G((z))/G[[z]].
\eeq
Here, $G[[z]]=G({\mathbb C}[[z]])$ and $G((z))=G({\mathbb C}((z)))$, where ${\mathbb C}[[z]]$ and ${\mathbb C}((z))$ are the rings of formal Taylor and Laurent series, respectively. Since any two $G[[z]]$ orbits in $G((z))$ differ by a finite number of terms, we can equivalently write 
${\rm Gr}_G= G[z, z^{-1}]/G[z].$

By a choice of gauge, we can take the bundle on $L_-$ to be trivial.  
Then, the bundle on $L_+$ has to have the singularity at $z=0$ which reflects the singular monopole charge. This means that possible $g(z)$ correspond to the orbit of $z^{-\mu}$ under left gauge transformations which are holomorphic at $z=0$, 
\beq\label{nearz}
g(z) \in  {\rm Gr}^{\mu}:= G[[z]] z^{-\mu}.
\eeq
${\rm Gr}^{\mu}$ is smooth and finite dimensional. Its dimension, viewed as orbit of $G[[z]]$ inside $G((z))$, is $2\langle \mu, ^L\rho\rangle.$  

\subsubsection{Monopole bubbling}
Unless $\mu$ is a minuscule co-weight, ${\rm Gr}^{\mu}$  is not compact. The source of non-compactness is monopole bubbling. One can compactify ${\rm Gr}^{\mu}$ by adding lower dimensional orbits. The orbit closure, which we denoted by ${\rm Gr}^{\mu^{\times}}$ in section \ref{s-one} is 
\beq\label{orbc}
{\rm Gr}^{\mu^{\times}} = \cup_{\nu \leq \mu} {\rm Gr}^{\nu},
\eeq
where $\nu$'s run over dominant co-weights of ${\fg}$, and $\nu < \mu$ means that $\mu-\nu$ is a sum of positive simple co-roots of ${\fg}$. The component ${\rm Gr}^{\nu}$ corresponds to a locus where, exactly $\mu-\nu$ smooth monopoles bubble off the singular monopole and disappear. 

\subsubsection{Fixing the total monopole charge}
Monopole bubbling changes the total monopole charge $\nu$. Rather than let it vary, in our problem, we want to keep $\nu$ fixed. 
The total the total monopole charge ${\nu}$ affects the behavior of $F$ and $\Psi$ at infinity. 

Let $G_1[z^{-1}]$ denote the subgroup of $G((z))$ consisting of elements that approach $1$ at $z\rightarrow \infty$, and 
$$Gr_{\nu} := G_1[z^{-1}]z^{-\nu},
 $$
the subgroup of elements that approach $z^{-\nu}$ at  $z\rightarrow \infty$. We get the moduli space of monopoles with a singular charge $\mu$ monopole at the origin of ${\mathbb R}^3$ and total monopole charge $\nu$ as the intersection
$${\cal X}^{\times}: = {\rm Gr}^{\mu^{\times}} \cap Gr_{\nu} = {{\rm Gr}^{{\mu}^{\times}}}_{\nu}.
$$
\subsubsection{Transversal slices}
Geometrically, ${\cal X}^{\times}$ is the transverse slice to ${\rm Gr}^\nu$ orbit inside ${\rm Gr}^{\nu^\times}$.  First, following \cite{Danilenko}, note that 
$G_1[z^{-1}]$ and $G[[z]]$ are transversal slices to $G((z))$ at the identity:  the tangent space to $G((z))$ at the identity, splits into 
$${\fg}((z)) = {\fg}[[z]] \oplus z^{-1} {\fg}[z^{-1}],
$$ 
which are the corresponding tangent spaces to $G[[z]]$ and to $G_1[z^{-1}]$. 
Similarly, the tangent space to $G((z))$ at $z^{-\nu}$ splits into the tangent space to ${\rm Gr}^{\nu} = G[[z]]z^{-\nu}$ and to 
 $$Gr_{\nu} = G_1[z^{-1}]z^{-\nu}.
 $$
Correspondingly, ${\rm Gr}_{\nu}$ intersects the orbit ${\rm Gr}^{\nu}$ in ${\rm Gr}^{{\mu}^{\times}}$ at
the single point $z^{-\nu}$. It follows  ${\cal X}^{\times}={{\rm Gr}^{\,{\mu}^\times}}_{\nu}$  is the transversal slice to the ${\rm Gr}^{\nu}$ orbit inside of ${{\rm Gr}^{\,{\mu}^\times}}$.

\subsection{Crystallography of affine Grassmannians}

We get a useful, ''crystallographic'' image of the geometry of ${\cal X}={{\rm Gr}^{{\vec\mu}}}_{\nu} $, by tracking fixed points of the ${\rm T}$-action on it. 
When ${\cal X}$ is smooth, the fixed points of the ${\rm T}$-action are isolated. This is the case for us for generic $y_i$, because we are assuming that the representations $V_i$ of $^L\fg$ are minuscule. 

The fixed points have a simple description. 
They are labeled by collections of weights ${\vec {\nu}} = (\nu_1, \ldots, \nu_i, \ldots, \nu_n)$, where $\nu_i$ is a weight in representation $V_i$, such that $\sum_i \nu_i = \nu$. Since ${V}_i$ is minuscule, every weight ${\nu}_i$ has multiplicity $1$.
The corresponding point in ${\cal X}$ comes from a point in the convolution Grassmannian ${\rm Gr}^{\vec \mu}$ given by the collection $(L_1, \ldots, L_i, \ldots , L_n)\in {\rm Gr}^n $, with $L_i={z^{-{\nu}_i}}$. 

The correspondence between the fixed points of the ${\rm T}$-action on ${\cal X}$ and weights of representation $V  = \bigotimes_i {V}_i$ that sum up to $\nu$ is a reflection of the geometric Satake correspondence in \eqref{GS}.
%Keeping track of the equivariant ${\rm T}$-action on $H^*({\cal X})$,  $H_{\rm T}^*({\cal X}_L)$ gets identified with ${\rm Hom}_{^L{\fg}}(\rho_{\lambda}, {\otimes_{i\in L}{V}_i }\otimes \,\rho_{\lambda-\nu_L}) $ and  $H_{\rm T}^*({\cal X}_R)$ with  ${\rm Hom}_{^L{\fg}}(\rho_{\lambda-\nu_L}, {\otimes_{i \in R} {V}_i} \otimes \,\rho_{\lambda-\nu}) .$ This tells us how equivariant actions on ${\cal X}_{L, R}$ match up. }

\subsubsection{Separating the monopoles}
Consider now the separating the monopoles into two groups, monopoles in $y< y_*$ and those in $y> y_*$. There are many different ways to do that depending on choices of splitting of $\vec\mu$ into a concatenation of ${\vec \mu}_L$ and ${\vec \mu}_R$, and $\nu$ into the sum $\nu = \nu_L+\nu_R$ with $0\leq \nu_{L, R} \leq \mu_{L,R}.$
A given choice of splitting corresponds to a local neighborhood in ${\cal X}$ of the form:
\beq\label{decompose}
 {\cal X} \;\supset \; {\cal X}_L\; {{ \times}}\;  {\cal X}_R =   {{\rm Gr}^{{\vec \mu}_L}}_{\nu_L}\; {{ \times}} \; {{\rm Gr}^{{\vec \mu}_R}}_{\nu_R}
\eeq
The region described by ${\cal X}_{L}{ { \times}} {\cal X}_R$
corresponds to widely separated monopoles, contained in the left and the right halves of ${\mathbb R} \times {\mathbb C}$, where the singular and total monopole charges split into $({\vec \mu}_L, \nu_L)$ and $({\vec \mu}_R, \nu_R)$.  ${\cal X}_{L} { { \times}} {\cal X}_R$ is an open neighborhood of ${\cal X}$ since their dimensions are the same (the complex dimension of ${{\rm Gr}^{{\vec \mu}}}_{\nu}$ is $\langle \mu - \nu, 2\rho\rangle$, and the weights add).
The neighborhood contains a subset of all the fixed points of the ${\rm T}$-action on ${\cal X}$, corresponding to a choice of a fixed point of ${\rm T}$-action on ${\cal X}_L$ and a fixed point in ${\cal X}_R$. Clearly, we can iterate this any number of  times.

\subsection{Finding eigensheaves} 

Consider bringing singular monopoles of charges $\mu_i$ and $\mu_j$ close together. As before, $\mu_i$ and ${\mu_j}$ are highest weight vectors of representations $V_i$ and $V_j$ coloring the vertex operators $\Phi_{V_i}(a_i)$ and $\Phi_{V_j}(a_j)$ corresponding to these monopoles. 
We would like to find an eigensheaf of braiding $\Phi_{V_i}(a_i)$ and $\Phi_{V_j}(a_j)$  corresponding to a representation $V_k$ in the tensor product $V_i\otimes V_j$. We raised the question of existence of such eigensheaves in section \ref{integrable}.

Suppose that we can iterate the decomposition of previous section to get a local neighborhood of ${\cal X}  =  {{\rm Gr}^{{\vec \mu}}}_{\nu}$ of the form
\beq\label{sing}
{\cal X}  \;\supset \; {{\rm Gr}^{\,{\vec \mu}_L}}_{\nu_{L}}\;\; { \times} \;\; {\rm Gr}^{ ({\mu}_i, {\mu}_j)}_{{\mu}_k} \;\;  { \times}\;\;  {{\rm Gr}^{\,{\vec \mu}_R}}_{\nu_{R}}.
\eeq
We recover ${\vec\mu}$ by concatenating ${\vec \mu}_L$, ${\vec \mu}_{ij} = ({\mu}_i, {\mu}_j)$ and ${\vec \mu}_R$, and we have that $\nu = \nu_L + {\mu}_k+ \nu_R.$
The first factor in \eqref{sing} corresponds to moduli of monopoles at $y< y_L$, the second to monopoles with $y\in [y_L, y_R]$ and the third to monopoles with $y>y_R$. Here, $y_L$ and $y_R$ are two arbitrary cutoffs, satisfying $y_L\ll y_i$ and $y_j \ll y_R$.

Recall from section \ref{s-three} that $F_k$ in 
\beq\label{cotA}
 {\rm Gr}^{ ({\mu}_i, {\mu}_j)}_{{\mu}_k}  = T^* F_k,
\eeq
is the vanishing cycle corresponding to the representation $V_k$ in the tensor product $V_i\otimes V_j$.  The weight $\mu_k$ is the highest weight of $V_k$. As we explained in section \ref{integrable}, the physical central charge ${\cal Z}^0$ is the eigenvector of the $U_{\fq}(^L\fg)$ action.  Moreover, the structure sheaf ${\mathscr O}_{F_k}$ of $F_k$ in $ {\rm Gr}^{ ({\mu}_i, {\mu}_j)}_{{\mu}_k}$ is the eigensheaf of braiding that takes $ {\rm Gr}^{ ({\mu}_i, {\mu}_j)}_{{\mu}_k}$ to $ {\rm Gr}^{ ({\mu}_j, {\mu}_i)}_{{\mu}_k}$ with order of $\mu_i$ and $\mu_j$ reversed, as we showed in section \ref{s_filtration}. It follows that, whenever ${\cal X}$ has a neighborhood of the form \eqref{sing}, ${\mathscr D}_{\cal X}$ has an eigensheaf of braiding $V_i\otimes V_j$, which in the local neighborhood of ${\cal X}$ of the form \eqref{sing}, restricts to a tensor product of an arbitrary coherent sheaf on the first and the third factor in \eqref{sing} with the structure sheaf ${\mathscr O}_{F_k}$ in the middle factor.
 
The question of existence of eigensheaves of braiding that exchanges $\Phi_{V_i}(a_i)$ and $\Phi_{V_j}(a_j)$ thus gets related to the question when a decomposition of the form in \eqref{sing} exists.

Note that fixed points of the ${\rm T}$-action on \eqref{sing} restrict to a weight $\mu_k$ subspace of representation $V_i \otimes V_j$. This is because all the fixed points of the ${\rm T}$-action on $ {\rm Gr}^{ ({\mu}_i, {\mu}_j)}_{{\mu}_k}$ are of this form \cite{Danilenko}. At the same time, $\mu_k$ is the highest weight of representation $V_k$ in \eqref{tensor}. Thus, one gets an eigensheaf of braiding corresponding to $V_k$ whenever the highest weight $\mu_k$ of $V_k$ contributes to $H^*({\cal X})$, as claimed in section \ref{s-three}.

\subsection{Minuscule and affine Grassmannians}																			
									
The identification in \eqref{cap}
$$
W^i =T^* U_i, \qquad  U_i = {\rm Gr}^{{\mu}_i}\cong  G/P_i,
$$
can be seen as follows. Recall $W^i$ is a resolution of a singularity
$$
m_i : W^i \rightarrow  W^{i\times}= {\rm Gr}^{\overline{\mu_i+ \mu_i^*}}_0.$$
The singularity contains a single torus fixed point, and the vanishing cycle is the fiber over it $U_i = m_i^{-1}(z^0)$.  The map $m_i$ projects ${\rm Gr}^{(\mu_i, \mu_i^*)}$ to  $ {\rm Gr}^{\overline{\mu_i+\mu_i^*}}$, by sending the convolution Grassmannian in \eqref{geomG} to its last term, and then intersecting it with the orbit through $z^0$. This means that 
$U_i$ is the following set:
$$
U_i = \{ (L_1, L_2) \in {\rm Gr}^2\,| \; L_0 \stackrel{{ \mu}_i}{ \longrightarrow} {L}_1\stackrel{{ \mu}_i^*}{ \longrightarrow} L_2,  {\textup{ such that }} L_2 = z^0\}.
$$
Thus $U_i$ is the space of two Hecke modifications, the first one of which is parameterized by points in ${\rm Gr}^{{{\mu}_i}^\times}$, the closure of $G[z] z^{{-\mu}_i}$ orbit.  Since ${\mu}_i$ is a minuscule weight ${\rm Gr}^{{\mu}_i}$ is already closed, so  ${\rm Gr}^{{{\mu}_i}^{\times}} ={\rm Gr}^{{\mu}_i}.$ For every point on ${\rm Gr}^{{{{\mu}_i}}}$ the second Hecke modification is completely fixed - it simply maps $L_1$ back to identity $L_2=z^0$. 
It follows that $U_i = {\rm Gr}^{{{{\mu}_i}}}$. The rest of the identification, the fact that ${\rm Gr}^{{{{\mu}_i}}} \cong G/P_i$ is, e.g, the lemma 2.1.13 in \cite{Zhu}. $G/P_{i}$ is known as the minuscule Grassmannian. 	

\subsection{Conventions}

Our conventions follow \cite{KW} and differ from \cite{yangian} by an overall complex conjugation, which is a symmetry. In \cite{yangian}, one defines ${\widetilde{\rm Gr}}^{\mu} := G[[z]]z^{\mu}$ and ${\widetilde{\rm Gr}}_{\nu}: = G[z^{-1}] z^{w_0 \nu}$, where $w_0$ is the longest element of the Weyl group, and $\mu$ and $\nu$ are dominant co-weights, and weights of $^L\fg$. Since for a dominant weight $\mu$,  its conjugate co-weight is $\mu^* = -w_0 \mu$, the notation in \cite{yangian} and ours is related by ${\widetilde{\rm Gr}}^{\mu}= G[[z]]z^{\mu} = G[[z]]z^{w_0 \mu} =  G[[z]]z^{- \mu^*} ={\rm Gr}^{\,\mu^*}$. In the second step we used the invariance of  ${\rm Gr}^{\mu}$ under the Weyl group action. In the same vain, we also have
  ${\widetilde{\rm Gr}}_{\nu} = G[z^{-1}] z^{w_0 \nu} =G[z^{-1}] z^{- \nu^*} ={\rm Gr}_{\nu^*}  $.

\end{document}